\documentclass[%
 floatfix,
 reprint,
groupedaddress,
nobibnotes,
 amsmath,amssymb,
 aps,
prb,
]{revtex4-2}

\usepackage{graphicx}
\usepackage{graphicx}
\usepackage{dcolumn}
\usepackage{bm}
\usepackage[T1]{fontenc}
\usepackage{natbib}

\usepackage{fancyhdr}
\usepackage{tikz}
\usepackage[export]{adjustbox}
\usepackage{amssymb}
\usepackage{xpatch}
\usepackage{hyperref}
\usepackage{color}
\usepackage{blindtext}
\usepackage{siunitx}
\DeclareSIUnit\bar{bar}
\DeclareSIUnit\day{d}
\usepackage{tcolorbox}
\usepackage{enumitem}
\usepackage{subcaption}
\usepackage{graphicx}
\usepackage{blindtext}
\usepackage{amsmath,bm}
\usepackage{lipsum}
\usepackage{float}
\usetikzlibrary{calc,patterns,decorations.pathmorphing,decorations.markings}   

\definecolor{KITgreen}{RGB}{0,150,130}
\definecolor{KITblue}{RGB}{70,100,170}
\definecolor{KITpurple}{RGB}{163,16,124}
\definecolor{KITred}{RGB}{162,34,35}
\definecolor{KITorange}{RGB}{223,155,27}
\definecolor{KITmaygreen}{RGB}{140,182,60}
\definecolor{Gray70}{RGB}{77,77,77}
\definecolor{Gray50}{RGB}{128,128,128}

\usepackage{pgfplots}
\pgfplotsset{compat=1.17}
\usepackage{tikz}
\usepackage{printlen}
\usepackage{stackengine}
\usepackage{soul}

\begin{document}

\title{Cryogenic payloads for the Einstein Telescope -- Baseline design with heat extraction, suspension thermal noise modelling and sensitivity analyses}

\def\andname{}
\author{Xhesika Koroveshi,${^{1,2}}$}
\email{xhesika.koroveshi@kit.edu}
\author{Lennard Busch,$^1$ Ettore Majorana,$^{3,4}$ Paola Puppo,$^{3}$ Piero Rapagnani,$^{3,4}$ Fulvio Ricci,$^{3,4}$ Paolo Ruggi,$^{5}$ and Steffen Grohmann$^{1,2}$}

\address{$^1$Institute of Technical Thermodynamics and Refrigeration - Organizational Unit: Refrigeration and Cryogenics, Karlsruhe Institute of Technology, 76131 Karlsruhe, Germany}
\address{$^2$Institute of Beam Physics and Technology, Karlsruhe Institute of Technology, 76344 Eggenstein-Leopoldshafen, Germany}
\address{$^{3}$INFN, Sezione di Roma,  I-00185 Roma, Italy}
\address{    $^{4}$Dipartimento di Fisica, Universit\`a degli studi di Roma "La Sapienza", I-00185 Roma, Italy}
\address{$^{5}$European Gravitational Observatory, I-56021 Cascina, Italy}

\date{August 11, 2023}

\begin{abstract}
The Einstein Telescope (ET) is a third generation gravitational wave detector that includes a room-temperature high-frequency (ET-HF) and a cryogenic low-frequency laser interferometer (ET-LF).
The cryogenic ET-LF is crucial for exploiting the full scientific potential of ET.
We present a new baseline design for the cryogenic payload that is thermally and mechanically consistent and compatible with the design sensitivity curve of ET.
The design includes two options for the heat extraction from the marionette, based on a monocrystalline high-conductivity marionette suspension fiber and a thin-wall titanium tube filled with static He-II, respectively. 
Following a detailed description of the design options and the suspension thermal noise (STN) modelling, we present the sensitivity curves of the two baseline designs, discuss the influence of various design parameters on the sensitivity of ET-LF and conclude with an outlook to future R\&D activities. 
\end{abstract}

\maketitle

\section{Introduction}

\begin{figure}[t]
\includegraphics[width=7.5cm, height=7cm]{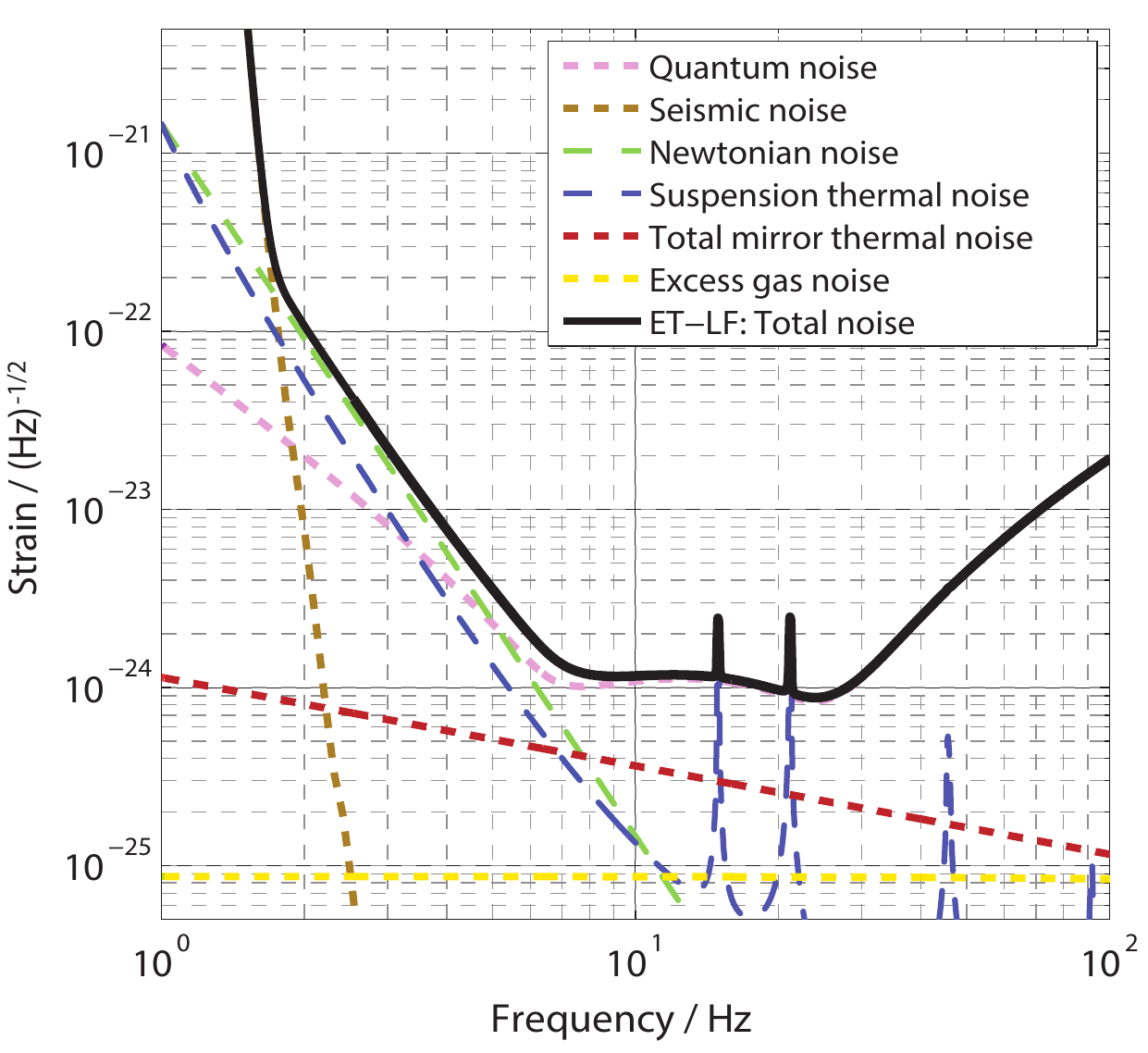}
\caption{\label{fig:Noises}ET-LF noise contributions in the ET-D sensitivity curve \cite{ET2011}.}
\end{figure}

The Einstein Telescope (ET) is a third generation gravitational wave (GW) detector with a xylophone design, combining a low-frequency (LF) and a high-frequency (HF) laser interferometer.
Sensitivities lie in the range of \SIrange[range-phrase={\text{~to~}}]{3}{30}{\hertz} (ET-LF) and \SI{30}{\hertz} to \SI{10}{\kilo\hertz} (ET-HF), respectively. 
The low-frequency sensitivity is crucial for exploiting the full scientific potential of ET, in particular with regard to:
\begin{itemize}
    \item the observation of binary neutron stars (BNS), staying long time in the bandwidth,
    \item pre-merger detection to probe the central engine of gamma ray bursts (GRB), particularly to understand the jet composition, the particle acceleration mechanism, the radiation and energy dissipation mechanisms,
    \item detecting a large number of kilonovae counterparts,
    \item detecting primordial black holes (PBH) at redshifts $z>30$, and
    \item detecting intermediate massive back holes (IMBH) in the range of $\num{e2}-\num{e4}\,M_\odot$ \cite{ETCoba}.
\end{itemize}
\hyperref[fig:Noises]{Figure~\ref*{fig:Noises}} shows the noise contributions to the sensitivity curve ET-D \cite{ET2011}, based on payload design parameters listed in \hyperref[tab:table1]{Table~\ref*{tab:table1}}.
Cryogenic operation of the payload is indispensable to suppress the suspension thermal noise (STN) to the level of gravity gradients, i.e.\ Newtonian noise (NN).
Both STN and NN are the fundamental noises that dominate the ET-LF noise budget at frequencies below \SI{10}{\hertz}.

\begin{table} [t]
\caption{\label{tab:table1}ET-LF payload design parameters from \cite{ET2011}, using a branched pendulum model as in Virgo.}
\begin{ruledtabular}
\begin{tabular}{lccc}
&Marionette &Recoil mass &Mirror\\
\hline
Mass (kg) & 422 & 211 & 211\\
Suspension length (m) & 2 & 2 & 2\\
Suspension diameter (mm) &3 &3 & 3\\
Suspension material (-)  & Ti6Al4V & Silicon & Silicon\\
Loss angle (-) &$\num{1E-5}$ & $\num{1E-8}$ & $\num{1E-8}$\\
Temperature (K) & 2 & 10 & 10\\
\end{tabular}
\end{ruledtabular}
\end{table}

The technical implementation of the parameters in \hyperref[tab:table1]{Table~\ref*{tab:table1}} is not straightforward \cite{payload-2011,puppo2}.
Therefore, in this paper we develop a baseline design of a cryogenic payload for ET-LF, which is consistent in terms of mechanical and thermal design as well as STN modelling.
It shall serve as a stepping stone for the cryostat design and for future payload design optimization, rather than assuming it ``final''.
The focus of this paper is purely on the payload, not yet including the impact of cooling interfaces, which is a subject of future R\&D.

\hyperref[sec:Baseline]{Section~\ref*{sec:Baseline}} introduces the baseline cryogenic payload design for ET-LF with two heat extraction concepts, which are further explained in \hyperref[sec:Monolithic]{Sections~\ref*{sec:Monolithic}} and \hyperref[sec:Heconcept]{\ref*{sec:Heconcept}}.
This is followed in \hyperref[sec:STN]{Section~\ref*{sec:STN}} by a detailed description of the STN modelling.
\hyperref[sec:BaselineSensitivity]{Section~\ref*{sec:BaselineSensitivity}} then presents the sensitivity curves of the baseline designs.
The influence of various design parameters on the sensitivity of ET-LF is analyzed in \hyperref[sec:ParameterStudy]{Section~\ref*{sec:ParameterStudy}}, before main conclusions and an outlook to future R\&D activities are presented in \hyperref[sec:Conclusions]{Section~\ref*{sec:Conclusions}}.

\section{Baseline design of a cryogenic payload for ET-LF}
\label{sec:Baseline}

\subsection{Overall operating conditions}

ET-LF shall be operated with a \SI{1550}{\nano\meter} wavelength laser at an arm power of \SI{18}{\kilo\watt}.
The baseline material for the mirror and its suspension fibers is monocrystalline silicon.
An alternative material for the mirror and its suspensions is sapphire.
The operating temperature of the mirror is between \SI{10}{\kelvin} and \SI{20}{\kelvin}.
While \SI{0.1}{\watt} heat load have been estimated in \cite{ET2011,ET2020}, we re-define an engineering design target of
\begin{equation}
\label{eq:HeatLoad}
    \dot{Q} = \SI{0.5}{\watt}
\end{equation}
total heat load on the ET-LF payload, considering the size and complexity of the cryostat and including the need for optical access.
This value entails a thermal safety margin and compares to a range of $\num{0.5}-\SI{1.0}{\watt}$ that KAGRA, the first cryogenic gravitational wave detector, assumes on its cryogenic test masses of $\SI{23}{\kilo\gram}$, partially caused by a higher absorption in its sapphire mirrors \cite{Yamamoto_Elba}.

\subsection{Conceptual design of the payload}

\begin{figure}[tb]
\includegraphics[width=6cm,height=8cm,scale=1.1]{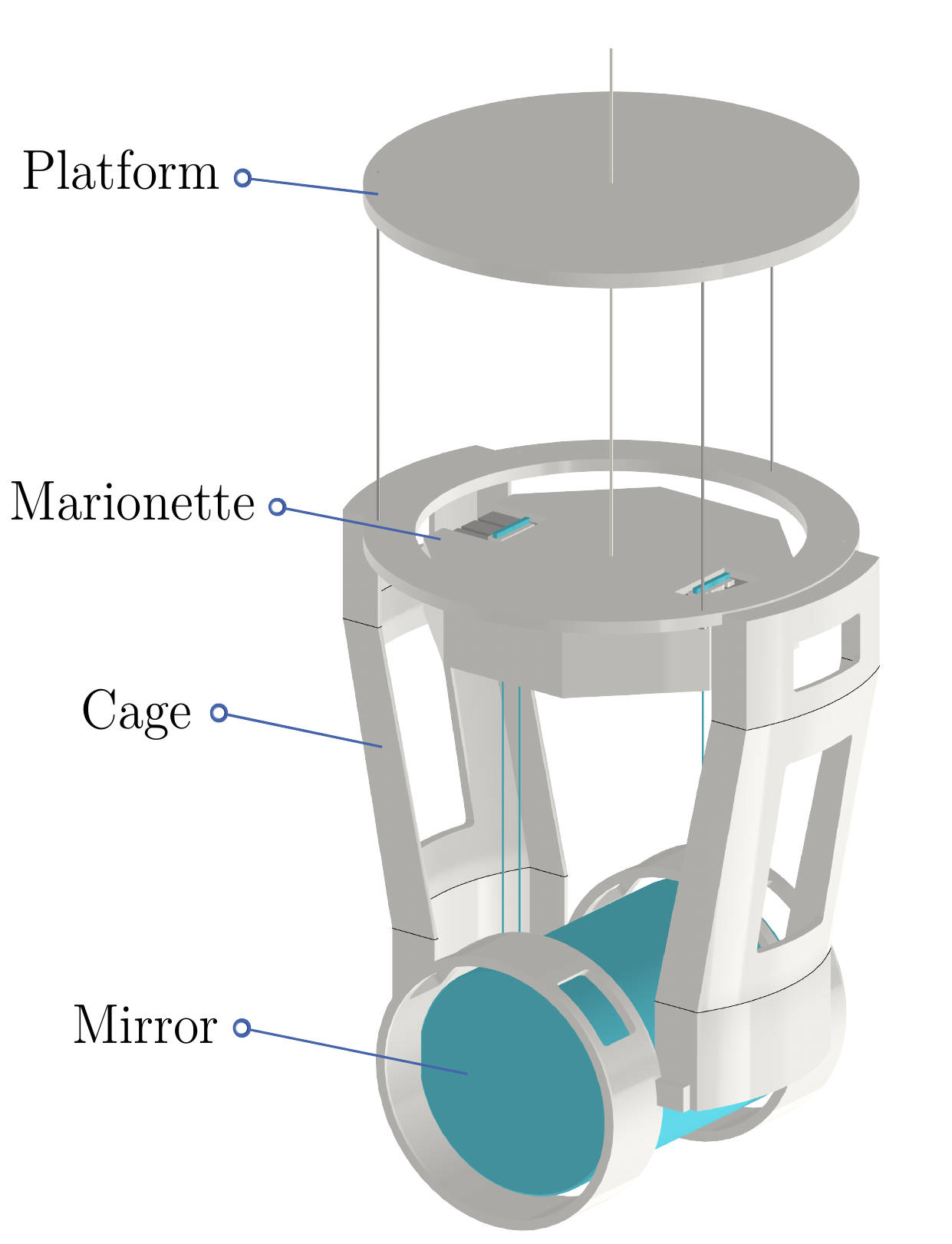}
\caption{\label{fig:payload} Baseline design of the ET-LF cryogenic payload based on the AdVirgo double pendulum design. }
\end{figure} 

\begin{table*}
\caption{\label{tab:baseline} Baseline design parameters of the ET-LF payload, including two marionette cooling concepts.}
\begin{ruledtabular}
\begin{tabular}{lccc|cc}
 &\multicolumn{3}{c|}{Marionette} &\multicolumn{2}{c}{Mirror} \\
Cooling concept  & Monolithic & Monolithic & He-II filled  & Silicon & Sapphire \\
\hline
Mass (kg) & 200 & 220 & 200 & 200 & 220 \\
Suspension length (m) & 1.0 & 1.0 & 1.0 & 1.2 & 1.2\\
Suspension diameter (mm) & 8.1 & 6.5 & 8.3 & 3.0 & 2.3 \\
Suspension material (-)  & Silicon & Sapphire & Ti, He-II & Silicon  & Sapphire \\
Bulk loss angle (-) &$\num{1E-9}$ &$\num{3E-9}$ &$\num{1E-6}$  &$\num{1E-9}$ & $\num{3E-9}$ \\
Temperature (K) & 15 &17 &2   & $15\ldots 20$ & $20\ldots 23$\\
\end{tabular}
\end{ruledtabular}
\end{table*}

The baseline design of the ET-LF cryogenic payload is derived from the double pendulum layout of the Advanced Virgo (AdVirgo) payload \cite{Naticchioni_2018}. 
It is depicted in \hyperref[fig:payload]{Fig.~\ref*{fig:payload}} and includes the following main components:
\begin{itemize}
\item\emph{Platform (PF)}, from which the marionette and the cage are suspended separately, using a single suspension for the marionette and three suspensions for the cage, respectively.
The platform is the first stage inside the ET-LF cryostat volume, being suspended from a warm super-attenuator system.

\item\emph{Marionette (MA)}, which coordinates the position of the mirror via four monocrystalline silicon suspension fibers.  
These suspensions are connected to so-called mirror ears that are attached via hydroxide catalysis bonding (HCB) onto the sides of the mirror.

\item\emph{Actuation cage (CA)}, which serves as a reaction mass for both the mirror and the marionette.  
In addition, various sensory devices are installed on this robust structure to avoid a direct contact with the sensitive optics.
In AdVirgo, the cage is rigidly attached to the PF, whereas a suspended cage is proposed here.

\item\emph{Mirror/Test mass (MI)}, which constitutes the core optical element of the interferometer.
\end{itemize}

The design of the cryogenic payload must consider thermal and mechanical feasibility, while fulfilling a low STN contribution compatible with the ET-D sensitivity curve \cite{ET2011}. 
The correct implementation of interfaces, temperature gradients and mechanical safety factors is essential. 
Further design aspects include the cryostat dimensions and space requirements for installation and auxiliary systems, the fabrication of long and high quality monocrystalline suspension fibers and the achievable marionette temperature based on the cooling concept.

The AdVirgo payload operating at room-temperature has a MA made of 316L stainless steel \cite{Naticchioni_2018}. 
In the ET-LF cryogenic payload, the material for the MA remains to be decided. 
In addition to mechanical functionality, the choice is influenced by thermal aspects, i.e.\ the transient cool-down behavior and the achievable temperatures in steady-state operation. 
Therefore, aluminum alloys offer an alternative material choice for the MA.

The combination of various constraints yields the baseline design parameters listed in \hyperref[tab:baseline]{Table~\ref*{tab:baseline}}.
While the mirror suspensions are generally made of monocrystalline silicon or sapphire fibers, we propose two alternative concepts for the heat extraction and the marionette suspension, respectively. 
The first concept presented in \hyperref[sec:Monolithic]{Sec.~\ref*{sec:Monolithic}} relies on a cooling interface on the CA and the PF, requiring a monocrystalline high-conductivity marionette suspension made of silicon or sapphire.
The second concept uses a thin-wall titanium tube as marionette suspension that is filled with superfluid He-II.
This concept provides cooling at \SI{2}{\kelvin} down to the marionette and is explained in \hyperref[sec:Heconcept]{Sec.~\ref*{sec:Heconcept}}.
In both heat extraction concepts, the cooling interface design will affect the sensitivity. 
Defining the interface deteriorative impact on sensitivity  and thermal resistivity requires a refined and experimental-based analysis.
The designs will result from experimental investigation and optimization, which are not yet advanced enough to be included in the scope of this paper.

\begin{table*} [tb]
\caption{\label{tab:physicaldata}Physical properties of silicon and sapphire at \SI{20}{\kelvin} and metals at \SI{2}{\kelvin}.  Some of the indicated references comprise  temperature dependencies, which are included in the STN model presented in the \hyperref[sec:BaselineSensitivity]{Sections~\ref*{sec:BaselineSensitivity}} and \hyperref[sec:ParameterStudy]{\ref*{sec:ParameterStudy}}.}
\begin{ruledtabular}
\begin{tabular}{lrr|rrr}
&Silicon &Sapphire & Ti6Al4V &Titanium 
&Al5056\footnote{Physical properties, except for $\phi_\mathrm{bulk},~ \lambda \text{~and~} c_\mathrm{p}$, are taken from Al5083 due to the lack of data at cryogenic temperatures for Al5056.}\\
$T$ (K) &20 &20 &2.0 &2.0 
&2.0\\
\hline
$\phi_\mathrm{bulk}$ (-) &$\num{1E-9}$ \cite{McGuigan1978MeasurementsOT}
\footnote{Value given at $T=\SI{10}{\kelvin}$ and applied in accordance with the silicon surface loss parameter reported in \cite{Nawrodt2013}.} 
&$\num{3E-9}$\cite{Tobar} 
&$\num{1E-4}$\cite{AMADORI2009340}
\footnote{Value given at $T=\SI{80}{\kelvin}$ due to lack of data at lower temperatures.} 
&$\num{1E-6}$\cite{DUFFY2000417} 
&$\num{2.5E-8}$\cite{DuffyAluminium} \\

$\sigma_\mathrm{y}$ (MPa) &230 \cite{Cumming_2013}\footnote{For brittle materials, the yield strength $\sigma_\mathrm{y}$ and the ultimate strength $\sigma_\mathrm{max}$ are nearly equivalent.  The data given for \SI{300}{\kelvin} tend to increase by about \SI{10}{\percent} at cryogenic temperatures.}
&$400~\mbox{\cite{Dobrovinskaya2009}}^\text{d}$ &1600 \cite{ekin}  &1200 \cite{ekin}  
&280 \cite{ekin} \\

$\lambda(T)$ (W/m/K)  &4940 \cite{TPRC} 
&6000 \cite{Khalaidovski_2014} &0.22 \cite{Cryocomp} &2.5 \cite{Cryocomp} 
&2.0 \cite{BAUDOUY20141}  \\

$c_\text{p}(T)$ (J/kg/K)  &3.40 \cite{TPRCcp} &0.69 \cite{White1997ThermophysicalPO} &0.01 \cite{Cryocomp} &0.12 \cite{Cryocomp} 
&0.10 \cite{BARUCCI20101452} \\

$\alpha(T)$ (1/K) &$\num{-2.9E-9}$\cite{ThermalexpansioncoeffSi} &$\num{1.3E-8}$\cite{TAYLOR1996Saalpha} &$\num{6.0E-6}$\cite{TPRCalpha} &$\num{5.5E-8}$ \cite{TPRCalpha} 
&$\num{14E-6}$\cite{ekin}\\

$\beta$ (1/K) &$\num{-7.9E-6}$\cite{GysinbetaSi}\footnote{Values in the range of \SIrange[]{80}{300}{\kelvin}; expected to decrease further at cryogenic temperatures.} &$\num{-4.4E-6}\mbox{\cite{Wachtman_betaSa}}^\text{e}$ &$\num{-4.6E-4}\mbox{\cite{Fukuhara1993ElasticMA}}^\text{e}$ &$\num{-4.6E-4}\mbox{\cite{Fukuhara1993ElasticMA}}^\text{e}$ 
&$\num{1.2E-4}$\cite{NIST}\\

$E$ (GPa) &130 \cite{ESiRef}\footnote{Data given for \SI{300}{\kelvin} tend to increase by about \SI{30}{\percent} at cryogenic temperatures.} &$360~\mbox{\cite{SaE_Mod}}^\text{f}$  & $127~\mbox{\cite{Boyer1994MaterialsPH}}^\text{f}$ & 130 \cite{ekin} 
& 81 \cite{NIST}\\

$\rho$ (\SI{}{\kilo\gram/\meter^3}) &2330 \cite{ET02709} &3980 \cite{ET02709} &4540 \cite{Cryocomp} &4540 \cite{Cryocomp} 
&2660 \cite{Cryocomp}\\

$\alpha_\mathrm{surf}$ (\SI{}{\meter})  &$\num{5E-13}$ \cite{Nawrodt2013}
&\num{5E-13}\footnote{For sapphire the same surface loss parameter as silicon is assumed due to the lack of experimental data.}  &0.0  &0.0 
&0.0\\
\end{tabular}
\end{ruledtabular}
\end{table*}

\section{Concept with monocrystalline marionette suspension}
\label{sec:Monolithic}

\subsection{Motivation}

The initial conceptual payload design with the parameters in \hyperref[tab:table1]{Table~\ref*{tab:table1}} is based on a heat extraction interface on the marionette, which is thermally insulated from the platform via a low-conductivity Ti6Al4V suspension \cite{ET2011}.
STN computations \cite{EttoreGWADW2021,PaolaGWADW2022} reveal that the cooling interface must be implemented on the CA and passed to the PF, as the direct connection of any high dissipation cooling path on the marionette would critically affect the thermal noise.
Hence, the suspension material must induce low STN and provide high thermal conductivity and mechanical strength.
Al6N can be used to connect the payload to the cryogenic system, but is not an option to suspend the marionette due to a very low yield strength \cite{Sumomogi2004MechanicalPO}, which is more than one order of magnitude smaller compared to crystalline silicon or sapphire.
The low STN requirement is achieved for crystalline sapphire and silicon at low temperatures thanks to their high quality factor $Q$ \cite{Nawrodt_2008}, where $Q$ is the inverse of the loss angle $\phi$ at resonance \cite{Saulson,Nowick1972}, cf.\ \hyperref[sec:STN]{Sec.~\ref*{sec:STN}}.

The overall results from the computations converge in the assessment that the marionette suspension mechanics must assure low thermal noise, i.e.\ mechanical dissipation, in order to preserve ET-LF sensitivity goals. 
The most advanced toy-model takes into account the presence of soft Al6N thermal links, similar to those implemented in KAGRA \cite{YAMADA2021103280,yamadaGW21,Ushiba_2021}, connecting the CA to the thermal shield, combined with a heat extraction through the PF via a crystalline marionette suspension fiber with high $Q$ and high thermal conductivity.
The payload and cryostat dimensions as well as the sensitivity goal in ET-LF differ from KAGRA, implying differences in the thermal and mechanical design requirements \cite{ET2020,Yamamoto_Elba}, which are not straight-forward to define.
\begin{figure}[tb!]
\includegraphics[width=8.5cm, height=6cm]{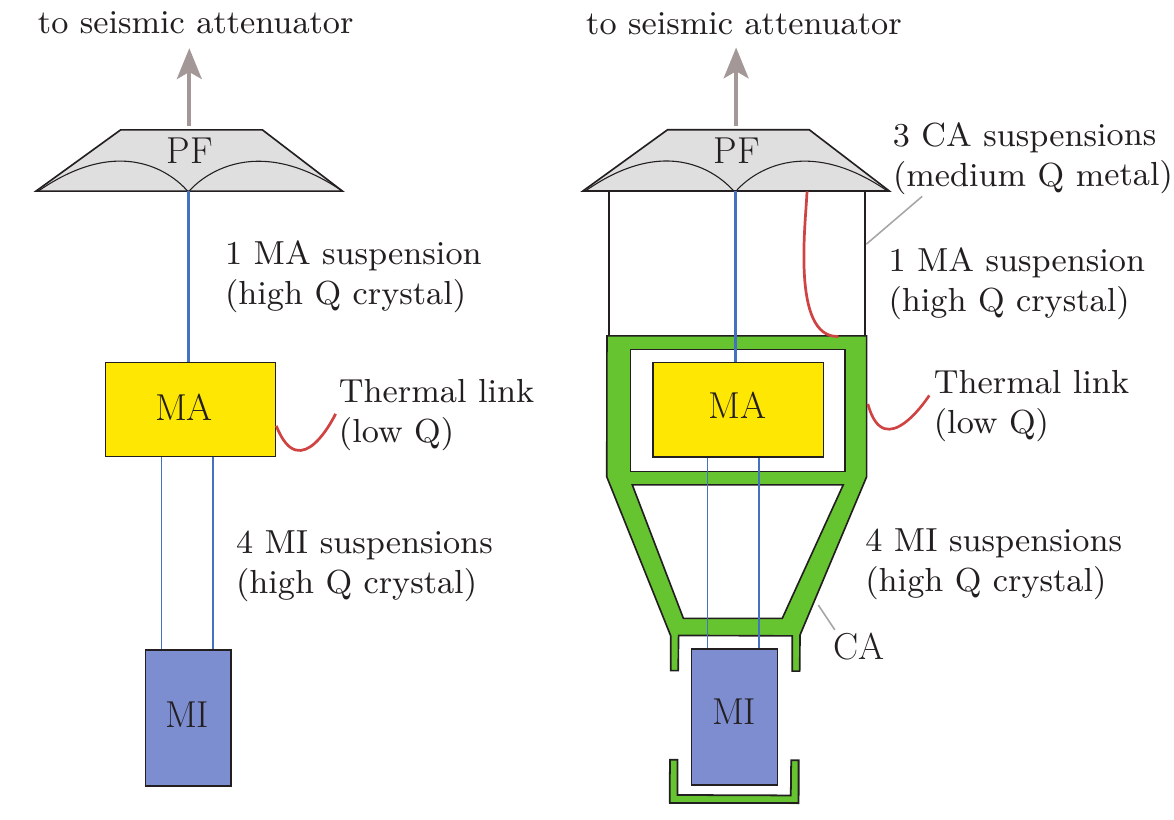}
\caption{\label{fig:schematicHL} Schemes of thermal link connection possibilities onto the cryogenic payload.}
\end{figure} 

\hyperref[fig:schematicHL]{Figure~\ref*{fig:schematicHL}} depicts the examined thermal link interface possibilities on the cryogenic payload.
A simple double stage payload with sapphire marionette and mirror suspensions is used as a reference. 
In the simplest case (left), the thermal link (TL) to the cryogenic system is modelled by a connection to the MA, showing up critical impact on the predicted sensitivity. 
A more realistic case (right) adopts a connection onto the CA and PF, ensuring sustainable mechanics of the TL connection.
\hyperref[fig:schematic2]{Figure~\ref*{fig:schematic2}} shows the effect of the soft thermal link on the STN for the cases depicted in \hyperref[fig:schematicHL]{Fig.~\ref*{fig:schematicHL}}.  
Imposing a nominal assumption for the STN roughly comparable with that of ET-LF, it can be realized that the TL should be connected far from the MI. 
{\hyperref[fig:schematic2]{Figure~\ref*{fig:schematic2}} demonstrates that the thermal links must be connected to the CA, being the minimum distance from the MI in order not to compromise the STN.
The criticality is reached with the parameters mentioned in the caption of \hyperref[fig:schematic2]{Fig.~\ref*{fig:schematic2}}. 
The computation is analytic, but FEA modelling provides similar results \cite{puppo2},\cite{RUGGI}.
A secondary but yet significant issue is the noise injection and the drag  through the thermal link, which has been estimated with a similar method as in \hyperref[fig:schematic2]{Fig.~\ref*{fig:schematic2}} \cite{EttoreGWADW2021}.

\begin{figure}[t]
\includegraphics[width=0.475\textwidth, scale=1.25]{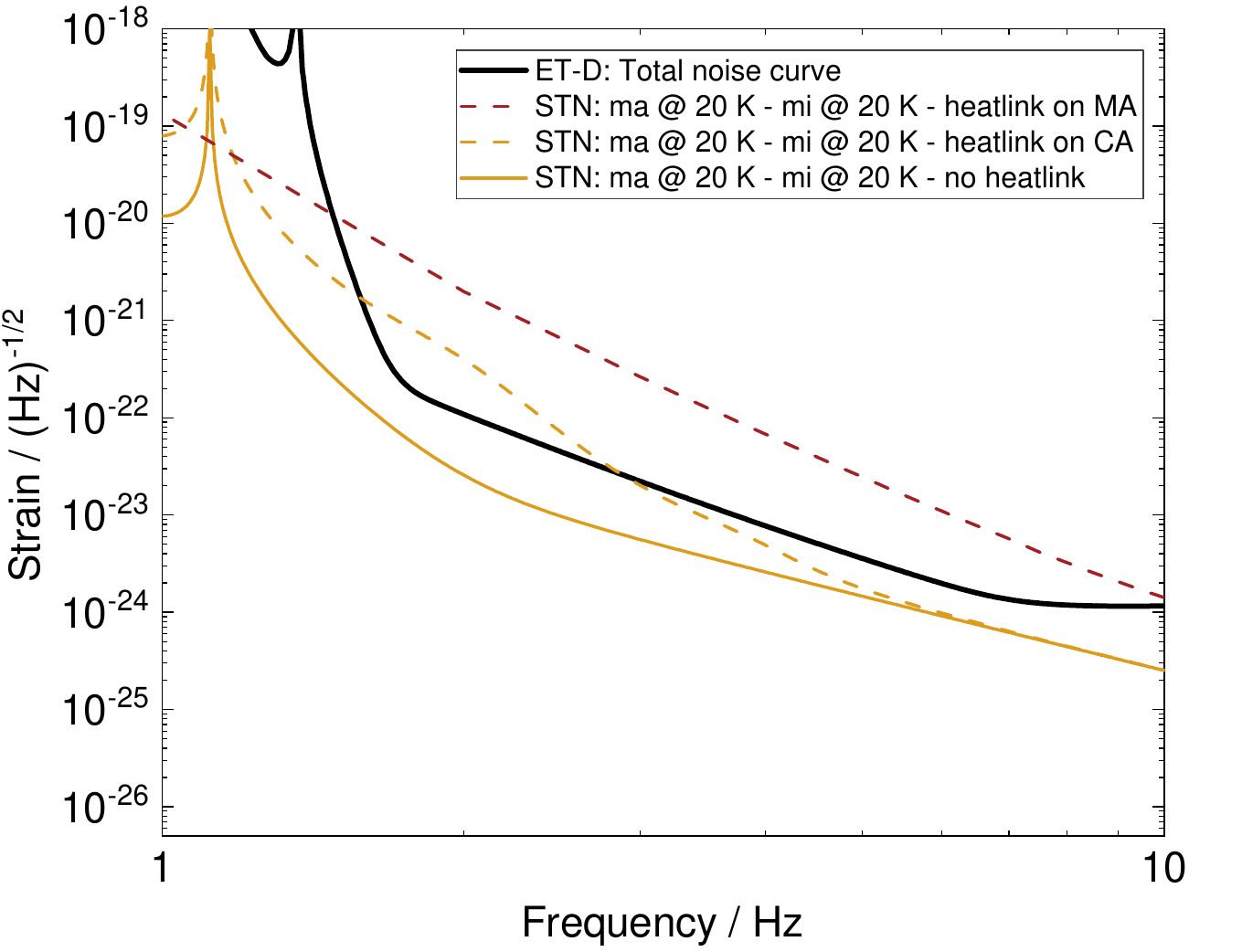}
\caption{\label{fig:schematic2} Impact on the STN due to a direct connection of a \SI{1}{\meter} thermal link (made of 28 braids, each composed by 49 Al6N wires with $d=\SI{150}{\micro\meter}$ \cite{YAMADA2021103280} and assuming $\phi_\mathrm{TL}=\num{0.5}$) on the MA and CA, respectively.}
\end{figure} 

The outline of the solid conduction cooling through thermal links is that in order to reduce the thermal noise (cf. \hyperref[sec:STN]{Sec.~\ref*{sec:STN}}) induced from the link, it must be connected to the CA and cannot reach the MA.
The implementation is feasible through a careful mechanical and thermal design of the payload in order to operate at $T_\mathrm{MI}\approx\SI{20}{\kelvin}$. 
The monocrystalline-based concept in this paper assumes using the same material (silicon or sapphire) for the mirror, the mirror suspensions and the marionette suspension.
Nonetheless, also a hydrid monocrystalline suspension application is being analyzed \cite{Dari_2010,Glasgow}.


\subsection{Mechanical dimensioning}

The marionette suspension is dimensioned for the total mechanical load of the MA and the MI, considering a safety factor SF $=3$ with regard to the ultimate strength $\sigma_\mathrm{max}$.
The material properties listed in \hyperref[tab:physicaldata]{Table~\ref*{tab:physicaldata}} for sapphire and silicon yield the dimensions in \hyperref[tab:baseline]{Table~\ref*{tab:baseline}}, which are used in the STN modelling in this paper.
For sapphire, a very conservative value of $400 ~{\rm MPa}$ is assumed, based upon \cite{Dobrovinskaya2009}. Significantly higher values of breaking strength at low temperatures, spread in the range of \SIrange[]{1000}{2600}{\mega\pascal}, have been recently measured \cite{Yamada23} and certified using samples produced in Japan by Shinkosha \cite{Shinkosha} and machined from a single ingot.

\subsection{Thermal behavior}

The thermal behavior depends on the thermal conductivity in the range of \SIrange[]{10}{30}{\kelvin}.
In thin suspension fibers, the phonon boundary scattering may significantly reduce the bulk conductivity \cite{Scurlock1966}.
Thermal conductivity data for high-purity monocrystals and monocrystalline fibers of silicon and sapphire are therefore compared in \hyperref[fig:lambda-Diagramm-ST500mW]{Fig.~\ref*{fig:lambda-Diagramm-ST500mW}}.
In silicon fibers, a marginal reduction of thermal conductivity is visible \cite{FlavioSiSuspensions}, whereas a significant reduction is reported for sapphire fibers \cite{Khalaidovski_2014}.
Further thermal conductivity measurements of silicon and sapphire fiber samples are planned within future R\&D activities.

The nominal heat load from \hyperref[eq:HeatLoad]{Eq.~(\ref*{eq:HeatLoad})} together with the dimensions in \hyperref[tab:baseline]{Table~\ref*{tab:baseline}} and the thermal conductivity data of monocrystalline fibers according \hyperref[fig:lambda-Diagramm-ST500mW]{Fig.~\ref*{fig:lambda-Diagramm-ST500mW}} yield temperature gradients along the marionette suspension of $\Delta{T_\mathrm{ma}}=\SI{3}{\kelvin}$ for silicon and $\Delta{T_\mathrm{ma}}=\SI{5}{\kelvin}$ for sapphire, respectively.


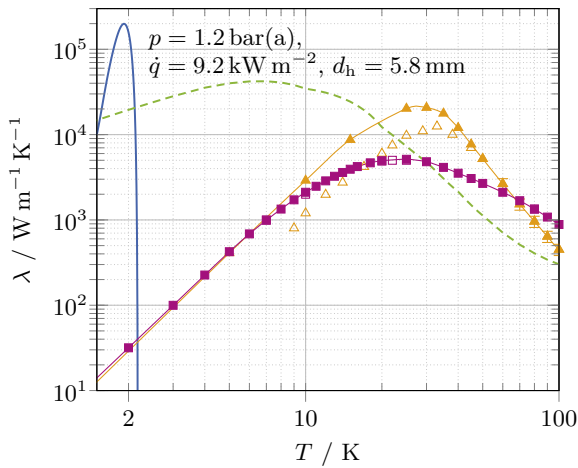
\begin{figure}
\begin{tikzpicture}
\pgfplotsset{
    xmin=1.5,
    xmax=100,
    minor x tick num=1,
    grid=major,
    width = 0.431\textwidth,
}
\begin{axis}[ymin=10, ymax=300000,
  xmode=log,
  ymode=log,
  ytick={10,100,1000,10000,100000,1000000}, ytick align=outside, ytick pos=left,
  x tick label style={major tick length=0pt},
  xticklabels=\empty,
  ylabel={$\lambda~/~\si{\watt\per\meter\per\kelvin}$},
  legend pos=south east,
    xminorgrids=true,
    minor grid style={line width=.3pt,draw=gray!50, densely dotted},
    yminorgrids=true,  
  legend style={draw=none,cells={align=left}},
  legend cell align={left}
  ]

\addplot [KITblue][line width=0.75pt] table {4He_raw.dat};\label{curve:4He}

\addplot [KITmaygreen][line width=0.75pt, dashed, densely dashed] table [x=x, y=y,y error=error, col sep=comma]{6NAl_raw.dat};\label{curve:6NAl}

\addplot+[
  KITorange, mark options={KITorange, scale=1.0},
  mark=triangle*,
  smooth, 
  error bars/.cd, 
    y fixed,
    y dir=both, 
    y explicit
] table [x=x, y=y,y error=error, col sep=comma] {Sp_raw.dat};
\label{curve:Sphighpure}

    \addplot+[
    only marks,
    mark=triangle,
    mark options={
         scale=1.0,
         draw=KITorange,
        },
]
    table {SpKAGRA_raw.dat};\label{scatterplot:SpKAGRA}
  
\addplot+[
  KITpurple, mark options={KITpurple, scale=0.75},
  smooth, 
  mark=square*,
  error bars/.cd, 
    y fixed,
    y dir=both, 
    y explicit
] table [x=x, y=y,y error=error, col sep=comma] {Si_raw.dat};
\label{curve:Sihighpure}

\addplot[
    only marks,
    mark=square,
    mark options={
         scale=0.75,
         draw=KITpurple,
        },
]
    table {SiTravasso_raw.dat};\label{scatterplot:SiTravasso}


\end{axis}

\begin{axis}[ymin=10, ymax=300000, 
  xmode=log,
  ymode=log,
  yticklabels=\empty,
  log ticks with fixed point, 
  xtick={2,10,100}, xtick align=outside, xtick pos=left,
  xlabel={$T~/~\si{\kelvin}$},
  grid=none,
  legend style={draw=none}]
 
  \end{axis}
  
\node at (2.8,4.5)[draw=none,align=left] {
$p=\SI{1.2}{bar(a)}$,
\\
$\dot{q}=\SI{9.2}{\kilo\watt\per\meter\squared}$, $d_\mathrm{h}=\SI{5.8}{\milli\meter}$
};
  
\end{tikzpicture}
\caption{{\label{fig:lambda-Diagramm-ST500mW}
Thermal conductivity of He-II at given design conditions \ref*{curve:4He} \cite{sato2006},\cite{HEPAK}, compared to 6N-aluminium (RRR = \num{E4}) \ref*{curve:6NAl} \cite{Cryocomp}, high-purity sapphire \ref*{curve:Sphighpure} \cite{TPRC2}, monocrystalline sapphire fibers \ref*{scatterplot:SpKAGRA} \cite{Khalaidovski_2014}, high-purity silicon \ref*{curve:Sihighpure} \cite{TPRC} and monocrystalline silicon fibers \ref*{scatterplot:SiTravasso} \cite{FlavioSiSuspensions}.
}}
\end{figure}

\section{Concept with He-II filled marionette suspension tube}
\label{sec:Heconcept}

\subsection{Motivation for using He-II}

\begin{figure}
\includegraphics[width=0.9\columnwidth]{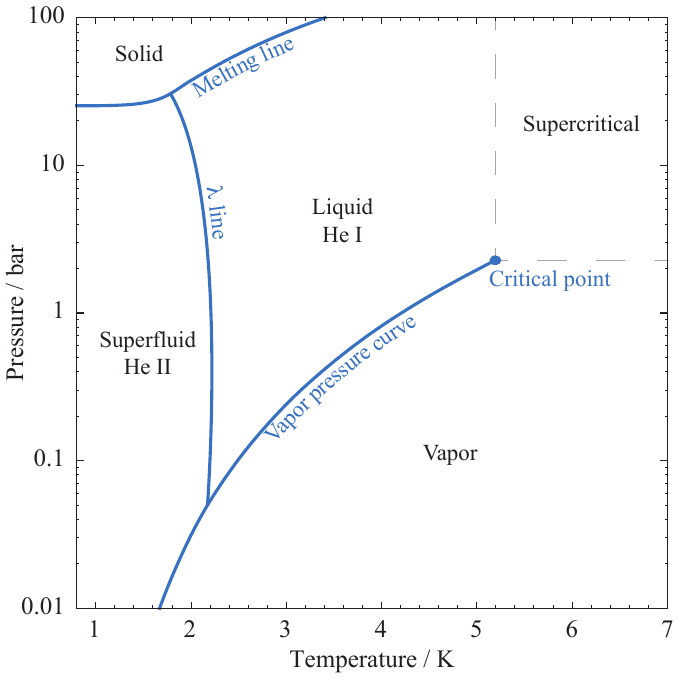}
\caption{\label{fig: He phase diagram} Phase diagram of \textsuperscript{4}He.}
\end{figure}

Cryogenic fluids have been extensively used to operate the second generation of resonant GW detectors \cite{Explorer} and later proposed for cooling the test masses of the GW interferometers \cite{GR-Cryo}. 
The use of He-II is motivated by the exceptional properties of superfluid helium, rather than its temperature around \SI{2}{\kelvin}.
The abundant \textsuperscript{4}He isotope can exist in two liquid forms, separated by the $\lambda$-line depicted in \hyperref[fig: He phase diagram]{Fig.~\ref*{fig: He phase diagram}}.
While liquid helium at $T > T_\mathrm{\lambda}$ (called He-I) exhibits normal fluid behavior, 
it becomes a quantum fluid (called He-II) at $T < T_\mathrm{\lambda}$ when fractions of the atoms condense in the ground-state as a Bose-Einstein condensate \cite{Vinen2004,Hebook}.
The He-II is composed of a normal and a superfluid component, as described by the two-fluid model \cite{landau,tisza}.
The second-order phase transition from He-I to He-II is associated with dramatic property changes.
Particularly relevant is the exceptional increase in thermal conductivity, yielding a thermal reservoir to absorb and conduct heat in the quietest possible manner.
This property enables the concept of heat extraction from the ET-LF payload via a static He-II column inside a thin-wall marionette suspension tube.
For the conditions given in \hyperref[fig:lambda-Diagramm-ST500mW]{Fig.~\ref*{fig:lambda-Diagramm-ST500mW}}, He-II can exceed the thermal conductivity of high-purity sapphire or silicon by at least one order of magnitude. 
This concept provides a temperature of \SI{2}{\kelvin} at the marionette, which is an essential parameter to reduce the STN as discussed in \hyperref[sec:BaselineSensitivity]{Sec.~\ref*{sec:BaselineSensitivity}}.

Related to the thermal conductivity, the quantum fluid properties may imply that thermal and mechanical dissipation in the static He-II column is very low, and that momentum transfer to/from the suspension tube may not take place due to superfluidity.
These hypotheses, however, require experimental validation, as the integration of a quantum fluid in suspensions of GW detectors has never been analysed and presents a new field of research.

\subsection{Conceptual layout}

The conceptual layout of the He-II marionette suspension is depicted in \hyperref[fig:Payload-ST]{Fig.~\ref*{fig:Payload-ST}}. 
In addition to the thin-wall marionette suspension tube, an internal guiding tube enables cool-down of the payload in counter-flow with supercritical helium ($p > \SI{2.3}{\bar}$) at adjustable supply temperatures.
The helium supply can be implemented at a cooling interface on the PF, using multiple thin-wall and `soft' capillaries attached to vibration isolation systems, similar to the heat link concept in KAGRA \cite{Yamada_2020,yamadaGW21}.
The capillaries connect the cooling interface to a cryogenic supply unit in the vicinity of the cryostat, cf.\ \cite{Busch_2022,Busch2022_Elba}.
Exemplary capillary dimensions are given in \cite{LennardGWADW2021}.

For steady-state operation, the normal He-I is transformed in a static He-II column \cite{Busch_2022}.
The internal guiding tube has no function in this case, i.e.\ heat conduction takes place via the entire He-II cross-section.
By contact with the He-II suspension, the marionette reaches a temperature of \SI{2}{\kelvin}. 
The silicon mirror temperature is around \SI{15}{\kelvin} due to the heat load and the temperature gradient in the monocrystalline mirror suspensions.

\begin{figure}
\includegraphics[width=7.5cm,height=9cm,center, scale=1.0]{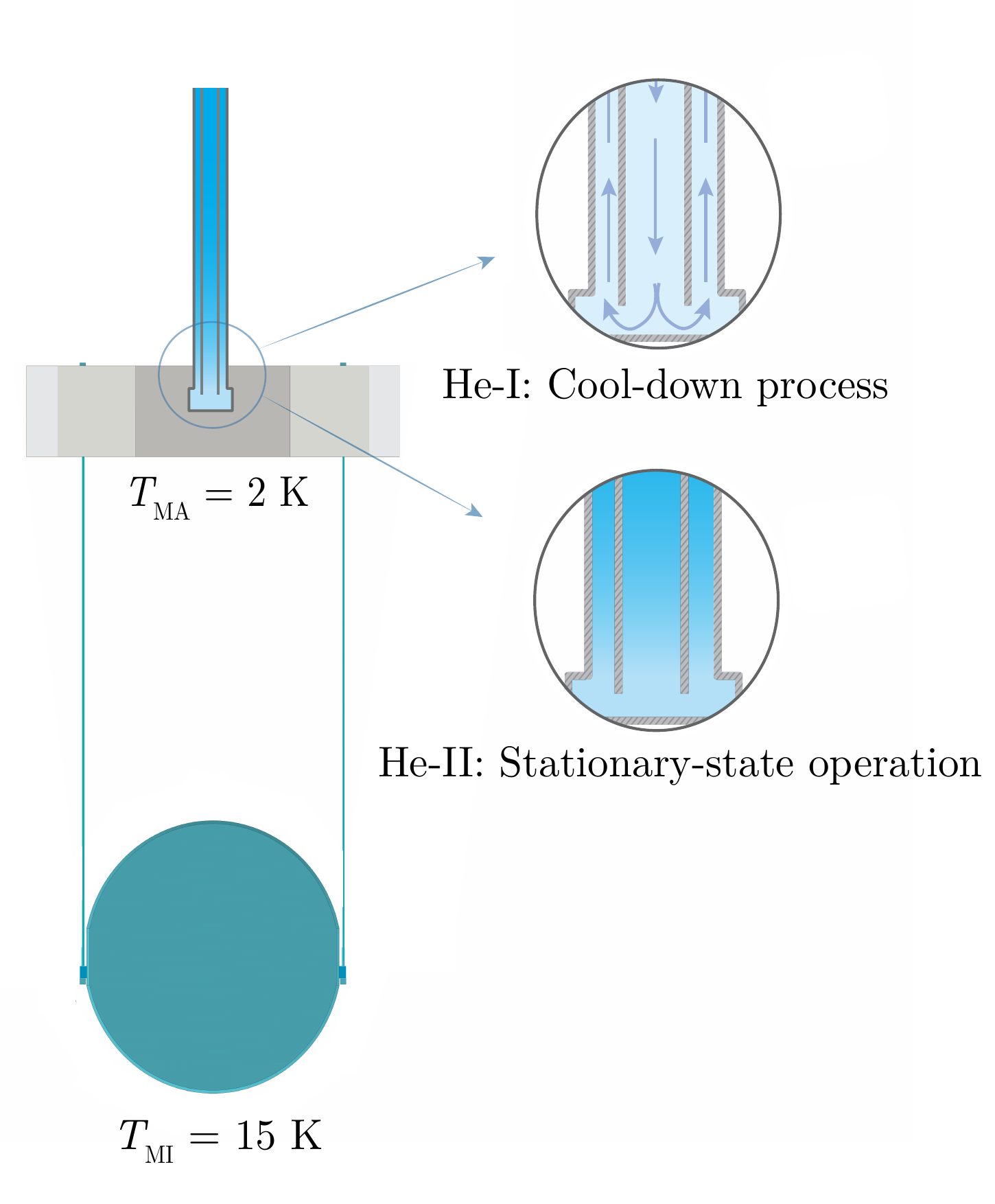}
\caption{\label{fig:Payload-ST} Conceptual layout of the He-II marionette suspension.}
\end{figure}

\subsection{Mechanical dimensioning}

The marionette suspension tube carries the mechanical load of the MA and the MI.
The dimensioning includes a mechanical safety factor $\mathrm{SF}=3$ with regard to the yield strength $\sigma_\mathrm{y}$ of the tube material (c.f.\ \hyperref[tab:physicaldata]{Table~\ref*{tab:physicaldata}} for material options).
Beside low-temperature ductility and mechanical strength, a decisive constrain in the material choice is related to suspension losses.
This yields a preference for titanium, as discussed in \hyperref[sec:BaselineSensitivity]{Sec.~\ref*{sec:BaselineSensitivity}}.

\subsection{Thermal dimensioning}

In this marionette suspension concept, the thermal dimensioning (i.e.\ the He-II cross-section) is independent from the mechanical dimensioning (i.e\ the suspension tube wall cross-section).
The two-fluid model \cite{tisza,landau} describes the heat transport in static He-II by a counter-flow between the normal and the superfluid components on a molecular level, i.e.\ there is no macroscopic movement of the bulk liquid.
The most efficient laminar regime is achieved only in narrow channels of $d < \SI{10}{\micro\meter}$, where the normal and superfluid components do not interact.
In channels of $d>\SI{1}{\milli\meter}$, an additional turbulent term starts dominating the temperature gradient by the excitation of rotons and a resulting mutual friction among the two components.
The mutual friction signifies a dissipative process that limits the heat transport \cite{Vinen2004}, but the thermal conductivity remains nonetheless higher than in pure solids as shown in \hyperref[fig:lambda-Diagramm-ST500mW]{Fig.~\ref*{fig:lambda-Diagramm-ST500mW}}.

\begin{figure} [t]

\includegraphics[width=6.5cm,height=6.0cm, scale=1.5]{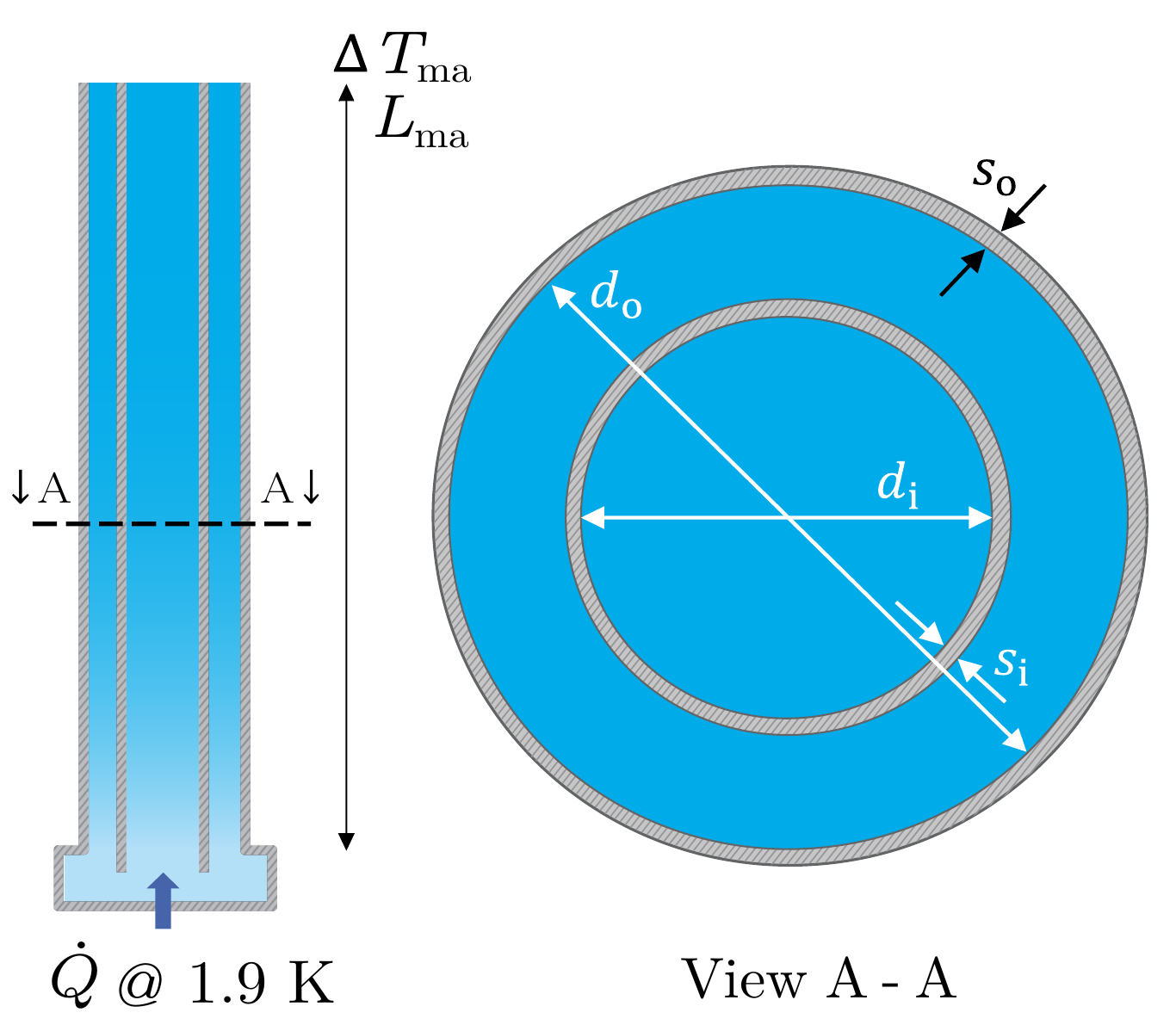}
\caption{\label{fig:ST} Suspension tube design.}
\end{figure}

The temperature gradient along the He-II column in the marionette suspension is given by:
\begin{equation}
\Delta T_\mathrm{ma}=\frac{32\,\eta\,L_\mathrm{ma}}{\left(d_\mathrm{h}\,\rho\,s\right)^2\,T} \dot{q} +\frac{L_\mathrm{ma}}{h\left(\frac{T}{T_{\lambda}}\right) g_\mathrm{peak}(p)}\dot{q}^{3.4} 
\end{equation}
where the left term signifies the analytic description of the laminar regime \cite{Hebook} and the right term uses the model from Sato et al.\ \cite{sato2006} for the turbulent regime.
$L_\mathrm{ma}$ denotes the marionette suspension length, $\eta$ the dynamic viscosity, $\rho$ the density and $s$ the entropy of the He-II, $d_\mathrm{h}$ refers to the hydraulic diameters of the circular and the annular cross-sections shown in \hyperref[fig:ST]{Fig.~\ref*{fig:ST}}, $\dot{q}$ is the heat flux, and $h(T)$ and $g_\mathrm{peak}(p)$ are empirical functions from Sato et al. \cite{sato2006}.
For the baseline design under nominal operating conditions, the contribution from the laminar term is negligibly small.

Defining a temperature gradient of $\Delta{T}_\mathrm{ma}=\SI{50}{\milli\kelvin}$ with regard to the overall He-II operating concept explained in \cite{Busch_2022}, the suspension tube design parameters are summarized in \hyperref[tab:tableST]{Table~\ref*{tab:tableST}}. 
The suspension tube lower end temperature is the highest temperature in the He-II system set to \SI{1.9}{\kelvin}, where the thermal conductivity peak is located.
The suspension tube outer diameter $d_\mathrm{o}$ results from the required He-II cross-section, whereas the wall thickness $s_\mathrm{o}$ from the mechanical design.
The inner guiding tube dimensions are chosen such that equal cross-sections of the inner tube and the annular gap yield similar flow velocities during cool-down.
\hyperref[fig:Qdot-do]{Figure~\ref*{fig:Qdot-do}} shows the relation between the required suspension tube diameter $d_\mathrm{o}$ and the heat load that can be extracted at $\Delta{T}_\mathrm{ma}=\SI{50}{\milli\kelvin}$ and $L_\mathrm{ma}=\SI{1}{\meter}$.
  
\begin{table} [t]
\caption{\label{tab:tableST} Suspension tube design parameters.}
\begin{tabular}{p{0.5\columnwidth}p{0.2\columnwidth}}
\hline
\hline
 Parameter & Value\\
 \hline
$L_\mathrm{ma}$  & \SI{1.0}{\meter} \\
$M_\mathrm{MA}$ & \SI{200}{\kilo\gram}\\
$M_\mathrm{MI}$ & \SI{200}{\kilo\gram} \\
\hline
\textbf{Constrains:}\\
Mechanical SF  &3.0\\
$T(y=L_\mathrm{ma})$  &\SI{1.9}{\kelvin}\\
$p_\mathrm{He-II,in}$   &\SI{1.2}{\bar (a)} \\
$\Delta T_\mathrm{ma}$  &\SI{50}{\milli\kelvin}\\
$\dot{Q}$  &\SI{0.5}{\watt}\\
\hline
\textbf{Design results:}\\
$d_\mathrm{o}$  &\SI{8.30}{\milli\meter}\\
$s_\mathrm{o}$ &\SI{0.36}{\milli\meter}  \\
$d_\mathrm{i}$   &\SI{5.80}{\milli\meter}\\
$s_\mathrm{i}$ &\SI{0.05}{\milli\meter}\\
\hline
\hline
\label{tabledimensioning}
\end{tabular}
\end{table}

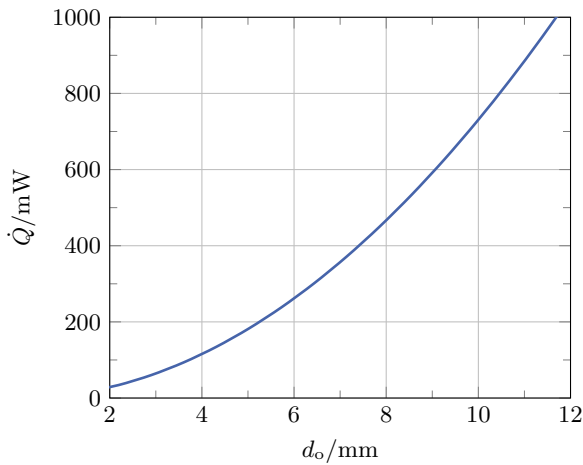
\begin{figure}\flushleft

\begin{tikzpicture}

    \begin{axis}[
    xmin = 2, xmax = 12,
    ymin = 0, ymax = 1000,
    xtick distance = 2,
    ytick distance = 200,
    xlabel=$d_\mathrm{o} / \SI{}{\milli\meter}$,
    ylabel=$\dot{Q} / \SI{}{\milli\watt}$,
    y tick label style={
    /pgf/number format/1000 sep={},
    },
    grid = major,
    minor tick num = 1,
    major grid style = {lightgray},
    width = 0.43\textwidth,
    legend style={at={(0.3,0.75), font=\fontsize{9}{7}\selectfont},anchor=south,
    legend cell align=right}
    ]
    
    \addplot [KITblue][line width=1pt, smooth] table{Q0Lma1m.txt}; 

    \end{axis}
\end{tikzpicture}

\caption{{\label{fig:Qdot-do}Cooling capacity of the He-II suspension as function of the outer tube diameter.}}
\end{figure}

\subsection{Cool-down with normal He-I flow}
\label{subsec:Cooldown}

One main advantage of this concept is the ability for convective cool-down of the ET-LF payload.
This is enabled by normal He-I flow through the double-walled marionette suspension tube as indicated in \hyperref[fig:Payload-ST]{Fig.~\ref*{fig:Payload-ST}}.
The heat flux $\dot{q}$ from the marionette to the helium flow is correlated by
\begin{equation}
\dot{q} = \alpha\left(T_\mathrm{wall}-T_\mathrm{He}\right), \\
\label{eq:heatfluxconv}
\end{equation}
where $\alpha$ denotes the heat transfer coefficient, $T_\mathrm{wall}$ the wall temperature and $T_\mathrm{He}$ the fluid temperature.
Using aluminum alloy 1200 as marionette material, the marionette temperature change is given by
\begin{equation}
\frac{\mathrm{d}T_\mathrm{MA}(t)}{\mathrm{d}t} = \frac{-\dot{q} A_\mathrm{HT}}{c_\mathrm{p,Al}(T_\mathrm{MA}(t)) M_\mathrm{MA}} \\
\label{eq:dTMARdt}
\end{equation}
where $A_\mathrm{HT}$ is the heat transfer area, $c_\mathrm{p,Al}(T)$ the specific heat capacity and $M_\mathrm{MA}$ the marionette mass.
This equation simplifies the marionette as a block capacitance, neglecting the influences of finite heat conductivity.
The cool-down process is therefore analyzed numerically by CFD simulation.
In the model development process, simulations were set up in ANSYS Fluent\textsuperscript{\textregistered} (finite volume method) and in COMSOL Multiphysics\textsuperscript{\textregistered} (finite element method), allowing validation of the numerical model independence.

Simulations are carried out for the marionette and suspension design parameters in \hyperref[tab:baseline]{Tables~\ref*{tab:baseline}} and \hyperref[tab:tableST]{\ref*{tab:tableST}}.
The suspension tube length is \SI{1.105}{\meter} in total, of which \SI{105}{\milli\meter} are centrally connected to the bottom half of the marionette, passing through a slightly wider bore in the upper half.
This insertion yields a heat transfer area of $A_\mathrm{HT}~\approx~ \SI{2750}{\milli\meter\squared}$.
\hyperref[tab:CFD]{Table~\ref*{tab:CFD}} lists additional simulation parameters and material properties in the relevant temperature range of \SIrange[]{3.0}{293.15}{\kelvin}.
The \SI{3}{\kelvin} denote the convective pre-cooling limit before the transformation to He-II operation.
\begin{table}[b]
\caption{\label{tab:CFD} CFD simulation parameters and material properties of the marionette convective cooling model.}
\begin{ruledtabular}
\begin{tabular}{lc}
 Parameter/property & Value/expression\\
\hline
$\dot{M}_\mathrm{He}$ & \SI{1.0}{\gram\per\second}\\
$p_\mathrm{He,out}$ & \SI{2.5}{\bar(a)}\\
$T_\mathrm{He,in}(t)$ & $\max \{\overline{T}_\mathrm{MA}(t) - \Delta T_\mathrm{MA-He,in}, \SI{3.0}{\kelvin}\}$\\
$\Delta T_\mathrm{MA-He,in}$ & $\SI{100}{\kelvin}$\\
\hline
$T_\mathrm{MA}(t=0)$ & \SI{293.15}{\kelvin}\\
$\overline{T}_\mathrm{MA}(t_\mathrm{end})$ & \SI{3.01}{\kelvin}\\
$d_\mathrm{MA}$ & \SI{700}{\milli\meter}\\
$h_\mathrm{MA}$ & \SI{210}{\milli\meter}\\
$A_\mathrm{HT}$ & \SI{2750}{\milli\meter\squared}\\
\hline
$\lambda_\mathrm{Al}(T)$ &  $\num{59.4}\ldots\SI{502}{\watt\per\meter\per\kelvin}$ \cite{Baudouy2011}\\
$c_\mathrm{p,Al}(T)$ &  $\num{0.29}\ldots\SI{942}{\joule\per\kilo\gram\per\kelvin}$ \cite{Cryocomp}\\
$\lambda_\mathrm{Ti}(T)$ &  $\num{4.03}\ldots\SI{36.0}{\watt\per\meter\per\kelvin}$ \cite{Cryocomp}\\
$c_\mathrm{p,Ti}(T)$&  $\num{0.20}\ldots\SI{520}{\joule\per\kilo\gram\per\kelvin}$ \cite{Cryocomp}
\end{tabular}
\end{ruledtabular}
\end{table}

In the numerical model, the geometry is simplified by axial symmetry, yielding a cylindrical marionette instead of the octagonal prism shape displayed in \hyperref[fig:payload]{Fig.~\ref*{fig:payload}}.
The helium properties are implemented via REFPROP \cite{Lemmon-RP9.1} in ANSYS Fluent\textsuperscript{\textregistered}, and by the Peng-Robinson (Twu) equation of state \cite{TWU199549} in COMSOL Multiphysics\textsuperscript{\textregistered}, respectively.
The operating conditions in \hyperref[tab:CFD]{Table~\ref*{tab:CFD}} yield exclusively turbulent flow regimes.
In order to solve the flow problems, the standard turbulence eddy viscosity $\mathrm{k-\epsilon}$ model with re-normalisation group (RNG) methods developed by Yakhot et al. \cite{yakhot1986} is applied for its accuracy regarding heat transfer \cite{Abrahamson2022}.
Scalable wall functions are implemented for the generated spatial discretization, since they enable an adequate resolution of thermally and fluid-dynamically induced effects close to the walls within the fluid domain.

The helium supply temperature $T_\mathrm{He,in}$ is set as function of the average marionette temperature for controlled cool-down.
A constant $\Delta{T}=\SI{100}{\kelvin}=\overline{T}_\mathrm{MA}(t)-T_\mathrm{He,in}$ is defined until the lowest helium supply temperature of $\SI{3}{\kelvin}$ is reached and held constant subsequently.
The low-pressure limit of $p_\mathrm{He,out} = \SI{2.5}{\bar(a)}$ ensures supercritical single-phase flow during the entire cool-down process.
The marionette and suspension tube surfaces are considered adiabatic, while the internal guiding tube is diabatic. 

\hyperref[fig:TMARdeltapPlot]{Figure~\ref*{fig:CFDvis}} shows exemplary results of the CFD simulation at an intermediate time step with $\overline{T}_\mathrm{MA}=\SI{103}{\kelvin}$ and $T_\mathrm{He,in} = \SI{3}{\kelvin}$.
\begin{figure}[b]
\includegraphics[width=0.4375\textwidth]{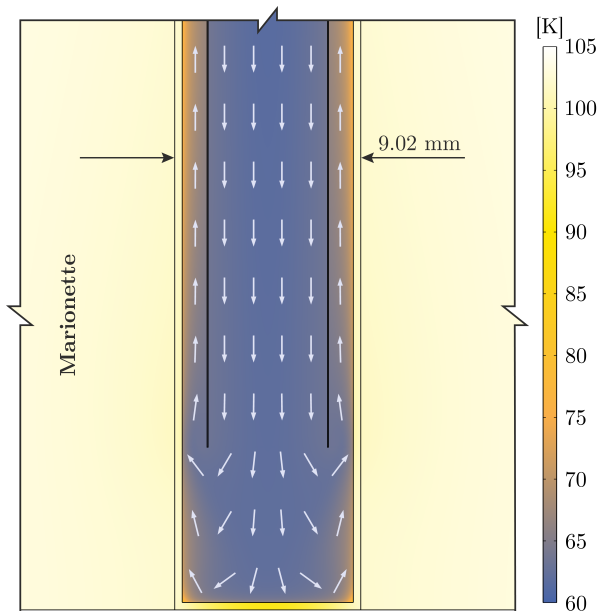}
\caption{\label{fig:CFDvis}Temperature contours and velocity field in the bottom section of the marionette; intermediate results at $\overline{T}_\mathrm{MA} \approx \SI{103}{\kelvin}$.}
\end{figure}
At the bottom end, the helium flow is returned from the inner guiding tube to the outer annular gap.
Due to internal heat exchange in the suspension, the helium supply flow is heated up by $\Delta{T}\approx\SI{57}{\kelvin}$ before entering the marionette heat transfer area $A_\mathrm{HT}$ at $T_\mathrm{He}\approx\SI{60}{\kelvin}$.
Yet, the temperature difference between marionette and helium is still around $\SI{40}{\kelvin}$, driving the heat extraction from the marionette.
In comparison, the temperature gradients within the marionette are small due to the high thermal conductivity of aluminum alloy 1200, especially at $T~<~\SI{100}{\kelvin}$ \cite{Baudouy2011}.

Results of the numerical simulation in terms of cool-down time and pressure loss are presented in \hyperref[fig:TMARdeltapPlot]{Fig.~\ref*{fig:TMARdeltapPlot}}.
\begin{figure}[t] 
\centering
\includegraphics[width=0.485\textwidth]{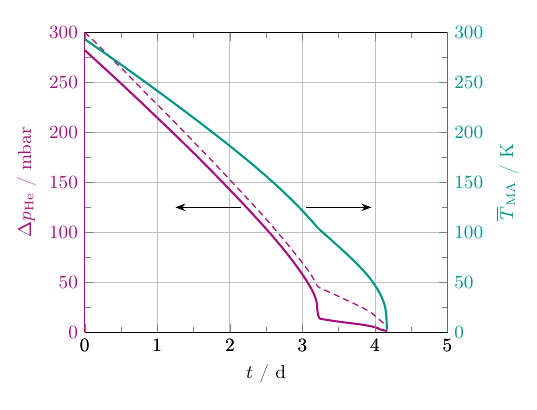}
\caption{\label{fig:TMARdeltapPlot}Helium pressure loss in the suspension tube and marionette temperature during cool-down for the conditions listed in \hyperref[tab:CFD]{Table~\ref*{tab:CFD}}.
}
\end{figure}
The pressure loss $\Delta{p}_\mathrm{He}=p_\mathrm{He,in}-p_\mathrm{He,out}$ decreases with temperature due to decreasing flow velocities at increasing densities.
A distinct point in the cool-down curve is found at $t \approx \SI{3.2}{\day}$, where $T_\mathrm{He,in}$ reaches the low temperature limit of \SI{3}{\kelvin}.
Towards the end of the cool-down at $t\ge\SI{4}{\day}$ and $T\le\SI{50}{\kelvin}$, the marionette temperature decreases rapidly due to the $T^3$-dependence of specific heat capacity.
The results in \hyperref[fig:TMARdeltapPlot]{Fig.~\ref*{fig:TMARdeltapPlot}} indicate that the marionette can be cooled from ambient to operating temperature in about \SI{4.2}{\day}.
With a helium mass flow rate of \SI{1}{\gram\per\second}, the pressure drop in the suspension is $\Delta{p}_\mathrm{He}<\SI{300}{\milli\bar}$, which is compatible with the helium supply system presented in \cite{Busch_2022}.
In addition to the numerical approach, the results are verified by implementing heat transfer and pressure drop correlations in a set of differential equations, yielding the dashed line for the pressure drop in \hyperref[fig:TMARdeltapPlot]{Fig.~\ref*{fig:TMARdeltapPlot}}. 
A more detailed discussion of this model exceeds the scope of this paper.

In a next step, the cool-down of a silicon test mass is investigated, using the data of the marionette cool-down model to define the temperatures of the mirror suspensions at their upper ends.
Radiative heat transfer is included in this model, as former studies have shown the need of combined convective and radiative cooling to achieve sufficient cool-down rates \cite{Busch2022_Elba}.
The simulation data are summarized in \hyperref[tab:TMcd]{Table~\ref*{tab:TMcd}}.
Conservative assumptions are made for the emissivity of silicon with $\epsilon_\mathrm{MI,Si}=0$ at $T<\SI{120}{\kelvin}$ and $\epsilon_\mathrm{MI,Si}=0.75$ at $T>\SI{260}{\kelvin}$, respectively, due to the lack of data.
Thermal radiation is implemented from the test mass to a surrounding black body at $T=\SI{5}{\kelvin}$ starting at $t=0$, representing the thermal shield operation around payload \cite{Busch_2022}.

\begin{table}[t]
\caption{\label{tab:TMcd} Simulation parameters used in the test mass cool-down model.}
\begin{ruledtabular}
\begin{tabular}{lc}
 Parameter & Value\\
\hline
$M_\mathrm{MI}$ & \SI{200}{\kilo\gram}\\
$d_\mathrm{MI}$ & \SI{450}{\milli\meter}\\
$h_\mathrm{MI}$ & \SI{570}{\milli\meter}\\
$d_\mathrm{mi}$ & \SI{3.0}{\milli\meter}\\
$L_\mathrm{mi}$ & \SI{1.2}{\meter}\\
$T_\mathrm{Shield}$ & \SI{5.0}{\kelvin}\\
$\epsilon_\mathrm{MI,Si}(T)$ & $\num{0.41}\ldots\num{0.75}$ \cite{CONSTANCIOJR2020}\footnote{Range represents values from \SIrange[]{120}{260}{\kelvin} for a sample with dimensions $\SI{70}{\milli\meter}\times\SI{30}{\milli\meter}\times\SI{103.5}{\milli\meter}$}\\
$\lambda_\mathrm{Si}(T)$ &  $\num{2330}\ldots\SI{5130}{\watt\per\meter\per\kelvin}$ \cite{TPRC}\\
$c_\mathrm{p,Si}(T)$&  $\num{0.28}\ldots\SI{707}{\joule\per\kilo\gram\per\kelvin}$ \cite{Desai1986}
\end{tabular}
\end{ruledtabular}
\end{table}
\begin{figure}[t]
\includegraphics{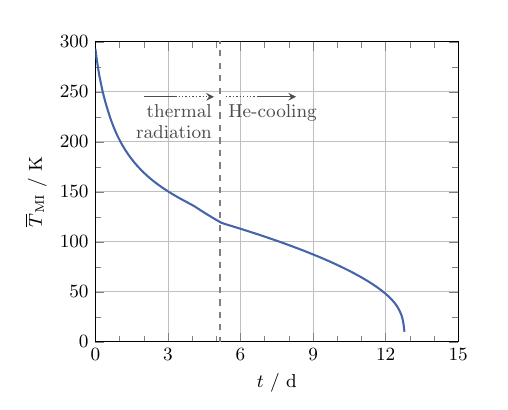}
\caption{\label{fig:TTMPlot}Cool-down of an ET-LF silicon test mass installed within a thermal shield at $T=\SI{5}{\kelvin}$ for the conditions listed in \hyperref[tab:CFD]{Tables~\ref*{tab:CFD}} and \hyperref[tab:TMcd]{\ref*{tab:TMcd}}.}
\end{figure}
\hyperref[fig:TTMPlot]{Figure~\ref*{fig:TTMPlot}} contains the simulation results, showing that the silicon test mass can be cooled from ambient to operating temperature in about \SI{12.8}{\day}.
The first phase at $t<\SI{5.2}{\day}$ is driven by thermal radiation, where temperature differences between the test mass and the shield are large, and the marionette cool-down is yet in progress.
In the second phase, thermal radiation is effectively disabled with $\epsilon_\mathrm{MI,Si}=0$.
The heat extraction from the test mass occurs exclusively via the test mass suspension fibers to the helium-cooled marionette.
Towards the cool-down end, this mechanism is amplified by increasing thermal conductivity values of silicon and the strongly decreasing heat capacity.

\section{Modelling of suspension thermal noise}
\label{sec:STN}

\subsection{Theoretical foundations}

Thermal noise is a thermally driven motion, which is directly related to the mechanical dissipative behavior of a system. 
Since thermal noise is a generalized type of Brownian motion with a random displacement of particles, it can be described using the Fluctuation-Dissipation Theorem (FDT) \cite{callen}.
It states that the dissipations in the system are the driving force for thermal fluctuations, which are presented in form of a displacement spectral density. 
The FDT is able to include the contributions of various dissipative sources, using an equivalent macroscopic mechanical model that describes the total impedance of the system in the frequency domain \cite{Saulsonbook}.
The theoretical foundations and further details of thermal noise modelling in mechanical systems, especially for suspensions used in gravitational wave detectors, are explained in \cite{Saulsonbook,Saulson,Levin,Gonzalez1994,Gonz_lez_2000, MAJORANA1997162}. 
The general approach can be summarized by the following steps:
\begin{enumerate}
\item Choice of the representative mechanical model and the dissipation mechanisms affecting the system.
\item Definition of the one-dimensional equation of motion for the mechanical displacement $x(t)$ and its Fourier transform $X(\omega)$.
\item Definition of the mechanical impedance $Z(\omega)$ and its inverse, the admittance $Y(\omega)$.
$Z(\omega)$ is the ratio between the Fourier transforms of the applied force and system velocity and is calculated by applying an external force on the system and inspecting its reaction based on its equation of motion:
\begin{equation}
Z(\omega) \equiv \frac{F}{i\omega X(\omega)} \quad \textrm{and} \quad Y(\omega) \equiv \left[Z(\omega)\right]^{-1}.
\label{ZY}
\end{equation}

\item Calculation of the thermal noise spectral density based on the FDT:
\begin{equation}
S_\mathrm{xx}(\omega)  = \frac{4k_\mathrm{B}T}{{\omega}^2}\cdot \text{Re}\left(\left[Z(\omega)\right]^{-1}\right)~,
\label{eq:FDT}
\end{equation}
where $k_\mathrm{B}$ is the Boltzmann constant, $T$ the suspension temperature and $\omega$ the angular frequency.
\end{enumerate}

\subsection{Dissipation in a pendulum system }
Energy dissipation in a mechanical system can arise from various sources, contributing directly to the generation of thermal noise \cite{Saulson}. 
The dissipation is quantified by the loss angle $\phi$, representing the ratio between the imaginary and the real restoring force in the system \cite{CAGNOLI200039}. 

The suspension thermal noise model in this paper considers the ET-LF payload as a double pendulum, with MA and MI as point-masses and suspensions with homogeneous mechanical losses.
The loss angle of a suspension includes the summation of different dissipation mechanisms as
\begin{equation}
\label{eq:phifiber}
\phi_\mathrm{susp}(\omega) = \phi_\mathrm{bulk} + \phi_\mathrm{therm}(\omega) + \phi_\mathrm{surf}+ \phi_\mathrm{join}~.
\end{equation}

\textbf{Bulk losses} $\bm{\phi_\mathrm{bulk}}$ designate intrinsic dissipations in the bulk material.
These losses arise from the structural composition and from defects in the material, depending on temperature and frequency \cite{Nowick1972}. 
The frequency-dependence is usually considered negligible and the temperature-dependence is determined experimentally \cite{Rowan2000}. 
For the ET-LF marionette suspension, several materials are proposed as depicted in \hyperref[tab:baseline]{Tables~\ref*{tab:baseline}} and \hyperref[tab:physicaldata]{~\ref*{tab:physicaldata}}. 
Their bulk losses at cryogenic temperatures are summarized in \hyperref[tab:phibulk]{Table~\ref*{tab:phibulk}}.
For the mirror suspensions, the proposed materials include monocrystalline silicon or sapphire. 
The bulk loss angles applied in this STN model are summarized in \hyperref[tab:physicaldata]{Table~\ref*{tab:physicaldata}}.

\begin{table}
\caption{\label{tab:phibulk} Bulk loss angles $\phi_\mathrm{bulk}$ of various materials at cryogenic temperatures.}
\begin{ruledtabular}
\begin{tabular}{lllr}
 Material &Type/treatment &$T$ (K) &$\phi_\mathrm{bulk}$ (-)\\
\hline
Silicon &Single crystal (100) \cite{McGuigan1978MeasurementsOT}  &3.5 &$\num{5E-10}$ \\
& &10 &$\num{1E-9}$  \\
& &20 &$\num{3E-9}$  \\
Silicon  & Single crystal (100) \cite{Nawrodt_2008}  &10 & $\num{5E-9}$ \\
& &20 &$\num{8E-9}$ \\
Silicon  &Single crystal (100) \cite{AnjaDiss} &18 & $\num{5E-9}$  \\
Silicon  &Single crystal (111) \cite{Nawrodt_2008} &10 & $\num{1.1E-8}$ \\
& &20 &$\num{1.2E-8}$ \\
Sapphire &Single crystal, annealed \cite{Tobar} &10 &$\num{2E-9}$ \\
& &20 &$\num{3E-9}$ \\
Sapphire &Single crystal, annelead \cite{Braginsky} &4.0 &$\num{2E-10}$\\
& &10 &$\num{1E-9}$ \\
& &20 &$\num{3E-9}$ \\
Sapphire &Hemlite Grade \cite{UCHIYAMA19995} &4.2 &$\num{4E-9}$  \\
& &10 &$\num{5E-9}$ \\
& &20 &$\num{5.6E-9}$ \\
Titanium & Grade 1, annealed \cite{DUFFY2000417}  &1...20 &$\num{6E-7}$  \\
Titanium & Grade 1, stress-relieved \cite{DUFFY2000417} &1...20 & $\num{1E-6}$ \\
Titanium & Grade 1, untreated \cite{DUFFY2000417} &1...20 &$\num{1E-6}$ \\
Titanium & Grade 2 \cite{Majorana_1992} &4.2 &$\num{5E-7}$  \\
& &20 &$\num{1E-6}$ \\
Ti6Al4V & Grade 5 \cite{AMADORI2009340} &80 &$\num{1E-4}$\\
Al5056 & untreated  \cite{DuffyAluminium90} &2.0 & $\num{6E-8}$  \\
Al5056 & annealed  \cite{DuffyAluminium90,coccia} &2.0 & $\num{2.5E-8}$  \\
Al5056 &- \cite{Explorerpiero} &2.0 & $\num{1.6E-7}$  \\
\end{tabular}
\end{ruledtabular}
\end{table}

\textbf{Thermoelastic losses $\bm{\phi_\mathrm{therm}}$} are frequency-dependent losses occurring in a suspension under tension, characterized by a broad maximum at a characteristic frequency \cite{Saulson,zener}. These losses originate from local temperature gradients generated by the compression and expansion at the suspension bending point. 
These gradients induce a heat flux that is accompanied with entropy generation (i.e.\ energy dissipation) \cite{Nawrodt2013,Cagnoli2002}.
For modelling, \cite{Cagnoli2002} proposes to consider both a contribution from the linear expansion coefficient $\alpha$, and a non-linear contribution from the temperature-dependence of the Young's modulus $E$ via the thermal elastic coefficient $\beta =\text{dln}E/\text{d}T$ 

\begin{equation} \label{eq:EqTE}
\phi_\mathrm{therm}(\omega)  = \frac{ET}{\rho c_\mathrm{p}} \left(\alpha-\sigma\frac{\beta}{E}\right)^2
\left(\frac{\omega\tau}{1+(\omega\tau)^2}\right)~,
\end{equation}
with $\sigma$ as the suspension tension and ${\tau}$ as the thermal diffusion time, for a circular suspension given as \cite{Nawrodt.2011}
\begin{equation} \label{eq:tau}
\tau = \frac{d^2 \rho c_\mathrm{p}}{13.55 \lambda}~,
\end{equation}
with $d$ as the suspension diameter and $\lambda$ as the thermal conductivity of the suspension material. 
This type of losses depends on geometry and tension (cf.\hyperref[eq:EqTE]{Eq.~(\ref*{eq:EqTE})}). 
Thus a reduction or nullification via an optimized suspension profile design, as applied in silica suspensions in current detectors, could be possible \cite{Cumming_2013}. 

\textbf{Surface losses $\bm{\phi_\mathrm{surf}}$} are mechanical losses in a thin surface layer $h_\mathrm{s}$, the dissipation depth, which differ from the bulk losses \cite{GRETARSSON2000108,Nawrodt2013}. 
These depend on the surface quality and on treatment techniques (e.g.\ polishing, dry or wet chemical etching), but are generally not yet fully understood \cite{Braginsky}. 
$\phi_\mathrm{surf}$ are determined from experimental data, using the surface loss parameter $\alpha_\mathrm{surf}=h_\mathrm{s}\phi_\mathrm{bulk}$. 
The relation between $\phi_\mathrm{surf}$, $\alpha_\mathrm{surf}$, the geometry factor ${\mu}$, the surface area $A_\mathrm{surf}$ and the volume $V$ is given by \cite{GRETARSSON2000108} and simplified for thin circular fibers with $\mu=2$ to
\begin{equation}
\phi_\mathrm{surf} = \alpha_\mathrm{surf}\frac{ \mu A_\mathrm{surf}}{V} = h_\mathrm{s}\phi_\mathrm{bulk} \frac{8}{d}
\label{eq:surf}
\end{equation}
This equation shows that surface losses become increasingly relevant with higher surface to volume ratio. 

 \textbf{Jointing losses} $\bm{\phi_\mathrm{join}}$ are additional mechanical losses resulting from the clamping between the suspensions and their anchors. 
 The minimization of these losses requires dedicated numerical simulations including the real payload geometry alongside experimental validation \cite{Cumming_2020}. 
 Equivalently to the ET conceptual design study \cite{ET2011} and the design report update \cite{ET2020}, this type of losses requires a more advanced design and is hence not yet considered in the model. 

\subsection{Dynamic behaviour of the pendulum system}

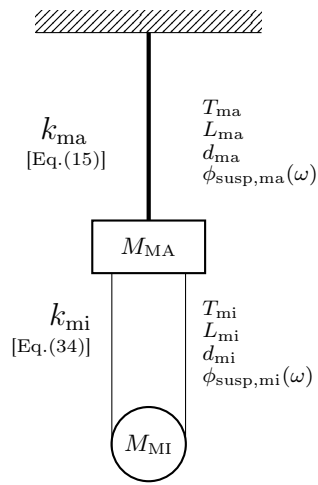
\begin{figure}
\begin{tikzpicture}[every node/.style={draw,outer sep=0pt,thick}]
\tikzstyle{spring}=[thick,decorate,decoration={aspect=0.5, segment length=0.5, amplitude=2mm,coil,pre length=0.3cm,post length=0.3cm,segment length=6}]

\tikzstyle{ground}=[fill,pattern=north east lines,draw=none,minimum width=3cm,minimum height=0.3cm]

\node (M) [minimum width=1.5cm,minimum height=0.7cm]{$M_\mathrm{MA}$};

\node (ground1) at (M.north)[ground, rotate=-180,yshift=-2.5cm,xshift=0.0cm,anchor=north] {};
\draw (ground1.north west) -- (ground1.north east);

\draw [ultra thick] (ground1.north) -- ($(M.north east)!(ground1.north)!(M.north west)$);

\draw [] (M.south) ++(-0.49,0cm) -- +(0,-2.35cm);
\draw [] (M.south) ++(0.49,0cm) -- +(0,-2.35cm);

\node[circle,draw] (c) at (0,-2.625) {$M_\mathrm{MI}$};

\node at (M.north) [draw=none,right=0.5cm,yshift=1.5cm]{\fontsize{8}{8}
$T_\mathrm{ma}$};
\node at (M.north) [draw=none,right=0.5cm,yshift=1.2cm]{\fontsize{8}{8}
$L_\mathrm{ma}$};
\node at (M.north) [draw=none,right=0.5cm,yshift=0.9cm]{\fontsize{8}{8}
$d_\mathrm{ma}$};
\node at (M.north) [draw=none,right=0.5cm,yshift=0.6cm]{\fontsize{8}{8} $\phi_\mathrm{susp,ma}(\omega)$};


\node at (M.north) [draw=none,left=0.7cm,yshift=1.2cm]{\fontsize{12}{12}
{$k_\mathrm{ma}$}};
\node at (M.north) [draw=none,left=0.45cm,yshift=0.8cm]{\fontsize{7}{7}$\text{[Eq.(\ref{Eq:kma})]}$};

\node at (c.north) [draw=none,right=0.5cm,yshift=1.3cm]{\fontsize{8}{8}
$T_\mathrm{mi}$};
\node at (c.north) [draw=none,right=0.5cm,yshift=1.0cm]{\fontsize{8}{8}
$L_\mathrm{mi}$};
\node at (c.north) [draw=none,right=0.5cm,yshift=0.7cm]{\fontsize{8}{8} $d_\mathrm{mi}$};
\node at (c.north) [draw=none,right=0.5cm,yshift=0.4cm]{\fontsize{8}{8} $\phi_\mathrm{susp,mi}(\omega)$};

\node at (c.north) [draw=none,left=0.65cm,yshift=1.2cm]{\fontsize{12}{12}
{$k_\mathrm{mi}$}};
\node at (c.north) [draw=none,left=0.65cm,yshift=0.8cm]{\fontsize{7}{7}$\text{[Eq.(\ref{eq:kmi})]}$};

\end{tikzpicture}
\caption{Scheme of the representative mechanical system used to model the STN of the ET-LF payload (ma = marionette suspension, MA = marionette, mi = mirror suspension, MI = mirror).}
\label{fig:Scheme}
\end{figure}

The suspension thermal noise modelling requires the mechanical impedance $Z(\omega)$ of the payload derived from the equations of motion as given in \hyperref[eq:FDT]{Eq.~(\ref*{eq:FDT})}. 
The double pendulum system representing the ET-LF payload is depicted in \hyperref[fig:Scheme]{Fig.~\ref*{fig:Scheme}}. 
It is modelled as a double mode oscillator with a stiffness constant for each pendulum stage \cite{Saulson,MAJORANA1997162}. 
The equations of motion in the frequency domain are given as
\begin{eqnarray}
0 & = & -M_\mathrm{MA}\omega^{2}X_\mathrm{ma} + k_\mathrm{ma}X_\mathrm{ma} + k_\mathrm{mi}(X_\mathrm{ma}-X_\mathrm{mi})\\
F & = & -M_\mathrm{MI}\omega^{2}X_\mathrm{mi} + k_\mathrm{mi}(X_\mathrm{mi} - X_\mathrm{ma})~,
\label{eq:eqsmotion}
\end{eqnarray}
where $F$ is an external force applied onto the mirror stage.
The spring constants $k_\mathrm{ma}$ and $k_\mathrm{mi}$ of the marionette and the mirror stages, respectively, are calculated via the representative mechanical system of each pendulum stage. 
Gonz{\'{a}}lez \cite{Gonz_lez_2000} provides a detailed summary of mechanical models applicable to suspensions used in gravitational wave detectors. 

\textbf{The marionette stage} is modelled using a simple pendulum with a suspended point-mass, considering a lossless gravitational potential and a lossy elastic potential as the only energy sources. 
The latter is a simplified treatment to introduce dissipation, given that the violin modes of the marionette suspension have been shown to have a negligible impact compared to the dominating ones of the mirror suspensions.
Hence, the marionette violin modes do not need to be considered in the dynamics of the representative system.
The marionette spring constant $k_\mathrm{ma}$ is obtained from the lossless gravitational spring constant $k_\mathrm{g}$ and the lossy elastic spring constant $k_\mathrm{el}$ via
\begin{equation}
k_\mathrm{ma} = k_\mathrm{g} + k_\mathrm{el}(1 + i\phi_\mathrm{susp}(\omega))~.\\
\label{kmageneral}
\end{equation}
Introducing the dilution factor $D$, which depicts the ratio between the system's elastic and gravitational potential energies \cite{CAGNOLI200039,Saulson}, as
\begin{equation} \label{eq:D}
D =\frac{k_\mathrm{el}}{k_\mathrm{g}} = \frac{n\sqrt{EI\sigma}}{2L^2}\frac{L}{Mg}= \frac{1}{2L}\sqrt{\frac{nEI}{ Mg}}~,
\end{equation}
with $M$ as the total mass suspended by $n$ wires, $\sigma$ as the tension in each wire,  $I$ as the area moment of inertia, $g$ as the gravitational acceleration, and $L$ as the suspension length, $k_\mathrm{ma}$ yields \cite{pppeffect}
\begin{equation}
k_\mathrm{ma} = k_\mathrm{g}(1 +D + i\phi_\mathrm{pend}(\omega))~. \\
\label{Eq:kma}
\end{equation}
The definition of the pendulum loss angle $\phi_\mathrm{pend}(\omega)$ in \hyperref[eq:phipend]{Eq.~(\ref*{eq:phipend})} shows that the pendulum losses are lower than the suspension losses $\phi_\mathrm{susp}$ according \hyperref[eq:phifiber]{Eq.~(\ref*{eq:phifiber})} due to dilution via $D$  
\begin{equation}
\phi_\mathrm{pend}(\omega) = \phi_\mathrm{susp}(\omega)D~.
\label{eq:phipend}
\end{equation}
The equation of motion for the marionette stage includes only a complex spring potential
\begin{equation}
-k_\mathrm{ma}x(t)= M_\mathrm{MA} \frac{\partial^{2}{x(t)}}{\partial{t^{2}}}~,
\end{equation}
yielding in the frequency domain ($x(t) \to X(\omega) \cdot e^{i \omega t}$):
\begin{equation}
k_\mathrm{ma}X(\omega) - M_\mathrm{MA} \omega^{2} X(\omega)=0~.
\end{equation}
\textbf{The mirror stage} is modelled using a pendulum consisting of four anelastic suspension fibers suspending a point mass. 
Thus, in addition to the pendulum's degree of freedom (DoF), from which the pendulum mode is extracted, also the degrees of freedom related to the transverse motion along the suspension are included in order to obtain the infinite series of violin modes associated to its bending \cite{Gonz_lez_2000,Gonzalez1994,pppeffect}.

The effective mirror spring constant $k_\mathrm{mi}$ associated to the suspension elasticity and gravitational restoring force is derived by solving the elastic equation for a slightly deflected suspension stretched by a tension $\sigma$ \cite{pppeffect,Gonz_lez_2000}
\begin{equation}
-E_\mathrm{cx}I\frac{\partial^{4}{x(y,t)}}{\partial{y^{4}}} + \sigma \frac{\partial^{2}{x(y,t)}}{\partial{y^{2}}}= \rho S \frac{\partial^{2}{x(y,t)}}{\partial{t^{2}}}~.
\end{equation}
The introduction of the Fourier transform of the displacement $X(y,\omega)$, yields 
\begin{equation}
E_\mathrm{cx}I\frac{\partial^{4}{X(y,\omega)}}{\partial{y^{4}}} - \sigma \frac{\partial^{2}{X(y,\omega)}}{\partial{y^{2}}} - \rho S \omega^{2} X(y,\omega)=0~,
\label{eq:beameq}
\end{equation}
\\
with $S$ as the cross-sectional area of the suspension. 
The complex Young's modulus $E_\mathrm{cx}$ introduces the dissipation into the system as 
\begin{equation}
E_\mathrm{cx} = E(1 + i\phi_\mathrm{susp}(\omega))~.
\label{Eq:Ecomplex}
\end{equation}
The general solution of \hyperref[eq:beameq]{Eq.~(\ref*{eq:beameq})} yields the displacement of the suspension $X(y,\omega)$ along the suspension axis $y$
\begin{equation}
X(y,\omega) = C_1 \sin(k_\mathrm{s}y) + C_2 \cos(k_\mathrm{s}y) + C_3 e^{k_\mathrm{e}y} +  C_4 e^{-k_\mathrm{e}y}
\label{eq:xy}
\end{equation}
with $k_\mathrm{s}$ as the wave number associated to the flexural stiffness of the suspension \cite{Gonzalez1994,somiya}
\begin{equation}
    k_\mathrm{s} = \sqrt{\frac{\sigma + \sqrt{\sigma^2+4E_\mathrm{cx}I\rho S\omega^2}}{2E_\mathrm{cx}I}}~,
\end{equation}
and $k_\mathrm{e}$ as the wave number of an elastic fiber   
\begin{equation}
    k_\mathrm{e} = \sqrt{\frac{-\sigma + \sqrt{\sigma^2+4E_\mathrm{cx}I\rho S\omega^2}}{2E_\mathrm{cx}I}}~,
\end{equation}
where the constants $C_1$ to $C_4$ are defined from the system boundary conditions.
For simplicity, henceforth the frequency dependency of $X(y,\omega)$ is dropped.

The first two boundary conditions result from the upper part of the suspension at $y=0$, where the fixed clamping on the marionette yields
\begin{equation}
    X(0)=0 \quad \textrm{and} \quad     \frac{\partial{X}}{\partial{y}}(0)=0~.
\end{equation}
The third and the fourth boundary conditions are associated to the bottom part at $y=L_\mathrm{mi}$, where the mirror is attached and foregoes a displacement of $X_{0}$
\begin{eqnarray}
    X(L_\mathrm{mi})& = & X_{0}~,\\ 
    \frac{\partial{X}}{\partial{y}}(L_\mathrm{mi}) & =& 0 \quad \textrm{or} \quad \frac{\partial^{2}{X}}{\partial{y^{2}}}(L_\mathrm{mi})=0~.
    \label{eq:4thRB}
\end{eqnarray}

Assuming the mirror as a lumped mass causes the suspension slope at the bottom to be a free parameter \cite{Gonz_lez_2000}. 
Somiya \cite{somiya} reports how this boundary condition can be defined for different mirror positioning approaches.
In case of the mirror facing in beam direction, the suspension bends at the attachment point yielding $\frac{\partial{X}}{\partial{y}}(L_\mathrm{mi})=0$ \cite{pppeffect}.
This condition is applied in this paper, equivalently to the ET-LF design in \cite{ET2020}.

After defining the suspension displacement function $X(y)$ according \hyperref[eq:xy]{Eq.~(\ref*{eq:xy})}, the effective mirror spring constant $ k_\mathrm{mi}$ can be derived by applying an external force $F$ on the suspended mass at $y=L_\mathrm{mi}$ \cite{pppeffect,Gonzalez1994}. 
From the equation of motion of a lumped mass
\begin{equation}
E_\mathrm{cx}I \frac{\partial^{3}{X}}{\partial{y^{3}}}(L_\mathrm{mi}) - \sigma \frac{\partial{X}}{\partial{y}}(L_\mathrm{mi}) - M_\mathrm{MI}\omega^2 X(L_\mathrm{mi}) = F~,
\label{eq:Fext}
\end{equation}
and $\frac{\partial{X}}{\partial{y}}(L_\mathrm{mi})=0$ as boundary condition, the mirror spring constant $k_\mathrm{mi}$ for 4 suspensions is given by \hyperref[eq:kmi]{Eq.~(\ref*{eq:kmi})}.

The mechanical impedance $Z_\mathrm{horz}(\omega)$, needed for the STN calculation in \hyperref[eq:FDT]{Eq.~(\ref*{eq:FDT})}, is defined by the equations of motion for the double pendulum system, presented in matrix form in \hyperref[Eq:matrix]{Eq.~(\ref*{Eq:matrix})}. 
Through this motion matrix, the mechanical impedance $Z_\mathrm{horz}(\omega)$ is obtained from \hyperref[EQ:ZMATRIX]{Eq.~(\ref*{EQ:ZMATRIX})}.
The spring constants $k_\mathrm{ma}$ and $k_\mathrm{mi}$ implemented in our STN model are taken from \hyperref[Eq:kma]{Eqs.~(\ref*{Eq:kma})} and \hyperref[eq:kmi]{(\ref*{eq:kmi})}.

The system dynamics described above refer to the horizontal DoF, representing the dominant source for the STN. 
Nonetheless, the vertical DoF delivers also a non-negligible contribution to the STN and is included in the model. 
The approach for modeling the vertical impedance $Z_\mathrm{vert}(\omega)$ is analogous to the algorithm above, whereby both the marionette and mirror stages are represented via simple pendulum systems, whose vertical spring constants are given by \cite{pppeffect}
\begin{eqnarray}
k_\mathrm{mi,vert} & = &\frac{4E_\mathrm{mi}S_\mathrm{mi}}{L_\mathrm{mi}}  (1+i\phi_\mathrm{susp,mi})~,\\
k_\mathrm{ma,vert} & = & (2\pi\cdot\SI{0.4}{\hertz})^2M_\mathrm{MA+MI}(1+i\phi_\mathrm{susp,ma})
\label{eq:kvertma}
\end{eqnarray}

The vertical spring constant for the marionette suspension can be evaluated as a set of two springs connected in series, namely the marionette suspension itself and the spring blades at its upper part. 
The resulting vertical spring constant is dominated by the soft magnetic spring blades of the super-attenuator system.
The value \SI{0.4}{\hertz} in \hyperref[eq:kvertma]{Eq.~(\ref*{eq:kvertma})} refers to the natural frequency measured for the magnetic anti-spring blades in AdVirgo \cite{pppeffect}.
Similar spring blades are assumed in this model for ET-LF.

Finally, the overall STN spectral density is 
\begin{equation}
S_\mathrm{xx}^\mathrm{total}(\omega) = S_\mathrm{xx}^\mathrm{horz}(\omega)+\theta_\mathrm{vh}^2S_\mathrm{xx}^\mathrm{vert}(\omega)~,
\end{equation}
\\
where $\theta_\mathrm{vh}$ is the vertical-to-horizontal coupling factor. 
Weak coupling of vertical motion into horizontal motion results from the non-parallel alignment of the test masses at the ends of the interferometer arms due to the Earth's curvature.
For a \SI{10}{\kilo\meter} ET-LF arm, $\theta_\mathrm{vh}$ yields
\begin{equation}
\theta_\mathrm{vh} = \frac{L_\mathrm{arm}}{d_\mathrm{Earth}} = \num{7.8E-4}~.
\end{equation}
\subsection{Implementation of temperature distribution}
For systems including non-uniform temperatures, the STN is usually modelled using the normal modal approach \cite{ET2011,ET2020,pppeffect}, as the standard FDT assumes a single homogeneous temperature in the whole system as seen in \hyperref[eq:FDT]{Eq.~(\ref*{eq:FDT})}. 
The modal approach can be a heavy computational task \cite{BONDU1998227} and includes only homogeneous dissipation.
To include inhomogeneuos losses, Levin \cite{Levin} introduces an extended formulation of the standard FDT.
Komori et al.\ \cite{PhysRevD.97.102001} propose a discrete version of this extended FDT, which can be applied for STN modelling of systems with inhomogeneous temperatures, such as cryogenic payloads.
This approach foresees the discretization of the system into elements, where each element is associated with a homogeneous temperature and an individual mechanical impedance, where the thermal noise spectral density sensed by the element $j$ of the system is given by \cite{PhysRevD.97.102001}
\begin{equation}
S_\mathrm{xx}(\omega)  = \frac{2k_\mathrm{B}}{{\omega}^2}\sum_{j} T_{j}Z^{-1}(Z_\mathrm{j}+Z_\mathrm{j}^{\dag})Z^{-1\dag}~.
\label{eq:Komori}
\end{equation}
The STN model in this paper uses the discrete FDT approach for modelling the cryogenic payload consisting of two elements (i.e., mirror stage and marionette stage).
Here homogeneous losses and a constant temperature along the suspensions, namely the highest temperatures at the lower ends, are assumed.
\begin{widetext} 
\begin{equation}
k_\mathrm{mi} = \frac{-E_\mathrm{cx}I\frac{\partial^{3}{X}}{\partial{y^{3}}}(L_\mathrm{mi})} {X(L_\mathrm{mi})}\\
= \frac{4E_\mathrm{cx}Ik_\mathrm{s}k_\mathrm{e}(k_\mathrm{s}^{3}\cos(k_\mathrm{e}L_\mathrm{mi})+k_\mathrm{s}^{2}k_\mathrm{e}\sin(k_\mathrm{e}L_\mathrm{mi})+k_\mathrm{e}^{3}\sin(k_\mathrm{e}L_\mathrm{mi}))}{(k_\mathrm{s}^{2}-k_\mathrm{e}^{2})\sin(k_\mathrm{e}L_\mathrm{mi})-2 k_\mathrm{s} k_\mathrm{e}\cos(k_\mathrm{e}L_\mathrm{mi})} \\
\label{eq:kmi}
\end{equation}
\begin{equation}
\left[
\begin{array}{cc}
k_\mathrm{ma} + k_\mathrm{mi} -M_\mathrm{MA}\omega^{2} &- k_\mathrm{mi}\\
- k_\mathrm{mi} &k_\mathrm{mi}-M_\mathrm{MI}\omega^{2}\\
\end{array}
\right]
\left[
\begin{array}{c}
X_\mathrm{ma}\\
X_\mathrm{mi}
\end{array}
\right] \\
=
\left[
\begin{array}{cc}
0\\
F
\end{array}
\right]
\label{Eq:matrix}
\end{equation}
\begin{equation}
Z_\mathrm{horz}(\omega)
=\frac{1}{i\omega}
\left[
\begin{array}{cc}
k_\mathrm{ma} + k_\mathrm{mi} -M_\mathrm{MA}\omega^{2} &- k_\mathrm{mi}\\
- k_\mathrm{mi} &k_\mathrm{mi}-M_\mathrm{MI}\omega^{2}\\
\end{array}
\right]
 \label{EQ:ZMATRIX}
\end{equation}
\end{widetext}

\hyperref[fig:Scheme]{Figure~\ref*{fig:Scheme}} illustrates the main modelling parameters implemented for the marionette and mirror suspensions, respectively. 
This conservative approach reduces computational effort, as results from KAGRA \cite{PhysRevD.97.102001} show that including the temperature gradients along the suspensions in the STN model has a negligible impact.

\section{Sensitivity of the baseline design}

\label{sec:BaselineSensitivity}

Using the STN model of \hyperref[sec:STN]{Sec.~\ref*{sec:STN}} with the parameters in \hyperref[tab:baseline]{Tables~\ref*{tab:baseline}} and \hyperref[tab:physicaldata]{~\ref*{tab:physicaldata}}, the STN curves of the baseline design options are depicted in \hyperref[fig:STNbaseline]{Fig.~\ref*{fig:STNbaseline}}.
Both the monocrystalline and the He-II filled marionette suspension concepts fulfill the sensitivity requirements of the ET-D curve \cite{ET2011}. 
The combination of a He-II filled marionette suspension with a sapphire mirror yields STN values similar to the monolithic sapphire marionette concept and is therefore not displayed.

When comparing the three STN curves of the baseline design in \hyperref[fig:STNbaseline]{Fig.~\ref*{fig:STNbaseline}} with the suspension losses from \hyperref[eq:phifiber]{Eq.~(\ref*{eq:phifiber})}, plotted for various materials in \hyperref[fig:lossTE]{Fig.~\ref*{fig:lossTE}}, two major conclusions can be drawn:
\begin{enumerate}
    \item The suspension loss angle $\phi_\mathrm{susp}$, especially of the mirror suspensions, has a crucial impact on the STN, yielding the difference between silicon and sapphire in the monocrystalline concepts. 
    
    \item The lower marionette suspension temperature $T_\mathrm{ma}$ compensates higher marionette suspension losses $\phi_\mathrm{susp}$ in the He-II filled titanium suspension tube, yielding results similar to the monocrystalline silicon suspension concept in the range of \SIrange[]{3}{30}{\hertz}.
\end{enumerate}
The latter effect can be deduced from the thermal noise of a simple harmonic oscillator far from its resonance (pendulum mode) at $\omega\gg\omega_\mathrm{0}$ \cite{Saulson} as
\begin{equation}
S_\mathrm{xx}(\omega) \propto T\phi_\mathrm{susp}~.
\end{equation}
When approaching the pendulum peak \mbox{$\omega \approx\omega_\mathrm{0}$}, however, the impact of $\phi_\mathrm{susp}$ dominantes the STN, which is visible at \SI{1}{\hertz} in \hyperref[fig:STNbaseline]{Fig.~\ref*{fig:STNbaseline}}.

\begin{figure}\includegraphics[width=0.475\textwidth, scale=1.25]{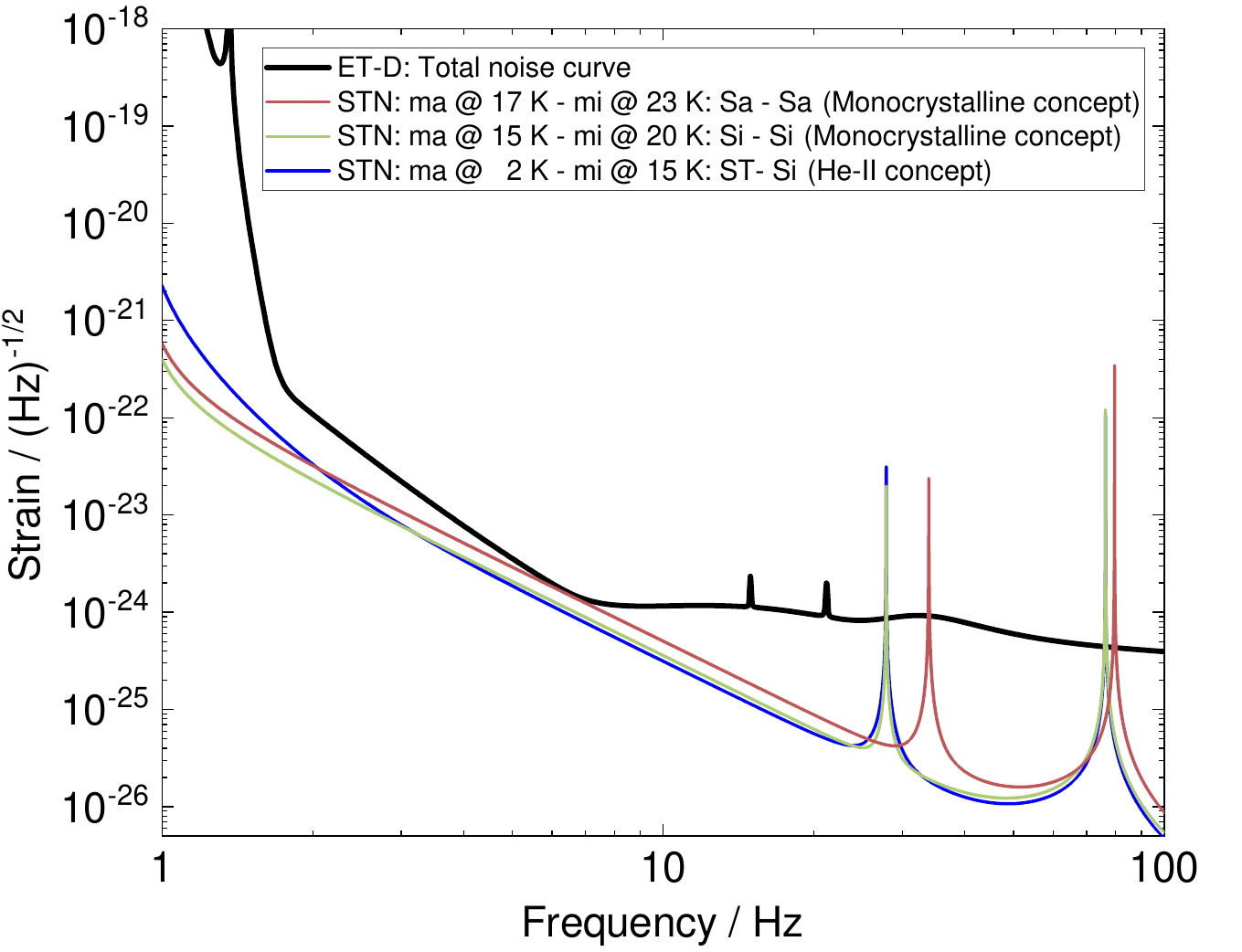}
\caption{\label{fig:STNbaseline} STN of the baseline design for the monocrystalline and He-II based marionette cooling concepts (Sa = sapphire, Si = silicon, ST = He-II suspension tube).
}
\end{figure} 

Various values for the bulk loss angle $\phi_\mathrm{bulk}$ of monocrystalline silicon and sapphire have been reported at cryogenic temperatures, see \hyperref[tab:phibulk]{Table~\ref*{tab:phibulk}}. Thus, implying the necessity for R\&D to refine the confidence interval of the data, which are immensely affected from the experimental setup.
In this work, $\phi_\mathrm{bulk}$ values of \num{1E-9} and \num{3E-9} have been applied for silicon and sapphire, respectively.
Given the crucial influence of this parameter, as also presented in \hyperref[sec:ParameterStudy]{Sec.~\ref*{sec:ParameterStudy}}, these values should be revised accordingly based on future R\&D.

The thermoelastic losses of silicon and sapphire at \mbox{$T\leqslant\SI{25}{\kelvin}$} are negligible compared to the dominant bulk losses in $\phi_\mathrm{susp}$, as depicted for \SI{20}{\kelvin} in \hyperref[fig:lossTE]{Fig.~\ref*{fig:lossTE}}.
Further investigations on surface losses of treated, strength-improved monolithic silicon and sapphire crystals are crucial, in order to use reliable values in the model, because especially in small-scale structures such as suspensions, these losses can be a significant source \cite{Cumming_2013,Nawrodt2013}.
For thin silicon flexures, Nawrodt et al.\ \cite{Nawrodt2013} report a surface loss parameter $\alpha_\mathrm{surf}$ of $\SI{5E-13}{\meter}$ at $T=\SI{10}{\kelvin}$, yielding a dissipation depth of $h_\mathrm{s}=\SI{5E-4}{\meter}$.
For sapphire, currently the surface loss parameter has not been investigated, hence it is assumed to be equal to that of silicon. 
In this model, for both silicon and sapphire suspensions a value of $\alpha_\mathrm{surf}=\SI{5E-13}{\meter}$ is applied.

In the He-II concept, only losses in the titanium suspension tube are being considered so far. 
An additional contribution may originate from the static superfluid. 
Though the He-II dissipation is expected to be minor, this may change when the relative velocity between the two fluid components exceeds a critical value \cite{Gorter-Mellink-1949,Feynman-1955}. 
Above this critical velocity, a tangle of quantized vortexes arises. 
Then an extra term, due to the interaction of the quantum vortexes with the normal fluid should appear in addition to that of the viscous normal component.  
Since the ratio between the superfluid and the normal component is a function of temperature,  the whole He-II contribution to the dissipation has to be investigated in future experiments, both in terms of frequency and temperature \cite{Koroveshi_Elba2}. 

In metals, $\phi_\mathrm{therm}$ represents the dominant loss contribution to $\phi_\mathrm{susp}$, cf.\ \hyperref[fig:lossTE]{Fig.~\ref*{fig:lossTE}}. 
Here, especially the parameters $\alpha$ and $\beta$ are decisive.
Compared to the other metals, titanium induces the lowest suspension losses, hence it is the proposed material for the suspension tube design. 
In this model,  $\beta_\mathrm{Ti}$ is conservatively set equal to $\beta_\mathrm{Ti6Al4V}=\num{-4.6E-4}$, instead of $\num{-1.9E-5}$ as reported in \cite{betaTi_Fisher}.
For the bulk losses, the conservative value of $\phi_\mathrm{bulk} = \num{1E-6}$ at \SI{2}{\kelvin} is used for the titanium suspension tube.
A contribution from $\phi_\mathrm{surf}$ is neglected, because the surface treatment and finishing technologies in metals are usually expected to provide a high quality surface and hence a minor $\phi_\mathrm{surf}$. 
The cross-sectional area of the marionette suspension tube
\begin{equation}
S_\mathrm{ST} = \pi\left(\frac{d_\mathrm{o}}{2}+s_\mathrm{o}\right)^2-\pi\left(\frac{d_\mathrm{o}}{2}\right)^2,    
\end{equation}
is implemented in the evaluation of the tension in \hyperref[eq:EqTE]{Eq.~(\ref*{eq:EqTE})}
\begin{equation}
 \sigma = \frac{ M_\mathrm{MA+MI}~g}{S_\mathrm{ST}}~,   
\end{equation}
and converted to an equivalent diameter
\begin{equation}
 d_\mathrm{ST}=\sqrt{\frac{4S_\mathrm{ST}}{\pi}} ~,  
\end{equation}
to be applied in \hyperref[eq:tau]{Eq.~(\ref*{eq:tau})}.
The suspension tube area moment of inertia used in \hyperref[eq:D]{Eq.~(\ref*{eq:D})} is
\begin{equation}
I_\mathrm{ST} = \frac{\pi}{4} \left[ 
\left(\frac{d_\mathrm{o}}{2}+s_\mathrm{o}\right)^4 - \left(\frac{d_\mathrm{o}}{2}\right)^4 \right].   
\end{equation}

\begin{figure} 
\includegraphics[width=0.475\textwidth,scale=1.25]{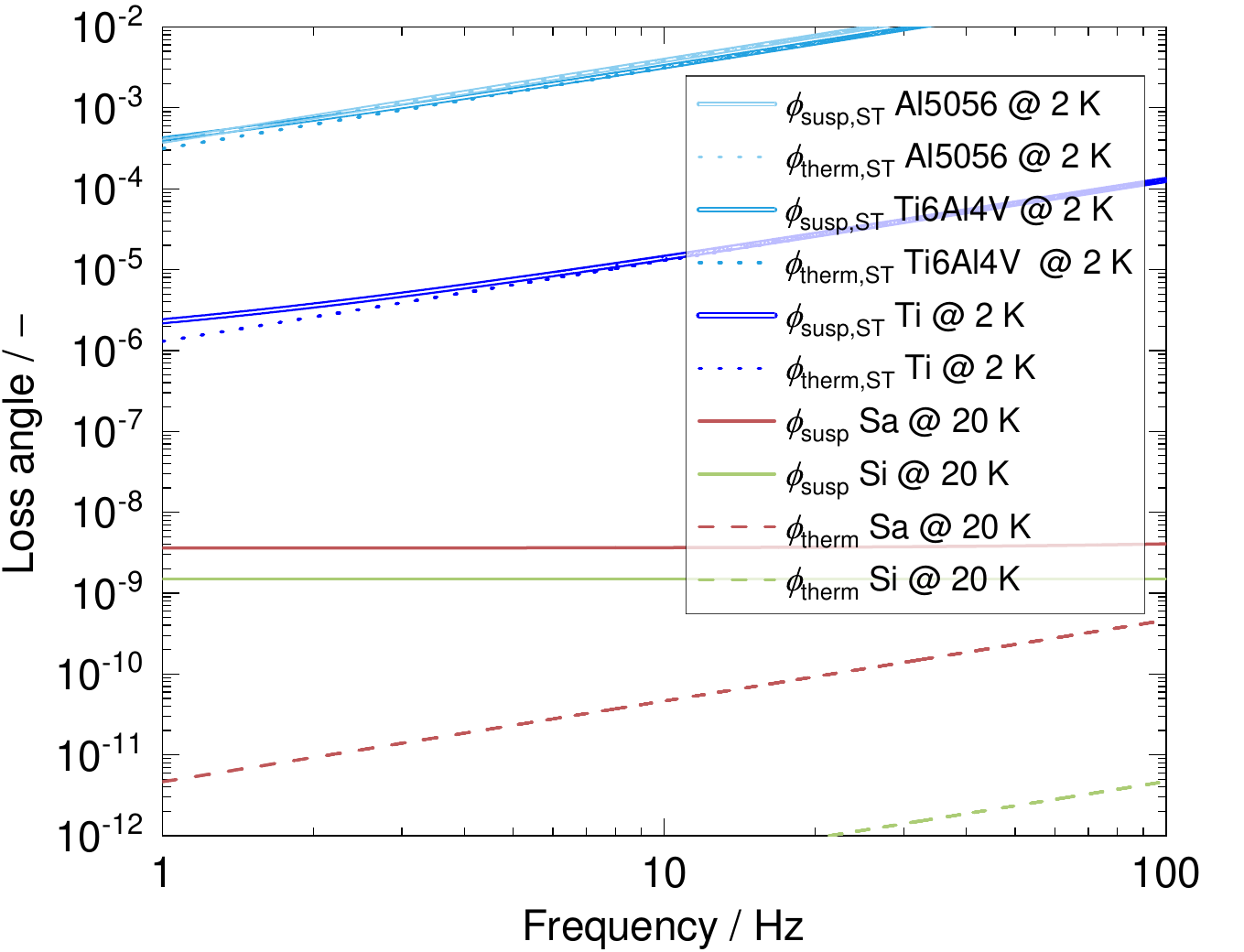}
\caption{\label{fig:lossTE} $\phi_\text{susp}$ \hyperref[eq:phifiber]{(Eq.~(\ref*{eq:phifiber}))} and $\phi_\text{therm}$ \hyperref[eq:EqTE]{(Eq.~(\ref*{eq:EqTE}))} of the marionette suspension for metallic suspension tubes (ST) and for monolithic silicon and sapphire suspensions, with design parameters from \hyperref[tab:baseline]{Table~\ref*{tab:baseline}}.
}
\end{figure}

\section{Parameter study}
\label{sec:ParameterStudy}
\subsection{General}

This section presents a study of various payload design parameters that influence the STN in the ET-LF frequency range. 
We use the He-II filled marionette suspension concept with a silicon mirror as a reference, because variations of other design parameters do not affect the temperature $T_\mathrm{ma}=T_\mathrm{MA}=\SI{2}{\kelvin}$ in this case.     
Therefore, effects of different mirror suspension designs can be better discriminated.
The applied physical property data are summarized in \hyperref[tab:physicaldata]{Table~\ref*{tab:physicaldata}}.
For a consistent comparison, the analysis considers the resulting mirror temperatures $T_\mathrm{mi}=T_\mathrm{MI}$ due to the parameter variations, i.e.\ a mechanical dimensioning and a thermal modelling is applied prior to each STN modelling. 
The results of the parameter study are visualized in \hyperref[fig:MIREff]{Figs.\ref*{fig:MIREff}} and \hyperref[fig:MAREff]{\ref*{fig:MAREff}} in the frequency range of \SIrange[]{0.3}{100}{\hertz} in order to include the impact on the pendulum modes below \SI{1}{\hertz}.

\begin{figure*}
\begin{subfigure}{\columnwidth}
    \includegraphics[width=\textwidth,scale=1.25]{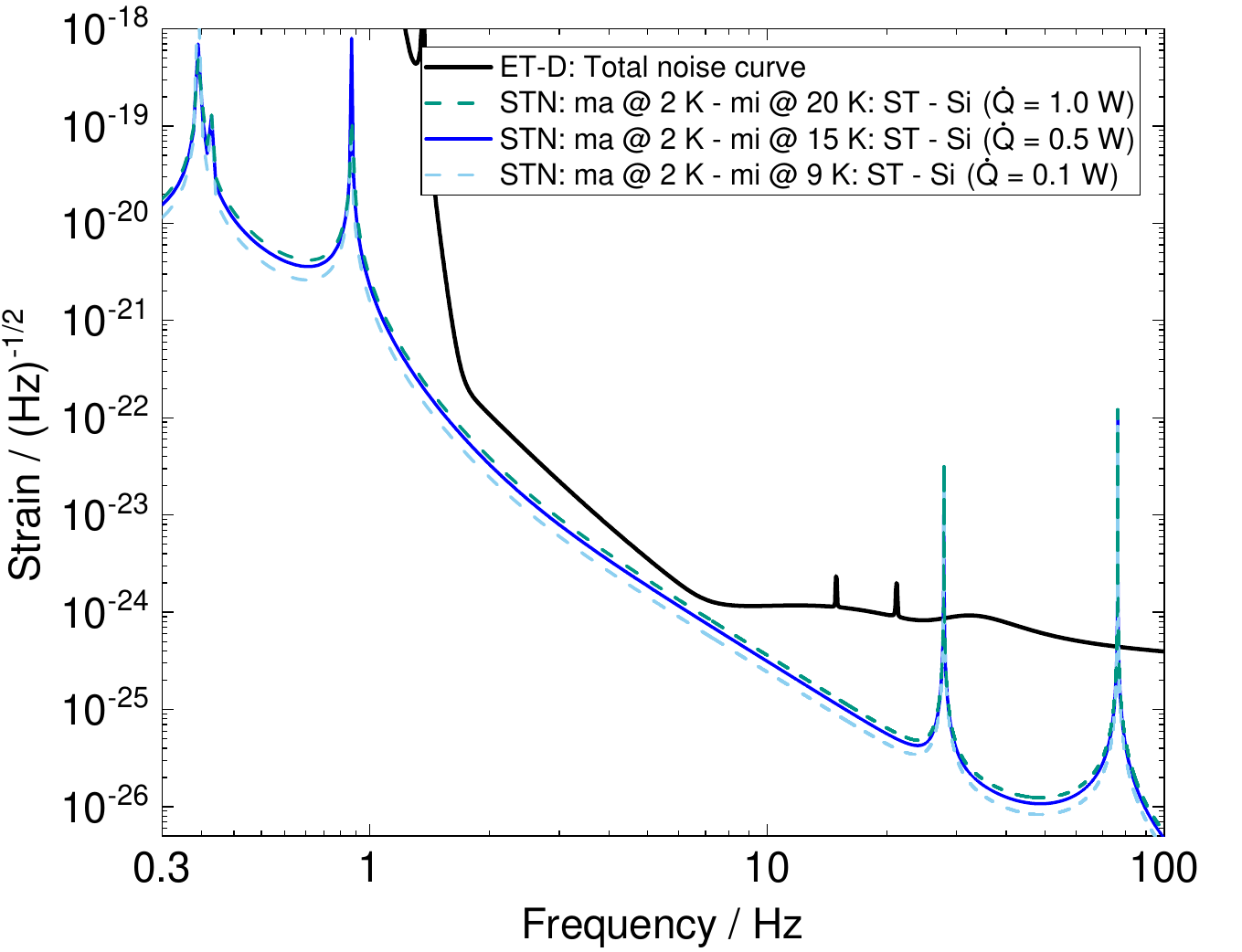}
    \caption{Impact of heat load/mirror temperature $T_\mathrm{mi}$}
    \label{fig:TMIR}
\end{subfigure}
\hfill
\begin{subfigure}{\columnwidth}
    \includegraphics[width=\textwidth,scale=1.25]{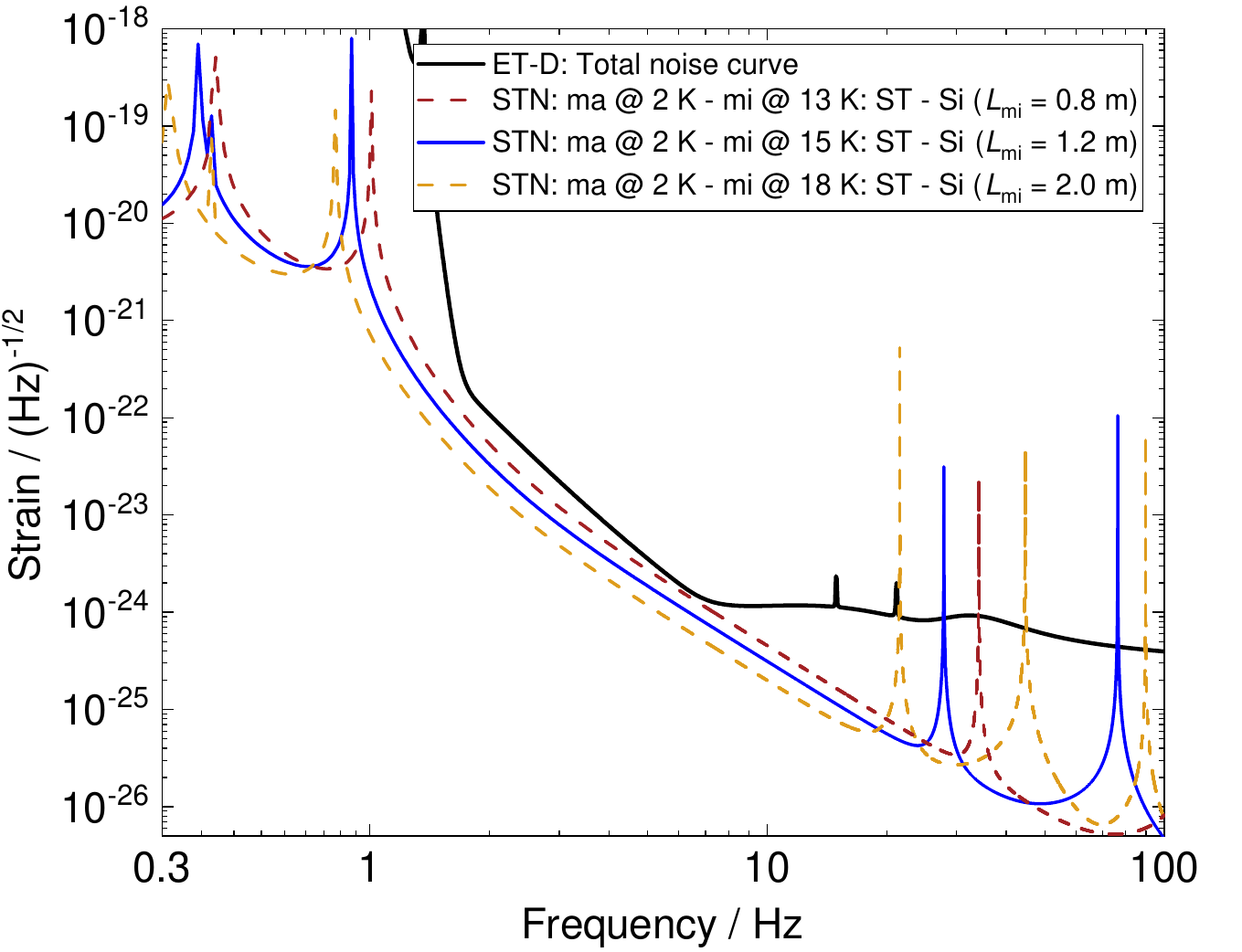}
    \caption{Impact of mirror suspension length $L_\mathrm{mi}$}
    \label{fig:LMIR}
\end{subfigure}
\hfill
\begin{subfigure}{\columnwidth}
    \includegraphics[width=\textwidth,scale=1.25]{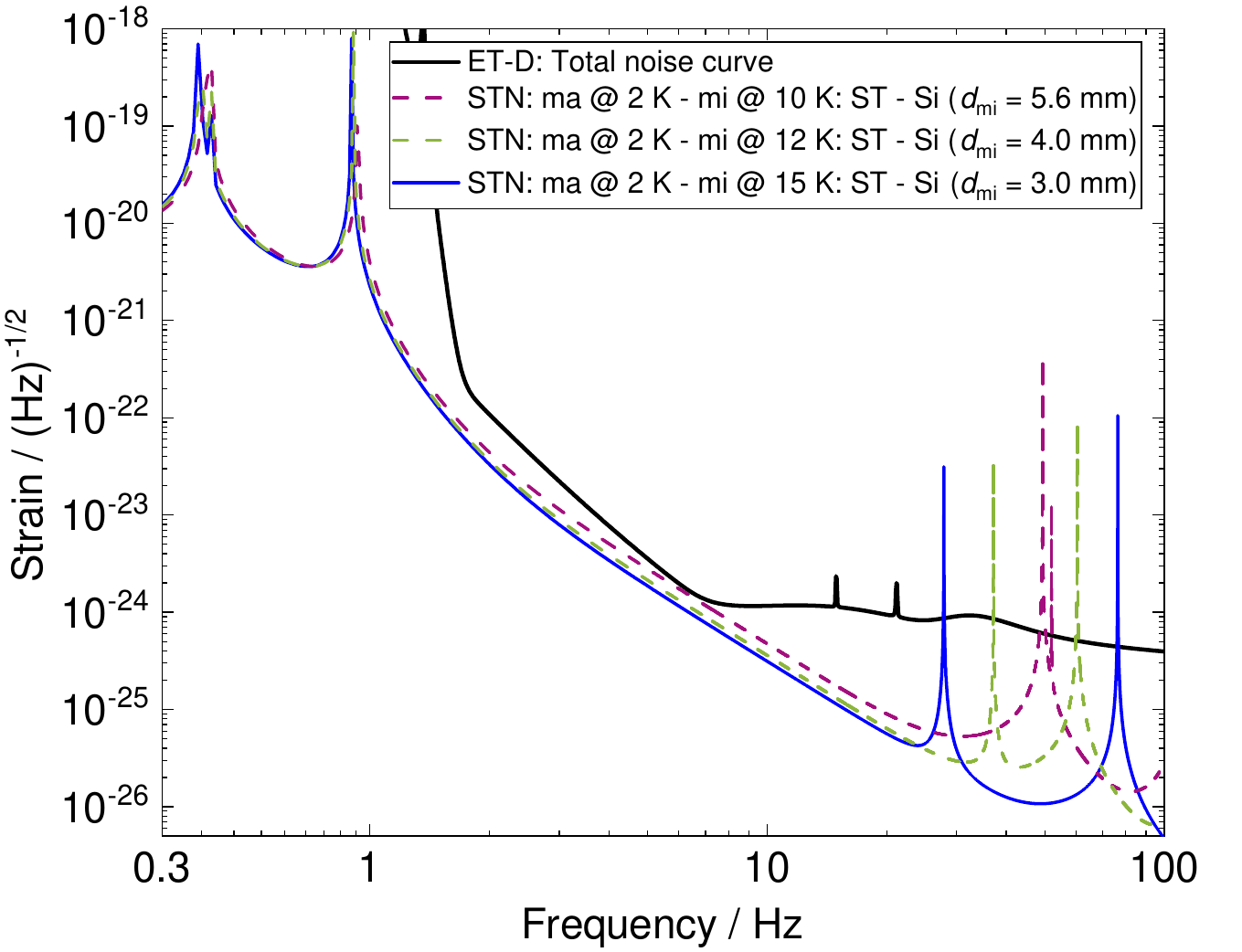}
    \caption{Impact of mirror suspension diameter $d_\mathrm{mi}$}
    \label{fig:dMIR}
\end{subfigure}
\hfill
\begin{subfigure}{\columnwidth}
    \includegraphics[width=\textwidth,scale=1.25]{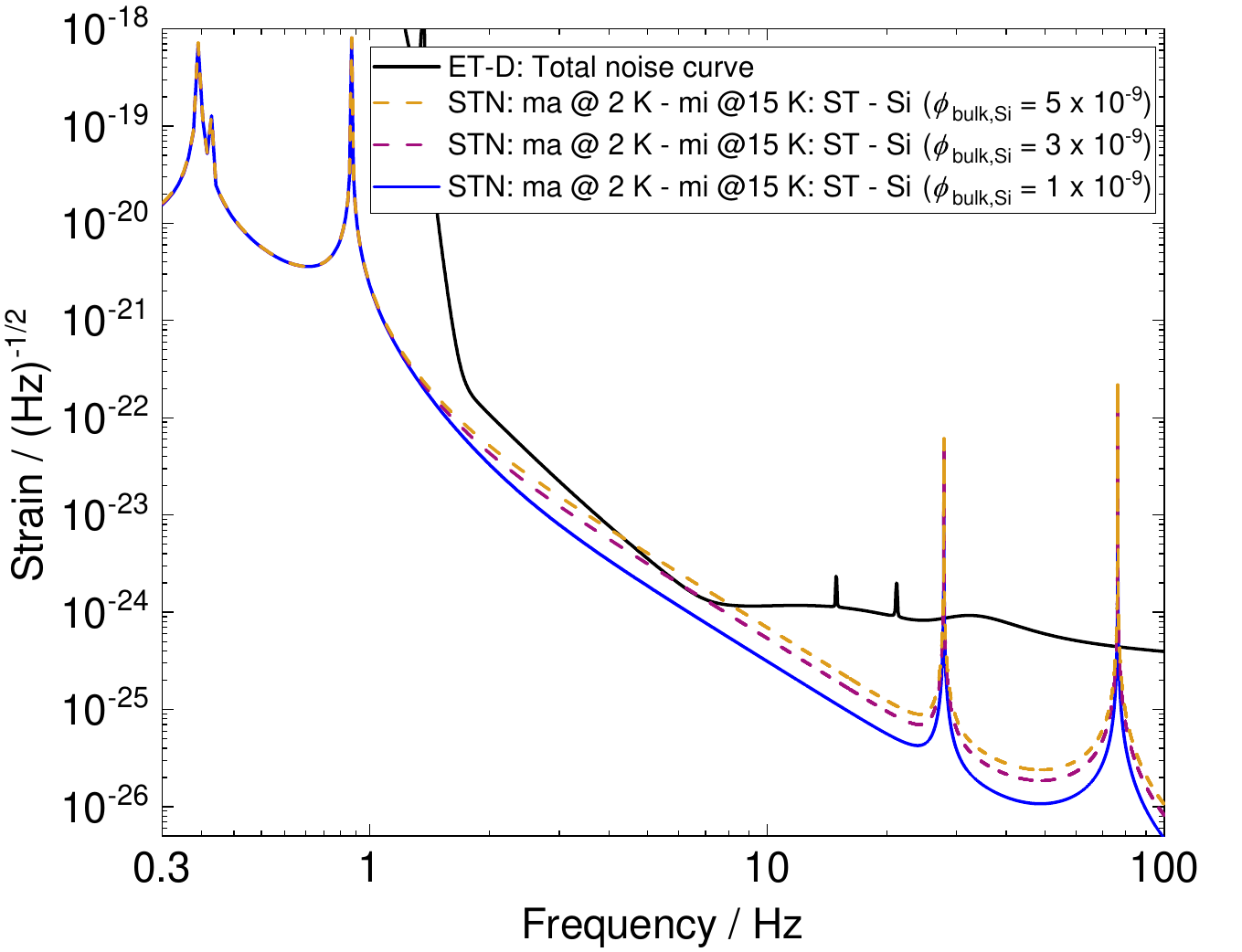}
    \caption{Impact of mirror suspension bulk loss angle $\phi_\mathrm{bulk,mi}$}
    \label{fig:phiMIR}
\end{subfigure}
\caption{Sensitivity analyses of the mirror suspension parameters $T_\mathrm{mi}$, $L_\mathrm{mi}$, $d_\mathrm{mi}$ and $\phi_\mathrm{bulk,mi}$ on the STN.}
\label{fig:MIREff}
\end{figure*}

\subsection{Influence of the mirror suspension design}

The mirror suspension design determines the STN in the frequency range above \SI{10}{\hertz}, especially due to the violin and the vertical modes, but impacts also the sensitivity at lower frequencies. 
Variations of the mirror suspension temperature, length, diameter and bulk losses are investigated.

With $T_\mathrm{ma}=T_\mathrm{MA}=\SI{2}{\kelvin}$, the temperature $T_\text{mi}=T_\text{MI}$ is a function of the heat load.
Around the design target from \hyperref[eq:HeatLoad]{Eq.~(\ref*{eq:HeatLoad})}, heat loads of \SI{0.1}{\watt}, \SI{0.5}{\watt} and \SI{1.0}{\watt} yield silicon mirror temperatures of \SI{9}{\kelvin}, \SI{15}{\kelvin} and \SI{20}{\kelvin}, respectively.
The corresponding STN curves in \hyperref[fig:TMIR]{Fig.~\ref*{fig:TMIR}} indicate a minor effect of the heat load on the STN. 
It must be noted, however, that the achievable mirror temperature strongly depends on the marionette temperature.

The length of the mirror suspensions is an essential design parameter, influencing both the STN and the cryostat design.
\hyperref[fig:LMIR]{Figure~\ref*{fig:LMIR}} shows the STN for mirror suspensions of \SI{2.0}{\meter}, \SI{1.2}{\meter} and \SI{0.8}{\meter} length, respectively, yielding mirror temperatures to \SI{18}{\kelvin}, \SI{15}{\kelvin} and \SI{13}{\kelvin} at \SI{0.5}{\watt} heat load.
A decreasing length $L_\mathrm{mi}$ yields a shift of all the modes to higher frequencies.
This is beneficial for the sensitivity regrading the violin and vertical modes, but it also implies a shift of the pendulum modes below \SI{1}{\hertz} to higher frequencies. 
The latter results in an STN increase between \SIrange[]{1}{20}{\hertz}. 
Therefore, $L_\mathrm{mi}$ is a design parameter to be optimized, considering constraints imposed by the ET-LF sensitivity, ongoing R\&D on high-quality fiber manufacturing \cite{FlavioSiSuspensions} and the ET-LF cryostat and tower dimensions.

The impact of the mirror suspension diameter $d_\mathrm{mi}$ is presented in \hyperref[fig:dMIR]{Fig.~\ref*{fig:dMIR}}, considering different ultimate strength values and mechanical safety factors.
Measured ultimate strength values of silicon at cryogenic temperatures lie between \SI{230}{\mega\pascal} \cite{Cumming_2013} and \SI{120}{\mega\pascal} (unpublished yet). 
In order to consider mechanical strength
uncertainties related to silicon jointing methods, the application of a safety factor of 3 or 6 is foreseen in the parameter study.
This yields $d_\mathrm{mi}=\SI{3}{\milli\meter}$ for $\sigma_\mathrm{max}=\SI{230}{\mega\pascal}$ with $\mathrm{SF}=3$, $d_\mathrm{mi}=\SI{4}{\milli\meter}$ for $\sigma_\mathrm{max}=\SI{120}{\mega\pascal}$ with $\mathrm{SF}=3$ and $d_\mathrm{mi}=\SI{5.6}{\milli\meter}$ for $\sigma_\mathrm{max}=\SI{120}{\mega\pascal}$ with $\mathrm{SF}=6$, respectively.  
Increasing suspension diameters $d_\mathrm{mi}$ result in higher STN values, despite a better heat extraction with lower temperatures $T_\mathrm{mi}$.
This is mainly caused by the shifting of the vertical and first violin modes towards each other.
Furthermore, $d_\mathrm{mi}$ determines the position of the mirror suspension bending points via $\lambda_\mathrm{bp}=\sqrt{EI/\sigma}$, which due to payload control related constrains must be aligned with the center of mass of the suspended mirror and marionette, respectively.
As a consequence, the overall length of the mirror suspensions must include these additional lengths in the upper and lower parts. 
For the baseline design parameters in \hyperref[tab:baseline]{Table~\ref*{tab:baseline}}, both sapphire and silicon yield a total additional length of \SI{6}{\centi\meter}.
This additional length has a negligible impact on the STN modelling, but is an important aspect to be considered in the suspension manufacturing and payload design, such as the calculation of the suspension system frequencies and temperature gradients.

The mirror suspension bulk loss angle  $\phi_\mathrm{bulk,mi}$ has a strong impact on the STN, given that it directly affects the overall mechanical dissipation of the suspensions, cf.\hyperref[eq:surf]{~Eq.~(\ref*{eq:phifiber})}.
\hyperref[fig:phiMIR]{Figure~\ref*{fig:phiMIR}} shows the STN for silicon mirror suspensions with $\phi_\mathrm{bulk,mi}$ of \num{1E-9}, \num{3E-9} and \num{5E-9}, respectively. 
The surface losses are calculated under the assumption of a constant dissipation depth of $h_s=\SI{5E-4}{\meter}$ according \hyperref[eq:surf]{Eq.~(\ref*{eq:surf})}, 
yielding total suspension losses $\phi_\mathrm{susp,mi}$ of \num{2.3E-9}, \num{7E-9} and \num{1.2E-8}, respectively.
Increasing $\phi_\mathrm{bulk,mi}$ induces a higher STN over the complete frequency range.
\begin{figure} 
\begin{adjustbox}{left}
\begin{subfigure}{\columnwidth}
    \includegraphics[width=\textwidth,scale=1.25]{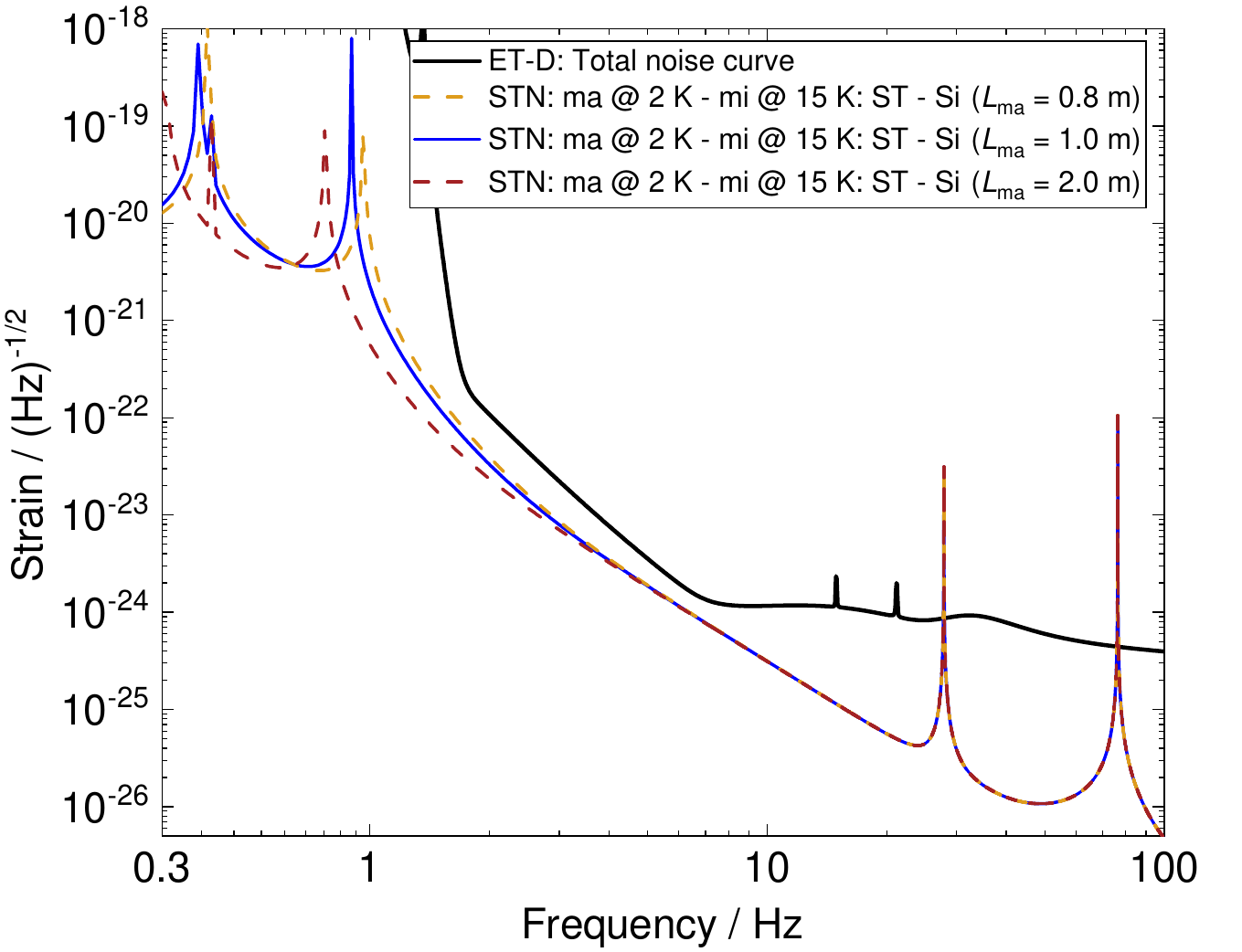}
    \caption{Impact of marionette suspension length $L_\mathrm{ma}$}
    \label{fig:LMAR}
\end{subfigure}
\end{adjustbox}
\hfill
\begin{adjustbox}{left}
\begin{subfigure}{\columnwidth}
    \includegraphics[width=\textwidth,scale=1.25]{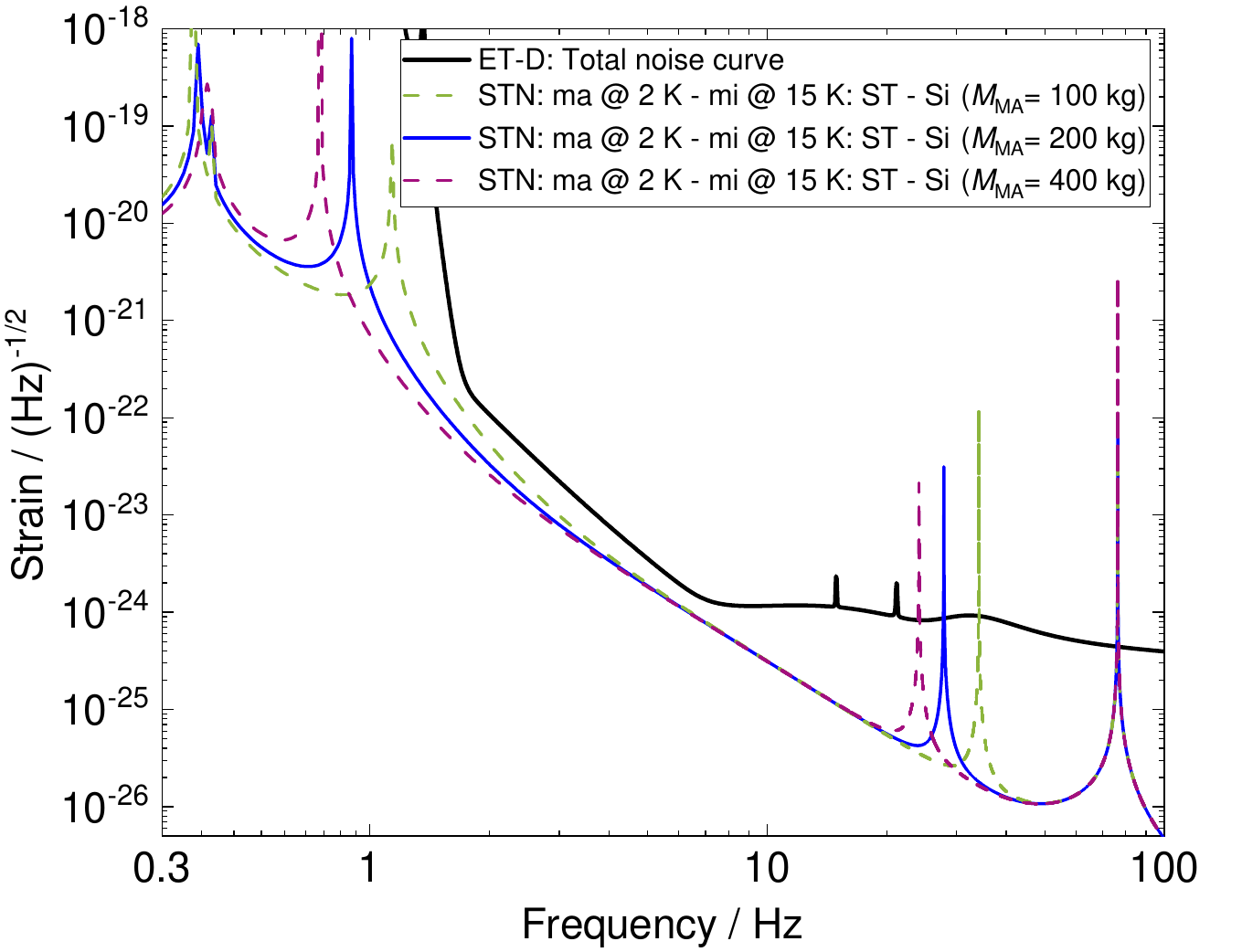}
    \caption{Impact of marionette mass $M_\mathrm{MA}$}
    \label{fig:MMAR}
\end{subfigure}
\end{adjustbox}
\hfill
\begin{adjustbox}{left}
\begin{subfigure}{\columnwidth}
    \includegraphics[width=\textwidth,scale=1.25]{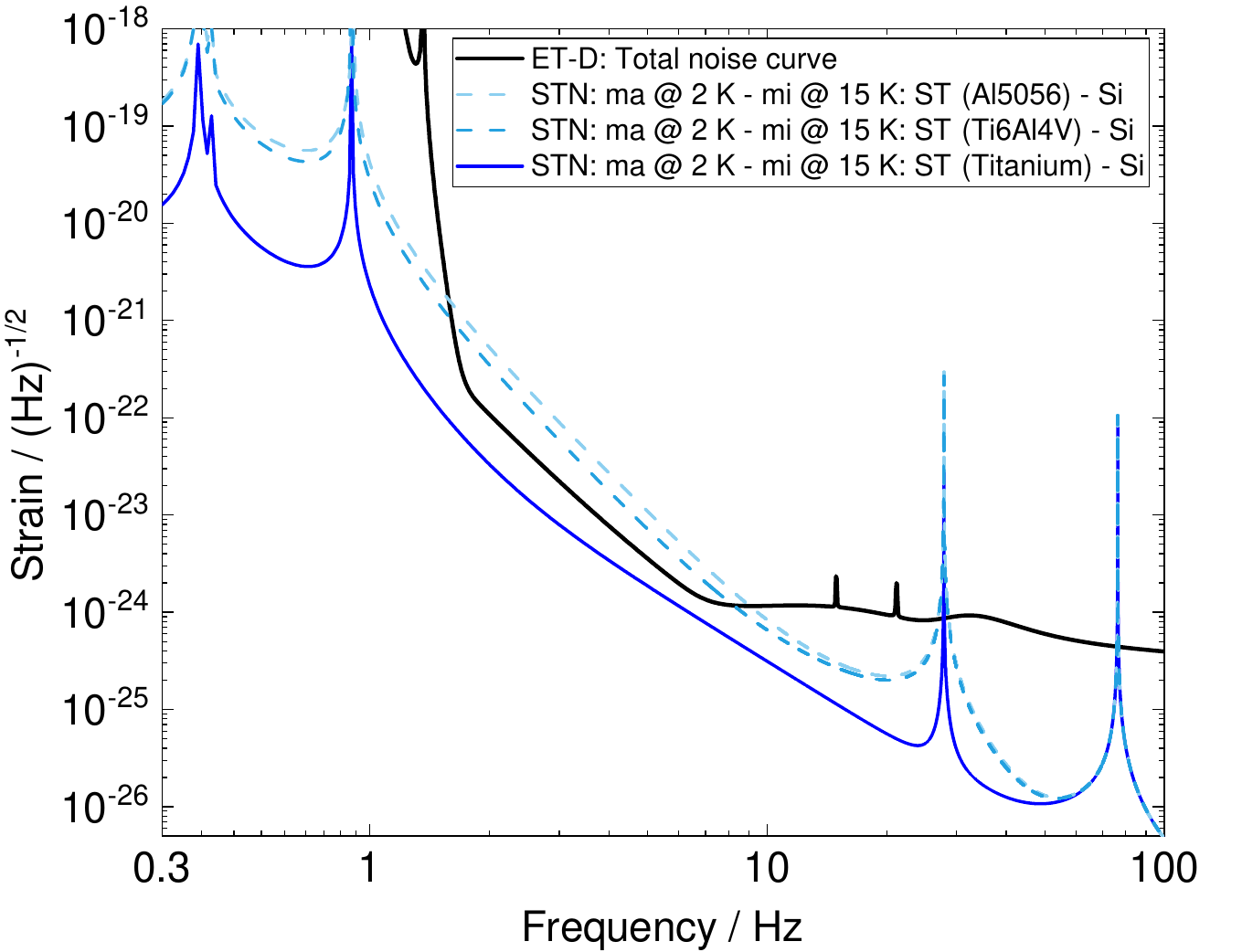}
    \caption{Impact of marionette suspension tube material}
    \label{fig:material}
\end{subfigure}
\end{adjustbox}
\caption{Parameter analysis of the marionette design parameters: $L_\mathrm{ma}$, $M_\mathrm{MA}$, $\phi_\mathrm{susp,ma}$ on the STN.}
\label{fig:MAREff}
\end{figure}

\subsection{Influence of the marionette suspension design}

The marionette suspension has a dominant impact on the STN at frequencies below \SI{10}{\hertz}.
Again, we use the He-II filled marionette suspension concept for reference, where $T_\mathrm{ma}=T_\mathrm{MA}=\SI{2}{\kelvin}$ are fixed on principle, and investigate the influence of the suspension length, the marionette mass and the suspension material.
The resulting trends may apply to monolithic marionette suspensions as well, but more detailed design studies including the cooling interface will be necessary in order to determine appropriate temperature values. 

\hyperref[fig:LMAR]{Figure~\ref*{fig:LMAR}} presents the STN modelled with marionette suspension lengths of $L_\mathrm{ma}=\SI{0.8}{\meter}$, \SI{1.0}{\meter} and \SI{2.0}{\meter}. 
A decrease in $L_\mathrm{ma}$ yields a shift of the pendulum modes to higher frequencies.
Increasing STN values, however, are only observed at \mbox{$f<\SI{3}{\hertz}$}.
The violin and the vertical modes remain unchanged, as they are defined solely by the mirror suspensions.

The variation of STN with marionettes of \SI{100}{\kilo\gram}, \SI{200}{\kilo\gram} and \SI{400}{\kilo\gram} is analyzed in \hyperref[fig:MMAR]{Fig.~\ref*{fig:MMAR}}.
A reduction of the marionette mass results in a shift of the pendulum and the vertical modes to higher frequencies, resulting in slightly higher STN values in the frequency range of \SIrange{3}{5}{\hertz}. 
The benefit of a lighter marionette, however, is a reduced cool-down time.
Additional restrictions may come from the payload control system, whereby the marionette should not weight less than the mirror. 

The marionette suspension tube material influences the STN via the suspension losses $\phi_\mathrm{susp}$ (cf.\ \hyperref[fig:lossTE]{Fig.~\ref*{fig:lossTE}}) and the wall thickness resulting from the mechanical dimensioning. 
\hyperref[fig:material]{Figure~\ref*{fig:material}} depicts the impact of different materials on the STN, showing that the ET-D sensitivity curve can only be reached with a titanium suspension tube.

\section{Conclusions and outlook}
\label{sec:Conclusions}
We presented a baseline design for the ET-LF cryogenic payload, which is thermally and mechanically consistent and fulfils the STN requirements given by the ET-D sensitivity curve.
Analytic and FEA simulations indicate that soft thermal links cannot be connected to the marionette.
Therefore, two possible heat extraction concepts are proposed, including a high-$Q$ and high-conductivity monocrystalline marionette suspension made of silicon or sapphire, and a He-II filled marionette suspension tube made of titanium, respectively.
In the latter case, the lower operating temperature of \SI{2}{\kelvin} compensates for the lower $Q$ of titanium.
The theoretical fundamentals of STN modelling applied to cryogenic payloads are described in detail and available sources for material data are compiled.
A parameter study is performed in order to identify the impact of various design parameters on the ET-LF sensitivity, illustrating the parameter space for future payload design optimizations.
The suspension losses are shown to have a decisive impact, highlighting the need for dedicated R\&D on bulk and surface losses under ET-LF operating conditions. 
A reduction of the mirror suspension length is shown to deteriorate the STN in the ET-LF frequency range, whereas the marionette suspension length has a less important impact.
Hence, a combined variation of these two parameters may be beneficial in future design studies.
The actual value of the heat load on the mirror is shown to have a marginal impact on the STN, assuming that the necessary cooling capacity is available.

Future R\&D on cryogenic payloads will be embedded in a wide context of activities outlined e.g.\ in \cite{ETfacilities-2022}.
For the monocrystalline concept foreseeing a silicon or sapphire marionette suspension, the cool-down behaviour and vibration transmission will be investigated in upcoming R\&D in the ET-Cryo facility of the Amaldi Research Center (ARC), devoted on testing and developing the main features of an ET-LF payload using a solid conductive cooling cryostat.
Thermal shielding, soft thermal links as well as high-$Q$ and high-conductivity monocrystalline suspensions for marionette and mirror will be tested.
Also, key relevant features concerning the cryostat design versus payload will be tested in order to envisage the actual impact of connecting the payload to the cryogenic system.
The ARC ET-Cryo Lab is ready and the design of the test cryostat is underway.
The alternative He-II concept is shown to fulfil the STN requirements as well, cooling the marionette to \SI{2}{\kelvin} and conducting the heat load through a static He-II column inside the marionette suspension tube.
This concept enables convective cool-down of the ET-LF payload by controlled He-I flow in about two weeks.  
Open questions related to the integration of a quantum fluid in a gravitational wave detector suspension, in particular the effect of He-II on mechanical dissipation and vibration transmission, will be addressed in future experiments by the authors at KIT.
A new facility for $Q$-measurements down to \SI{2}{\kelvin} is presently being planned, allowing both investigations of solid and He-II filled suspensions.
The scope of this facility includes R\&D on the mechanical integration of the cooling interface on the platform, the supply capillaries and their vibration attenuation system in order to investigate the noise propagation from the cooling system into the payload.

\section*{Acknowledgments}
The authors would like to acknowledge the support from the German Ministry for Education and Research (BMBF, Gr 05A20VK4), and from the Karlsruhe School of Elementary Particle and Astroparticle Physics: Science and Technology (KSETA).
The study in this paper has been developed within the frameworks of Italian PRIN2020, cod. 2020BSYXCB LoVeC-ET (Low-frequency Versus Cryogenics for ET), the EC exchange programme NEWS - H2020-MSCA-RISE-2016 GA no. 734303, and ETIC - Einstein Telescope Infrastructure Consortium (IR0000004) - MUR call n. 3264 PNRR, Miss.4 - Comp. 2, Line 3.1. 
We are indebted to KAGRA colleagues, for the precious discussions concerning solid conduction cooling-down of payloads.


\pagestyle{plain} 
\newpage
\section*{Symbol list}
\begin{table} [h!]
\begin{tabular}{llc}
\hline
\hline
Symbol  &Definition\\
\hline
$\alpha$ & Linear expansion coefficient \\
$\alpha$ & Heat transfer coefficient \\
$\alpha_\mathrm{surf}$ & Surface loss parameter \\
$\beta$ & Thermal elastic coefficient \\
$\epsilon$ & Effective emissivity \\
$\lambda$ & Thermal conductivity \\
$\lambda_\mathrm{bp}$ & Bending point position\\
$\mu$ & Geometry factor \\
$\rho$ &Density \\
$\eta$ & Dynamic viscosity \\
$\omega$ &Angular frequency \\
$\sigma$ &Tension \\
$\sigma_\mathrm{max}$ &Ultimate tensile strength \\
$\sigma_\mathrm{y}$ & Yield strength \\
$\tau$ &Thermal diffusion time \\
$\phi$ &Loss angle \\
$A$ & Area \\
$A_\mathrm{surf}$ & Surface area \\
$C$ & Constant\\
$c_\mathrm{p}$ & Specific heat capacity \\
$d$ & Diameter \\
$D$ & Dilution factor \\
$E$ & Young's modulus \\
$f$ & Frequency \\
$F$ & Force \\
$g$ & Standard gravitational acceleration \\
$h$ & Height \\
$h_\mathrm{s}$ & Dissipation depth \\
$I$ & Area moment of inertia \\
$k$ & Spring constant \\
$k_\mathrm{B}$ & Boltzmann constant \\
$k_\mathrm{e}$ & Elastic fiber wave number \\
$k_\mathrm{s}$ & Flexural stiffness wave number \\
$L$ & Length \\
$\dot{M}$ & Mass flow \\
$M$ & Mass \\
$n$ & Number of fibers \\
$p$ & Pressure \\
$\dot{q}$ & Heat flux \\
$\dot{Q}$ & Cooling power \\
$Q$ & Quality factor \\
$s$ & Specific entropy \\
$s$ & Wall thickness \\
$S$ & Cross-sectional area \\
$S_\mathrm{xx}$  & Displacement spectral density \\
$t$ & Time \\
$T$ & Temperature \\
$V$ & Volume \\
$x$, $X$  & Displacement in the time and frequency-domain\\
$y$  & Longitudinal coordinate \\
$Y$  & Mechanical admittance \\
$Z$  & Mechanical impedance \\
\hline
\hline
\end{tabular}
\end{table}

\textcolor{white}{DUMMY FORMAT TEXT The authors would like to acknowledge the support from the German Ministry for Education and Research.}

\section*{Abbreviation list}
\textcolor{white}{DUMMY FORMAT TEXT }
\begin{table}[h] 
\begin{tabular}{llc}
\hline
\hline
Abbreviation  &Definition\\
\hline
$\mathrm{Al}$ & Aluminium alloy 1200\\
$\mathrm{CA}$ & Cage\\
$\mathrm{CFD}$ & Computational fluid dynamics\\
$\mathrm{DoF}$ & Degree of freedom\\
$\mathrm{el}$ & Elastic\\
$\mathrm{cx}$ & Complex\\
$\mathrm{ET}$ & Einstein Telescope\\
$\mathrm{FDT}$ & Fluctuation Dissipation Theorem\\
$\mathrm{g}$ & Gravitational\\
$\mathrm{GW}$ & Gravitational Wave\\
$\mathrm{HCB}$ &Hydroxide catalysis bonding \\
$\mathrm{h}$ & Hydraulic\\
$\mathrm{horz}$ & Horizontal\\
$\mathrm{HF}$ & High Frequency\\
$\mathrm{HT}$ & Heat transfer\\
$\mathrm{i}$ & Inner\\
$\mathrm{in}$ & Inlet\\
$\mathrm{join}$ & Jointing\\
$\mathrm{LF}$ & Low Frequency\\
$\mathrm{MA}$ & Marionette\\
$\mathrm{ma}$ & Marionette suspension\\
$\mathrm{MI}$ & Mirror\\
$\mathrm{mi}$ & Mirror suspension\\
$\mathrm{PF}$ & Platform\\
$\mathrm{pend}$ & Pendulum\\
$\mathrm{o}$ & Outer\\
$\mathrm{out}$ & Outlet\\
$\mathrm{RRR}$ & Residual resistivity ratio\\
$\mathrm{Sa}$ & Sapphire\\
$\mathrm{Si}$ & Silicon\\
$\mathrm{SF}$ & Safety factor\\
$\mathrm{surf}$ & Surface\\
$\mathrm{susp}$ & Suspension\\
$\mathrm{ST}$ & Suspension tube\\
$\mathrm{STN}$ & Suspension thermal noise\\
$\mathrm{therm}$ & Thermoelastic\\
$\mathrm{Ti}$ & Titanium\\
$\mathrm{TL}$ & Thermal link\\
$\mathrm{vert}$ & Vertical\\
\hline
\hline
\end{tabular}
\end{table}
\textcolor{white}{DUMMY FORMAT TEXT The authors would like to acknowledge the support from the German Ministry for Education and Research.}

\newpage

\begin{thebibliography}{98}%
\makeatletter
\providecommand \@ifxundefined [1]{%
 \@ifx{#1\undefined}
}%
\providecommand \@ifnum [1]{%
 \ifnum #1\expandafter \@firstoftwo
 \else \expandafter \@secondoftwo
 \fi
}%
\providecommand \@ifx [1]{%
 \ifx #1\expandafter \@firstoftwo
 \else \expandafter \@secondoftwo
 \fi
}%
\providecommand \natexlab [1]{#1}%
\providecommand \enquote  [1]{``#1''}%
\providecommand \bibnamefont  [1]{#1}%
\providecommand \bibfnamefont [1]{#1}%
\providecommand \citenamefont [1]{#1}%
\providecommand \href@noop [0]{\@secondoftwo}%
\providecommand \href [0]{\begingroup \@sanitize@url \@href}%
\providecommand \@href[1]{\@@startlink{#1}\@@href}%
\providecommand \@@href[1]{\endgroup#1\@@endlink}%
\providecommand \@sanitize@url [0]{\catcode `\\12\catcode `\$12\catcode
  `\&12\catcode `\#12\catcode `\^12\catcode `\_12\catcode `\%12\relax}%
\providecommand \@@startlink[1]{}%
\providecommand \@@endlink[0]{}%
\providecommand \url  [0]{\begingroup\@sanitize@url \@url }%
\providecommand \@url [1]{\endgroup\@href {#1}{\urlprefix }}%
\providecommand \urlprefix  [0]{URL }%
\providecommand \Eprint [0]{\href }%
\providecommand \doibase [0]{https://doi.org/}%
\providecommand \selectlanguage [0]{\@gobble}%
\providecommand \bibinfo  [0]{\@secondoftwo}%
\providecommand \bibfield  [0]{\@secondoftwo}%
\providecommand \translation [1]{[#1]}%
\providecommand \BibitemOpen [0]{}%
\providecommand \bibitemStop [0]{}%
\providecommand \bibitemNoStop [0]{.\EOS\space}%
\providecommand \EOS [0]{\spacefactor3000\relax}%
\providecommand \BibitemShut  [1]{\csname bibitem#1\endcsname}%
\let\auto@bib@innerbib\@empty
\bibitem [{\citenamefont {{M. Branchesi and others}}(2023)}]{ETCoba}%
  \BibitemOpen
  \bibfield  {author} {\bibinfo {author} {\bibnamefont {{M. Branchesi and
  others}}},\ }\href {https://arxiv.org/abs/2303.15923} {\bibinfo {title}
  {{Science with the Einstein Telescope: a comparison of different designs}}}
  (\bibinfo {year} {2023})\BibitemShut {NoStop}%
\bibitem [{\citenamefont {{ET Science Team}}(2011)}]{ET2011}%
  \BibitemOpen
  \bibfield  {author} {\bibinfo {author} {\bibnamefont {{ET Science Team}}},\
  }\href {https://apps.et-gw.eu/tds/?content=3&r=8709} {\bibinfo {title}
  {{Einstein gravitational wave Telescope conceptual design study}}} (\bibinfo
  {year} {2011})\BibitemShut {NoStop}%
\bibitem [{\citenamefont {Basti}\ \emph {et~al.}(2011)\citenamefont {Basti},
  \citenamefont {Frasconi}, \citenamefont {Majorana}, \citenamefont
  {Naticchioni}, \citenamefont {Perciballi}, \citenamefont {Puppo},
  \citenamefont {Rapagnani},\ and\ \citenamefont {Ricci}}]{payload-2011}%
  \BibitemOpen
  \bibfield  {author} {\bibinfo {author} {\bibfnamefont {F.}~\bibnamefont
  {Basti}}, \bibinfo {author} {\bibfnamefont {F.}~\bibnamefont {Frasconi}},
  \bibinfo {author} {\bibfnamefont {E.}~\bibnamefont {Majorana}}, \bibinfo
  {author} {\bibfnamefont {L.}~\bibnamefont {Naticchioni}}, \bibinfo {author}
  {\bibfnamefont {M.}~\bibnamefont {Perciballi}}, \bibinfo {author}
  {\bibfnamefont {P.}~\bibnamefont {Puppo}}, \bibinfo {author} {\bibfnamefont
  {P.}~\bibnamefont {Rapagnani}},\ and\ \bibinfo {author} {\bibfnamefont
  {F.}~\bibnamefont {Ricci}},\ }\bibfield  {title} {\bibinfo {title} {A
  cryogenic payload for the 3rd generation of gravitational wave
  interferometers},\ }\href
  {https://doi.org/10.1016/j.astropartphys.2011.05.004  } {\bibfield  {journal}
  {\bibinfo  {journal} {Astroparticle Physics}\ }\textbf {\bibinfo {volume}
  {35}},\ \bibinfo {pages} {67} (\bibinfo {year} {2011})}\BibitemShut {NoStop}%
\bibitem [{\citenamefont {Puppo}\ \emph {et~al.}(2022)\citenamefont {Puppo},
  \citenamefont {Koroveshi}, \citenamefont {Majorana}, \citenamefont
  {Rapagnani},\ and\ \citenamefont {Grohmann}}]{puppo2}%
  \BibitemOpen
  \bibfield  {author} {\bibinfo {author} {\bibfnamefont {P.}~\bibnamefont
  {Puppo}}, \bibinfo {author} {\bibfnamefont {X.}~\bibnamefont {Koroveshi}},
  \bibinfo {author} {\bibfnamefont {E.}~\bibnamefont {Majorana}}, \bibinfo
  {author} {\bibfnamefont {P.}~\bibnamefont {Rapagnani}},\ and\ \bibinfo
  {author} {\bibfnamefont {S.}~\bibnamefont {Grohmann}},\ }\href
  {https://agenda.infn.it/event/28336/contributions/177984/} {\bibinfo {title}
  {{Update on the suspension thermal noise modelling of the ET-LF cryogenic
  payload}}},\ \bibinfo {howpublished} {Talk held at ECLOUD and GWDVac'22
  Workshops, Portoferraio, Italy} (\bibinfo {year} {2022})\BibitemShut
  {NoStop}%
\bibitem [{\citenamefont {{ET Steering Committee}}(2020)}]{ET2020}%
  \BibitemOpen
  \bibfield  {author} {\bibinfo {author} {\bibnamefont {{ET Steering
  Committee}}},\ }\href {https://apps.et-gw.eu/tds/ql/?c=15418} {\bibinfo
  {title} {{Design Report Update 2020 for the Einstein Telescope}}} (\bibinfo
  {year} {2020})\BibitemShut {NoStop}%
\bibitem [{\citenamefont {Yamamoto}(2022)}]{Yamamoto_Elba}%
  \BibitemOpen
  \bibfield  {author} {\bibinfo {author} {\bibfnamefont {K.}~\bibnamefont
  {Yamamoto}},\ }\href
  {https://agenda.infn.it/event/28336/contributions/179456/} {\bibinfo {title}
  {{Playload design at KAGRA and its impact to vacuum and cryogenics}}},\
  \bibinfo {howpublished} {Talk held at ECLOUD and GWDVac'22 Workshops,
  Portoferraio, Italy} (\bibinfo {year} {2022})\BibitemShut {NoStop}%
\bibitem [{\citenamefont {Naticchioni}\ and\ \citenamefont {on~behalf of~the
  Virgo~Collaboration}(2018)}]{Naticchioni_2018}%
  \BibitemOpen
  \bibfield  {author} {\bibinfo {author} {\bibfnamefont {L.}~\bibnamefont
  {Naticchioni}}\ and\ \bibinfo {author} {\bibnamefont {on~behalf of~the
  Virgo~Collaboration}},\ }\bibfield  {title} {\bibinfo {title} {{The payloads
  of Advanced Virgo: current status and upgrades}},\ }\href
  {https://doi.org/10.1088/1742-6596/957/1/012002 } {\bibfield  {journal}
  {\bibinfo  {journal} {Journal of Physics: Conference Series}\ }\textbf
  {\bibinfo {volume} {957}},\ \bibinfo {pages} {012002} (\bibinfo {year}
  {2018})}\BibitemShut {NoStop}%
\bibitem [{\citenamefont {McGuigan}\ \emph {et~al.}(1978)\citenamefont
  {McGuigan}, \citenamefont {Lam}, \citenamefont {Gram}, \citenamefont
  {Hoffman}, \citenamefont {Douglass},\ and\ \citenamefont
  {Gutche}}]{McGuigan1978MeasurementsOT}%
  \BibitemOpen
  \bibfield  {author} {\bibinfo {author} {\bibfnamefont {D.~F.}\ \bibnamefont
  {McGuigan}}, \bibinfo {author} {\bibfnamefont {C.~C.}\ \bibnamefont {Lam}},
  \bibinfo {author} {\bibfnamefont {R.~Q.}\ \bibnamefont {Gram}}, \bibinfo
  {author} {\bibfnamefont {A.~W.}\ \bibnamefont {Hoffman}}, \bibinfo {author}
  {\bibfnamefont {D.~H.}\ \bibnamefont {Douglass}},\ and\ \bibinfo {author}
  {\bibfnamefont {H.~W.}\ \bibnamefont {Gutche}},\ }\bibfield  {title}
  {\bibinfo {title} {Measurements of the mechanical {Q} of single-crystal
  silicon at low temperatures},\ }\href {https://doi.org/10.1007/BF00116202 }
  {\bibfield  {journal} {\bibinfo  {journal} {Journal of Low Temperature
  Physics}\ }\textbf {\bibinfo {volume} {30}},\ \bibinfo {pages} {621}
  (\bibinfo {year} {1978})}\BibitemShut {NoStop}%
\bibitem [{\citenamefont {Nawrodt}\ \emph {et~al.}(2013)\citenamefont {Nawrodt}
  \emph {et~al.}}]{Nawrodt2013}%
  \BibitemOpen
  \bibfield  {author} {\bibinfo {author} {\bibfnamefont {R.}~\bibnamefont
  {Nawrodt}} \emph {et~al.},\ }\bibfield  {title} {\bibinfo {title}
  {Investigation of mechanical losses of thin silicon flexures at low
  temperatures},\ }\href {https://doi.org/10.1088/0264-9381/30/11/115008}
  {\bibfield  {journal} {\bibinfo  {journal} {Fraunhofer IOF}\ }\textbf
  {\bibinfo {volume} {30}} (\bibinfo {year} {2013})}\BibitemShut {NoStop}%
\bibitem [{\citenamefont {Locke}\ \emph {et~al.}(2002)\citenamefont {Locke},
  \citenamefont {Tobar},\ and\ \citenamefont {Ivanov}}]{Tobar}%
  \BibitemOpen
  \bibfield  {author} {\bibinfo {author} {\bibfnamefont {C.~R.}\ \bibnamefont
  {Locke}}, \bibinfo {author} {\bibfnamefont {M.~E.}\ \bibnamefont {Tobar}},\
  and\ \bibinfo {author} {\bibfnamefont {E.~N.}\ \bibnamefont {Ivanov}},\
  }\bibfield  {title} {\bibinfo {title} {Properties of a monolithic sapphire
  parametric transducer: prospects of measuring the standard quantum limit},\
  }\href {https://doi.org/10.1088/0264-9381/19/7/388} {\bibfield  {journal}
  {\bibinfo  {journal} {Classical and Quantum Gravity}\ }\textbf {\bibinfo
  {volume} {19}},\ \bibinfo {pages} {1877} (\bibinfo {year}
  {2002})}\BibitemShut {NoStop}%
\bibitem [{\citenamefont {Amadori}\ \emph {et~al.}(2009)\citenamefont
  {Amadori}, \citenamefont {Bonetti}, \citenamefont {Pasquini}, \citenamefont
  {Deodati}, \citenamefont {Donnini}, \citenamefont {Montanari},\ and\
  \citenamefont {Testani}}]{AMADORI2009340}%
  \BibitemOpen
  \bibfield  {author} {\bibinfo {author} {\bibfnamefont {S.}~\bibnamefont
  {Amadori}}, \bibinfo {author} {\bibfnamefont {E.}~\bibnamefont {Bonetti}},
  \bibinfo {author} {\bibfnamefont {L.}~\bibnamefont {Pasquini}}, \bibinfo
  {author} {\bibfnamefont {P.}~\bibnamefont {Deodati}}, \bibinfo {author}
  {\bibfnamefont {R.}~\bibnamefont {Donnini}}, \bibinfo {author} {\bibfnamefont
  {R.}~\bibnamefont {Montanari}},\ and\ \bibinfo {author} {\bibfnamefont
  {C.}~\bibnamefont {Testani}},\ }\bibfield  {title} {\bibinfo {title} {Low
  temperature anelasticity in {Ti6Al4V alloy and Ti6Al4V–SiCf} composite},\
  }\href {https://doi.org/https://doi.org/10.1016/j.msea.2008.09.156}
  {\bibfield  {journal} {\bibinfo  {journal} {Materials Science and
  Engineering: A}\ }\textbf {\bibinfo {volume} {521-522}},\ \bibinfo {pages}
  {340} (\bibinfo {year} {2009})}\BibitemShut {NoStop}%
\bibitem [{\citenamefont {Duffy}(2000)}]{DUFFY2000417}%
  \BibitemOpen
  \bibfield  {author} {\bibinfo {author} {\bibfnamefont {W.}~\bibnamefont
  {Duffy}},\ }\bibfield  {title} {\bibinfo {title} {{Acoustic quality factor of
  titanium from 50 mK to 300 K}},\ }\href
  {https://doi.org/https://doi.org/10.1016/S0011-2275(00)00053-9  } {\bibfield
  {journal} {\bibinfo  {journal} {Cryogenics}\ }\textbf {\bibinfo {volume}
  {40}},\ \bibinfo {pages} {417} (\bibinfo {year} {2000})}\BibitemShut
  {NoStop}%
\bibitem [{\citenamefont {Duffy}(2002)}]{DuffyAluminium}%
  \BibitemOpen
  \bibfield  {author} {\bibinfo {author} {\bibfnamefont {W.}~\bibnamefont
  {Duffy}},\ }\bibfield  {title} {\bibinfo {title} {Acoustic quality factor of
  aluminium and selected aluminium alloys from {50 mK to 300 K}},\ }\href
  {https://doi.org/10.1016/S0011-2275(02)00021-8   } {\bibfield  {journal}
  {\bibinfo  {journal} {Cryogenics}\ }\textbf {\bibinfo {volume} {42}},\
  \bibinfo {pages} {245} (\bibinfo {year} {2002})}\BibitemShut {NoStop}%
\bibitem [{\citenamefont {Cumming}\ \emph {et~al.}(2013)\citenamefont {Cumming}
  \emph {et~al.}}]{Cumming_2013}%
  \BibitemOpen
  \bibfield  {author} {\bibinfo {author} {\bibfnamefont {A.~V.}\ \bibnamefont
  {Cumming}} \emph {et~al.},\ }\bibfield  {title} {\bibinfo {title} {Silicon
  mirror suspensions for gravitational wave detectors},\ }\href
  {https://doi.org/10.1088/0264-9381/31/2/025017   } {\bibfield  {journal}
  {\bibinfo  {journal} {Classical and Quantum Gravity}\ }\textbf {\bibinfo
  {volume} {31}},\ \bibinfo {pages} {025017} (\bibinfo {year}
  {2013})}\BibitemShut {NoStop}%
\bibitem [{\citenamefont {Dobrovinskaya}\ \emph {et~al.}(2009)\citenamefont
  {Dobrovinskaya}, \citenamefont {Lytvynov},\ and\ \citenamefont
  {Pishchik}}]{Dobrovinskaya2009}%
  \BibitemOpen
  \bibfield  {author} {\bibinfo {author} {\bibfnamefont {E.~R.}\ \bibnamefont
  {Dobrovinskaya}}, \bibinfo {author} {\bibfnamefont {L.~A.}\ \bibnamefont
  {Lytvynov}},\ and\ \bibinfo {author} {\bibfnamefont {V.}~\bibnamefont
  {Pishchik}},\ }\bibinfo {title} {Properties of sapphire},\ in\ \href
  {https://doi.org/10.1007/978-0-387-85695-7_2   } {\emph {\bibinfo {booktitle}
  {Sapphire: Material, Manufacturing, Applications}}}\ (\bibinfo  {publisher}
  {Springer US},\ \bibinfo {address} {Boston},\ \bibinfo {year} {2009})\ pp.\
  \bibinfo {pages} {55--176}\BibitemShut {NoStop}%
\bibitem [{\citenamefont {Ekin}(2006)}]{ekin}%
  \BibitemOpen
  \bibfield  {author} {\bibinfo {author} {\bibfnamefont {J.~W.}\ \bibnamefont
  {Ekin}},\ }\href@noop {} {\emph {\bibinfo {title} {Experimental Techniques
  for Low-Temperature Measurements}}}\ (\bibinfo  {publisher} {Oxford
  University Press, Oxford, UK},\ \bibinfo {year} {2006})\BibitemShut {NoStop}%
\bibitem [{\citenamefont {Touloukian}\ \emph {et~al.}(1970)\citenamefont
  {Touloukian}, \citenamefont {Powell}, \citenamefont {Ho},\ and\ \citenamefont
  {Klemens}}]{TPRC}%
  \BibitemOpen
  \bibfield  {author} {\bibinfo {author} {\bibfnamefont {Y.~S.}\ \bibnamefont
  {Touloukian}}, \bibinfo {author} {\bibfnamefont {R.~W.}\ \bibnamefont
  {Powell}}, \bibinfo {author} {\bibfnamefont {C.~Y.}\ \bibnamefont {Ho}},\
  and\ \bibinfo {author} {\bibfnamefont {P.~G.}\ \bibnamefont {Klemens}},\
  }\href@noop {} {\bibinfo {title} {{Thermophysical properties of matter - the
  TPRC data series. Volume 1. Thermal conductivity - metallic elements and
  alloys. Data book}}} (\bibinfo {year} {1970})\BibitemShut {NoStop}%
\bibitem [{\citenamefont {Khalaidovski}\ \emph {et~al.}(2014)\citenamefont
  {Khalaidovski} \emph {et~al.}}]{Khalaidovski_2014}%
  \BibitemOpen
  \bibfield  {author} {\bibinfo {author} {\bibfnamefont {A.}~\bibnamefont
  {Khalaidovski}} \emph {et~al.},\ }\bibfield  {title} {\bibinfo {title}
  {{Evaluation of heat extraction through sapphire fibers for the GW
  observatory KAGRA}},\ }\href {https://doi.org/10.1088/0264-9381/31/10/105004}
  {\bibfield  {journal} {\bibinfo  {journal} {Classical and Quantum Gravity}\
  }\textbf {\bibinfo {volume} {31}},\ \bibinfo {pages} {105004} (\bibinfo
  {year} {2014})}\BibitemShut {NoStop}%
\bibitem [{\citenamefont {{CryoData Inc.}}(1999)}]{Cryocomp}%
  \BibitemOpen
  \bibfield  {author} {\bibinfo {author} {\bibnamefont {{CryoData Inc.}}},\
  }\href {https://trc.nist.gov/cryogenics/materials/materialproperties.htm}
  {\bibinfo {title} {{CryoComp, Data from: NIST - Properties of solid materials
  from cryogenic- to room-temperatures}}} (\bibinfo {year} {1999})\BibitemShut
  {NoStop}%
\bibitem [{\citenamefont {Baudouy}\ and\ \citenamefont
  {Four}(2014)}]{BAUDOUY20141}%
  \BibitemOpen
  \bibfield  {author} {\bibinfo {author} {\bibfnamefont {B.}~\bibnamefont
  {Baudouy}}\ and\ \bibinfo {author} {\bibfnamefont {A.}~\bibnamefont {Four}},\
  }\bibfield  {title} {\bibinfo {title} {Low temperature thermal conductivity
  of aluminum alloy 5056},\ }\href
  {https://doi.org/https://doi.org/10.1016/j.cryogenics.2013.12.008} {\bibfield
   {journal} {\bibinfo  {journal} {Cryogenics}\ }\textbf {\bibinfo {volume}
  {60}},\ \bibinfo {pages} {1} (\bibinfo {year} {2014})}\BibitemShut {NoStop}%
\bibitem [{\citenamefont {Touloukian}\ and\ \citenamefont
  {Buyco}(1971)}]{TPRCcp}%
  \BibitemOpen
  \bibfield  {author} {\bibinfo {author} {\bibfnamefont {Y.~S.}\ \bibnamefont
  {Touloukian}}\ and\ \bibinfo {author} {\bibfnamefont {E.~H.}\ \bibnamefont
  {Buyco}},\ }\href@noop {} {\bibinfo {title} {{Thermophysical properties of
  matter - the TPRC data series. Volume 4. Specific heat - metallic elements
  and alloys. Data book}}} (\bibinfo {year} {1971})\BibitemShut {NoStop}%
\bibitem [{\citenamefont {White}\ and\ \citenamefont
  {Minges}(1997)}]{White1997ThermophysicalPO}%
  \BibitemOpen
  \bibfield  {author} {\bibinfo {author} {\bibfnamefont {G.~K.}\ \bibnamefont
  {White}}\ and\ \bibinfo {author} {\bibfnamefont {M.~L.}\ \bibnamefont
  {Minges}},\ }\bibfield  {title} {\bibinfo {title} {Thermophysical properties
  of some key solids: An update},\ }\href
  {https://link.springer.com/content/pdf/10.1007/BF02575261  .pdf} {\bibfield
  {journal} {\bibinfo  {journal} {International Journal of Thermophysics}\
  }\textbf {\bibinfo {volume} {18}},\ \bibinfo {pages} {1269} (\bibinfo {year}
  {1997})}\BibitemShut {NoStop}%
\bibitem [{\citenamefont {Barucci}\ \emph {et~al.}(2010)\citenamefont
  {Barucci}, \citenamefont {Ligi}, \citenamefont {Lolli}, \citenamefont
  {Marini}, \citenamefont {Martelli}, \citenamefont {Risegari},\ and\
  \citenamefont {Ventura}}]{BARUCCI20101452}%
  \BibitemOpen
  \bibfield  {author} {\bibinfo {author} {\bibfnamefont {M.}~\bibnamefont
  {Barucci}}, \bibinfo {author} {\bibfnamefont {C.}~\bibnamefont {Ligi}},
  \bibinfo {author} {\bibfnamefont {L.}~\bibnamefont {Lolli}}, \bibinfo
  {author} {\bibfnamefont {A.}~\bibnamefont {Marini}}, \bibinfo {author}
  {\bibfnamefont {V.}~\bibnamefont {Martelli}}, \bibinfo {author}
  {\bibfnamefont {L.}~\bibnamefont {Risegari}},\ and\ \bibinfo {author}
  {\bibfnamefont {G.}~\bibnamefont {Ventura}},\ }\bibfield  {title} {\bibinfo
  {title} {Very low temperature specific heat of al 5056},\ }\href
  {https://doi.org/10.1016/j.physb.2009.11.013 } {\bibfield  {journal} {\bibinfo
   {journal} {Physica B: Condensed Matter}\ }\textbf {\bibinfo {volume}
  {405}},\ \bibinfo {pages} {1452} (\bibinfo {year} {2010})}\BibitemShut
  {NoStop}%
\bibitem [{\citenamefont {Swenson}(1983)}]{ThermalexpansioncoeffSi}%
  \BibitemOpen
  \bibfield  {author} {\bibinfo {author} {\bibfnamefont {C.~A.}\ \bibnamefont
  {Swenson}},\ }\bibfield  {title} {\bibinfo {title} {{Recommended Values for
  the Thermal Expansivity of Silicon from 0 to 1000 K}},\ }\href
  {https://doi.org/10.1063/1.555681 } {\bibfield  {journal} {\bibinfo  {journal}
  {Journal of Physical and Chemical Reference Data}\ }\textbf {\bibinfo
  {volume} {12}},\ \bibinfo {pages} {179} (\bibinfo {year} {1983})}\BibitemShut
  {NoStop}%
\bibitem [{\citenamefont {Taylor}\ \emph {et~al.}(1996)\citenamefont {Taylor},
  \citenamefont {Notcutt}, \citenamefont {Wong}, \citenamefont {Mann},\ and\
  \citenamefont {Blair}}]{TAYLOR1996Saalpha}%
  \BibitemOpen
  \bibfield  {author} {\bibinfo {author} {\bibfnamefont {C.}~\bibnamefont
  {Taylor}}, \bibinfo {author} {\bibfnamefont {M.}~\bibnamefont {Notcutt}},
  \bibinfo {author} {\bibfnamefont {E.}~\bibnamefont {Wong}}, \bibinfo {author}
  {\bibfnamefont {A.}~\bibnamefont {Mann}},\ and\ \bibinfo {author}
  {\bibfnamefont {D.}~\bibnamefont {Blair}},\ }\bibfield  {title} {\bibinfo
  {title} {Measurement of the coefficient of thermal expansion of a cryogenic,
  all-sapphire, {Fabry-Perot optical} cavity},\ }\href
  {https://doi.org/https://doi.org/10.1016/0030-4018(96)00293-3  } {\bibfield
  {journal} {\bibinfo  {journal} {Optics Communications}\ }\textbf {\bibinfo
  {volume} {131}},\ \bibinfo {pages} {311} (\bibinfo {year}
  {1996})}\BibitemShut {NoStop}%
\bibitem [{\citenamefont {Touloukian}\ \emph {et~al.}(1975)\citenamefont
  {Touloukian}, \citenamefont {Kirby}, \citenamefont {Taylor},\ and\
  \citenamefont {Desai}}]{TPRCalpha}%
  \BibitemOpen
  \bibfield  {author} {\bibinfo {author} {\bibfnamefont {Y.~S.}\ \bibnamefont
  {Touloukian}}, \bibinfo {author} {\bibfnamefont {R.~K.}\ \bibnamefont
  {Kirby}}, \bibinfo {author} {\bibfnamefont {R.~E.}\ \bibnamefont {Taylor}},\
  and\ \bibinfo {author} {\bibfnamefont {P.~D.}\ \bibnamefont {Desai}},\
  }\href@noop {} {\bibinfo {title} {{Thermophysical properties of matter - the
  TPRC data series. Volume 12. Thermal expansion metallic elements and alloys.
  Data book}}} (\bibinfo {year} {1975})\BibitemShut {NoStop}%
\bibitem [{\citenamefont {Gysin}\ \emph {et~al.}(2004)\citenamefont {Gysin},
  \citenamefont {Rast}, \citenamefont {Ruff}, \citenamefont {Meyer},
  \citenamefont {Lee}, \citenamefont {Vettiger},\ and\ \citenamefont
  {Gerber}}]{GysinbetaSi}%
  \BibitemOpen
  \bibfield  {author} {\bibinfo {author} {\bibfnamefont {U.}~\bibnamefont
  {Gysin}}, \bibinfo {author} {\bibfnamefont {S.}~\bibnamefont {Rast}},
  \bibinfo {author} {\bibfnamefont {P.}~\bibnamefont {Ruff}}, \bibinfo {author}
  {\bibfnamefont {E.}~\bibnamefont {Meyer}}, \bibinfo {author} {\bibfnamefont
  {D.~W.}\ \bibnamefont {Lee}}, \bibinfo {author} {\bibfnamefont
  {P.}~\bibnamefont {Vettiger}},\ and\ \bibinfo {author} {\bibfnamefont
  {C.}~\bibnamefont {Gerber}},\ }\bibfield  {title} {\bibinfo {title}
  {Temperature dependence of the force sensitivity of silicon cantilevers},\
  }\href {https://doi.org/10.1103/PhysRevB.69.045403  } {\bibfield  {journal}
  {\bibinfo  {journal} {Phys. Rev. B}\ }\textbf {\bibinfo {volume} {69}},\
  \bibinfo {pages} {045403} (\bibinfo {year} {2004})}\BibitemShut {NoStop}%
\bibitem [{\citenamefont {{Wachtman Jr.}}\ \emph {et~al.}(1961)\citenamefont
  {{Wachtman Jr.}}, \citenamefont {Tefft}, \citenamefont {{Lam Jr.}},\ and\
  \citenamefont {Apstein}}]{Wachtman_betaSa}%
  \BibitemOpen
  \bibfield  {author} {\bibinfo {author} {\bibfnamefont {J.~B.}\ \bibnamefont
  {{Wachtman Jr.}}}, \bibinfo {author} {\bibfnamefont {W.~E.}\ \bibnamefont
  {Tefft}}, \bibinfo {author} {\bibfnamefont {D.~G.}\ \bibnamefont {{Lam
  Jr.}}},\ and\ \bibinfo {author} {\bibfnamefont {C.~S.}\ \bibnamefont
  {Apstein}},\ }\bibfield  {title} {\bibinfo {title} {Exponential temperature
  dependence of young's modulus for several oxides},\ }\href
  {https://doi.org/10.1103/PhysRev.122.1754  } {\bibfield  {journal} {\bibinfo
  {journal} {Phys. Rev.}\ }\textbf {\bibinfo {volume} {122}},\ \bibinfo {pages}
  {1754} (\bibinfo {year} {1961})}\BibitemShut {NoStop}%
\bibitem [{\citenamefont {Fukuhara}\ and\ \citenamefont
  {Sanpei}(1993)}]{Fukuhara1993ElasticMA}%
  \BibitemOpen
  \bibfield  {author} {\bibinfo {author} {\bibfnamefont {M.}~\bibnamefont
  {Fukuhara}}\ and\ \bibinfo {author} {\bibfnamefont {A.}~\bibnamefont
  {Sanpei}},\ }\bibfield  {title} {\bibinfo {title} {{Elastic moduli and
  internal frictions of Inconel 718 and Ti-6Al-4V as a function of
  temperature}},\ }\href {https://link.springer.com/article/10.1007/BF00420541 }
  {\bibfield  {journal} {\bibinfo  {journal} {Journal of Materials Science
  Letters}\ }\textbf {\bibinfo {volume} {12}},\ \bibinfo {pages} {1122}
  (\bibinfo {year} {1993})}\BibitemShut {NoStop}%
\bibitem [{\citenamefont {{National Institute of Standards and Technology
  (NIST)}}()}]{NIST}%
  \BibitemOpen
  \bibfield  {author} {\bibinfo {author} {\bibnamefont {{National Institute of
  Standards and Technology (NIST)}}},\ }\href
  {https://trc.nist.gov/cryogenics/materials/materialproperties.htm} {\bibinfo
  {title} {Properties of solid materials from cryogenic- to
  room-temperatures}}\BibitemShut {NoStop}%
\bibitem [{\citenamefont {Hopcroft}\ \emph {et~al.}(2010)\citenamefont
  {Hopcroft}, \citenamefont {Nix},\ and\ \citenamefont {Kenny}}]{ESiRef}%
  \BibitemOpen
  \bibfield  {author} {\bibinfo {author} {\bibfnamefont {M.~A.}\ \bibnamefont
  {Hopcroft}}, \bibinfo {author} {\bibfnamefont {W.~D.}\ \bibnamefont {Nix}},\
  and\ \bibinfo {author} {\bibfnamefont {T.~W.}\ \bibnamefont {Kenny}},\
  }\bibfield  {title} {\bibinfo {title} {{What is the Young's Modulus of
  Silicon?}},\ }\href {https://doi.org/10.1109/JMEMS.2009.2039697 } {\bibfield
  {journal} {\bibinfo  {journal} {Journal of Microelectromechanical Systems}\
  }\textbf {\bibinfo {volume} {19}},\ \bibinfo {pages} {229} (\bibinfo {year}
  {2010})}\BibitemShut {NoStop}%
\bibitem [{\citenamefont {{Wachtman Jr.}}\ and\ \citenamefont {{Lam
  Jr.}}(1959)}]{SaE_Mod}%
  \BibitemOpen
  \bibfield  {author} {\bibinfo {author} {\bibfnamefont {J.~B.}\ \bibnamefont
  {{Wachtman Jr.}}}\ and\ \bibinfo {author} {\bibfnamefont {D.~G.}\
  \bibnamefont {{Lam Jr.}}},\ }\bibfield  {title} {\bibinfo {title} {Young's
  modulus of various refractory materials as a function of temperature},\
  }\href {https://doi.org/https://doi.org/10.1111/j.1151-2916.1959.tb15462.x}
  {\bibfield  {journal} {\bibinfo  {journal} {Journal of the American Ceramic
  Society}\ }\textbf {\bibinfo {volume} {42}},\ \bibinfo {pages} {254}
  (\bibinfo {year} {1959})}\BibitemShut {NoStop}%
\bibitem [{\citenamefont {Boyer}\ \emph {et~al.}(1994)\citenamefont {Boyer},
  \citenamefont {Welsch},\ and\ \citenamefont
  {Collings}}]{Boyer1994MaterialsPH}%
  \BibitemOpen
  \bibfield  {author} {\bibinfo {author} {\bibfnamefont {R.~R.}\ \bibnamefont
  {Boyer}}, \bibinfo {author} {\bibfnamefont {G.}~\bibnamefont {Welsch}},\ and\
  \bibinfo {author} {\bibfnamefont {E.~W.}\ \bibnamefont {Collings}},\
  }\href@noop {} {\bibinfo {title} {Materials properties handbook: Titanium
  alloys}} (\bibinfo {year} {1994})\BibitemShut {NoStop}%
\bibitem [{\citenamefont {Nawrodt}\ \emph {et~al.}(2009)\citenamefont {Nawrodt}
  \emph {et~al.}}]{ET02709}%
  \BibitemOpen
  \bibfield  {author} {\bibinfo {author} {\bibfnamefont {R.}~\bibnamefont
  {Nawrodt}} \emph {et~al.},\ }\href {https://apps.et-gw.eu/tds/ql/?c=7556}
  {\bibinfo {title} {{ Mirror thermal noise calculation for ET}}} (\bibinfo
  {year} {2009})\BibitemShut {NoStop}%
\bibitem [{\citenamefont {Majorana}(2021)}]{EttoreGWADW2021}%
  \BibitemOpen
  \bibfield  {author} {\bibinfo {author} {\bibfnamefont {E.}~\bibnamefont
  {Majorana}},\ }\href
  {https://agenda.infn.it/event/26121/contributions/136321/attachments/81472/106807/GWDAW21_majorana_1.pdf}
  {\bibinfo {title} {{Outline of cryogenic payload compliance with Einstein
  Telescope LF}}},\ \bibinfo {howpublished} {Talk held at GWADW} (\bibinfo
  {year} {2021})\BibitemShut {NoStop}%
\bibitem [{\citenamefont {Puppo}(2022)}]{PaolaGWADW2022}%
  \BibitemOpen
  \bibfield  {author} {\bibinfo {author} {\bibfnamefont {P.}~\bibnamefont
  {Puppo}},\ }\href {https://apps.et-gw.eu/tds/ql/?c=16309} {\bibinfo {title}
  {{FEA models for the ET payload: status and preliminary results}}},\ \bibinfo
  {howpublished} {Talk held at GWADW} (\bibinfo {year} {2022})\BibitemShut
  {NoStop}%
\bibitem [{\citenamefont {Sumomogi}\ \emph {et~al.}(2004)\citenamefont
  {Sumomogi}, \citenamefont {Masashi~Yoshida}, \citenamefont {Osono},\ and\
  \citenamefont {Kino}}]{Sumomogi2004MechanicalPO}%
  \BibitemOpen
  \bibfield  {author} {\bibinfo {author} {\bibfnamefont {T.}~\bibnamefont
  {Sumomogi}}, \bibinfo {author} {\bibfnamefont {M.~N.}\ \bibnamefont
  {Masashi~Yoshida}}, \bibinfo {author} {\bibfnamefont {H.}~\bibnamefont
  {Osono}},\ and\ \bibinfo {author} {\bibfnamefont {T.}~\bibnamefont {Kino}},\
  }\bibfield  {title} {\bibinfo {title} {Mechanical properties of ultra
  high-purity aluminum},\ }\href {https://doi.org/10.2320/jinstmet.68.958 }
  {\bibfield  {journal} {\bibinfo  {journal} {Journal of The Japan Institute of
  Metals}\ }\textbf {\bibinfo {volume} {68}},\ \bibinfo {pages} {958} (\bibinfo
  {year} {2004})}\BibitemShut {NoStop}%
\bibitem [{\citenamefont {Nawrodt}\ \emph {et~al.}(2008)\citenamefont {Nawrodt}
  \emph {et~al.}}]{Nawrodt_2008}%
  \BibitemOpen
  \bibfield  {author} {\bibinfo {author} {\bibfnamefont {R.}~\bibnamefont
  {Nawrodt}} \emph {et~al.},\ }\bibfield  {title} {\bibinfo {title} {High
  mechanical {Q-factor} measurements on silicon bulk samples},\ }\href
  {https://doi.org/10.1088/1742-6596/122/1/012008 } {\bibfield  {journal}
  {\bibinfo  {journal} {Journal of Physics: Conference Series}\ }\textbf
  {\bibinfo {volume} {122}},\ \bibinfo {pages} {012008} (\bibinfo {year}
  {2008})}\BibitemShut {NoStop}%
\bibitem [{\citenamefont {Saulson}(1990)}]{Saulson}%
  \BibitemOpen
  \bibfield  {author} {\bibinfo {author} {\bibfnamefont {P.~R.}\ \bibnamefont
  {Saulson}},\ }\bibfield  {title} {\bibinfo {title} {Thermal noise in
  mechanical experiments},\ }\href {https://doi.org/10.1103/PhysRevD.42.2437  }
  {\bibfield  {journal} {\bibinfo  {journal} {Phys. Rev. D}\ }\textbf {\bibinfo
  {volume} {42}},\ \bibinfo {pages} {2437} (\bibinfo {year}
  {1990})}\BibitemShut {NoStop}%
\bibitem [{\citenamefont {Nowick}\ and\ \citenamefont
  {Berry}(1972)}]{Nowick1972}%
  \BibitemOpen
  \bibfield  {author} {\bibinfo {author} {\bibfnamefont {A.~S.}\ \bibnamefont
  {Nowick}}\ and\ \bibinfo {author} {\bibfnamefont {B.~S.}\ \bibnamefont
  {Berry}},\ }\href@noop {} {\emph {\bibinfo {title} {Anelastic Relaxation in
  Crystalline Solids}}}\ (\bibinfo  {publisher} {Academic Press},\ \bibinfo
  {year} {1972})\BibitemShut {NoStop}%
\bibitem [{\citenamefont {Yamada}\ \emph {et~al.}(2021)\citenamefont {Yamada}
  \emph {et~al.}}]{YAMADA2021103280}%
  \BibitemOpen
  \bibfield  {author} {\bibinfo {author} {\bibfnamefont {T.}~\bibnamefont
  {Yamada}} \emph {et~al.},\ }\bibfield  {title} {\bibinfo {title} {High
  performance thermal link with small spring constant for cryogenic
  applications},\ }\href
  {https://doi.org/https://doi.org/10.1016/j.cryogenics.2021.103280   } {\bibfield
   {journal} {\bibinfo  {journal} {Cryogenics}\ }\textbf {\bibinfo {volume}
  {116}},\ \bibinfo {pages} {103280} (\bibinfo {year} {2021})}\BibitemShut
  {NoStop}%
\bibitem [{\citenamefont {Yamada}(2021)}]{yamadaGW21}%
  \BibitemOpen
  \bibfield  {author} {\bibinfo {author} {\bibfnamefont {T.}~\bibnamefont
  {Yamada}},\ }\href
  {https://agenda.infn.it/event/26121/contributions/136324/attachments/81566/106947/gwadw2021_tyamada_reduced.pdf}
  {\bibinfo {title} {{Reduction of vibration transfer via heat links in KAGRA
  cryogenic mirror suspension system}}},\ \bibinfo {howpublished} {Talk held at
  GWADW} (\bibinfo {year} {2021})\BibitemShut {NoStop}%
  \bibitem [{\citenamefont {{Ushiba}}\ \emph {et~al.}(2021)\citenamefont {{Ushiba}} \emph {et~al.}}]{Ushiba_2021}%
  \BibitemOpen
  \bibfield  {author} {\bibinfo {author} {\bibfnamefont {T.}~\bibnamefont {{Ushiba}}} \emph {et~al.},\ }\bibfield  {title} {\bibinfo {title} {{Cryogenic suspension design for a kilometer-scale gravitational-wave detector}},\ }\href
  {https://doi.org/10.1088/1361-6382/abe9f3} {\bibfield
  {journal} {\bibinfo  {journal} {Classical and Quantum Gravity}\ }\textbf {\bibinfo {volume}
  {38}},\ \bibinfo {number} {8} (\bibinfo {year} {2021})}\BibitemShut {NoStop}%
\bibitem [{\citenamefont {Ruggi}(2022)}]{RUGGI}%
  \BibitemOpen
  \bibfield  {author} {\bibinfo {author} {\bibfnamefont {P.}~\bibnamefont
  {Ruggi}},\ }\href {https://agenda.infn.it/event/28968/contributions/176786/}
  {\bibinfo {title} {Mechanical noise in gravitational wave detectors}},\
  \bibinfo {howpublished} {Talk held at Amaldi Research Center Summer School,
  Paestum, Italy} (\bibinfo {year} {2022})\BibitemShut {NoStop}%
\bibitem [{\citenamefont {Dari}\ \emph {et~al.}(2010)\citenamefont {Dari},
  \citenamefont {Travasso}, \citenamefont {Vocca},\ and\ \citenamefont
  {Gammaitoni}}]{Dari_2010}%
  \BibitemOpen
  \bibfield  {author} {\bibinfo {author} {\bibfnamefont {A.}~\bibnamefont
  {Dari}}, \bibinfo {author} {\bibfnamefont {F.}~\bibnamefont {Travasso}},
  \bibinfo {author} {\bibfnamefont {H.}~\bibnamefont {Vocca}},\ and\ \bibinfo
  {author} {\bibfnamefont {L.}~\bibnamefont {Gammaitoni}},\ }\bibfield  {title}
  {\bibinfo {title} {Breaking strength tests on silicon and sapphire bondings
  for gravitational wave detectors},\ }\href
  {https://doi.org/10.1088/0264-9381/27/4/045010   } {\bibfield  {journal}
  {\bibinfo  {journal} {Classical and Quantum Gravity}\ }\textbf {\bibinfo
  {volume} {27}},\ \bibinfo {pages} {045010} (\bibinfo {year}
  {2010})}\BibitemShut {NoStop}%
\bibitem [{\citenamefont {Phelps}\ \emph {et~al.}(2018)\citenamefont {Phelps},
  \citenamefont {Reid}, \citenamefont {Douglas}, \citenamefont {van Veggel},
  \citenamefont {Mangano}, \citenamefont {Haughian}, \citenamefont
  {Jongschaap}, \citenamefont {Kelly}, \citenamefont {Hough},\ and\
  \citenamefont {Rowan}}]{Glasgow}%
  \BibitemOpen
  \bibfield  {author} {\bibinfo {author} {\bibfnamefont {M.}~\bibnamefont
  {Phelps}}, \bibinfo {author} {\bibfnamefont {M.~M.}\ \bibnamefont {Reid}},
  \bibinfo {author} {\bibfnamefont {R.}~\bibnamefont {Douglas}}, \bibinfo
  {author} {\bibfnamefont {A.-M.}\ \bibnamefont {van Veggel}}, \bibinfo
  {author} {\bibfnamefont {V.}~\bibnamefont {Mangano}}, \bibinfo {author}
  {\bibfnamefont {K.}~\bibnamefont {Haughian}}, \bibinfo {author}
  {\bibfnamefont {A.}~\bibnamefont {Jongschaap}}, \bibinfo {author}
  {\bibfnamefont {M.}~\bibnamefont {Kelly}}, \bibinfo {author} {\bibfnamefont
  {J.}~\bibnamefont {Hough}},\ and\ \bibinfo {author} {\bibfnamefont
  {S.}~\bibnamefont {Rowan}},\ }\bibfield  {title} {\bibinfo {title} {Strength
  of hydroxide catalysis bonds between sapphire, silicon, and fused silica as a
  function of time},\ }\href {https://doi.org/10.1103/PhysRevD.98.122003  }
  {\bibfield  {journal} {\bibinfo  {journal} {Phys. Rev. D}\ }\textbf {\bibinfo
  {volume} {98}},\ \bibinfo {pages} {122003} (\bibinfo {year}
  {2018})}\BibitemShut {NoStop}%
\bibitem [{\citenamefont {Yamada}(2023)}]{Yamada23}%
  \BibitemOpen
  \bibfield  {author} {\bibinfo {author} {\bibfnamefont {T.}~\bibnamefont
  {Yamada}},\ }\href
  {https://gwdoc.icrr.u-tokyo.ac.jp/cgi-bin/DocDB/ShowDocument?docid=14883}
  {\bibinfo {title} {{Sapphire bending tests, Public and Internal note database
  at Institute for Cosmic Ray Research (ICRR) University of Tokyo,
  JGW-T2314883-v1}}} (\bibinfo {year} {2023})\BibitemShut {NoStop}%
\bibitem [{Shi()}]{Shinkosha}%
  \BibitemOpen
  \href@noop {} {\bibinfo {title} {{SHINKOSHA CO., LTD. 2-4-1 Kosugaya,
  Sakae-ku, Yokohama, Kanagawa 247-0007 Japan}}}\BibitemShut {NoStop}%
\bibitem [{\citenamefont {Scurlock}(1966)}]{Scurlock1966}%
  \BibitemOpen
  \bibfield  {author} {\bibinfo {author} {\bibfnamefont {R.~G.}\ \bibnamefont
  {Scurlock}},\ }\href@noop {} {\emph {\bibinfo {title} {{Low Temperature
  Behaviour of Solids}}}}\ (\bibinfo  {publisher} {Routledge and Kegan Paul
  PLC},\ \bibinfo {year} {1966})\BibitemShut {NoStop}%
\bibitem [{\citenamefont {Travasso}\ \emph {et~al.}(2019)\citenamefont
  {Travasso} \emph {et~al.}}]{FlavioSiSuspensions}%
  \BibitemOpen
  \bibfield  {author} {\bibinfo {author} {\bibfnamefont {F.}~\bibnamefont
  {Travasso}} \emph {et~al.},\ }\bibfield  {title} {\bibinfo {title} {{Towards
  a silicon monolithic suspension}},\ }\href
  {https://doi.org/10.5281/zenodo.3820523  } {\bibfield  {journal} {\bibinfo
  {journal} {Proceedings of the 2nd GRavitational-waves Science \& technology
  Symposium (GRASS), Padova, Italy}\ } (\bibinfo {year} {2019})}\BibitemShut
  {NoStop}%
\bibitem [{\citenamefont {Sato}\ \emph {et~al.}(2006)\citenamefont {Sato},
  \citenamefont {Maeda}, \citenamefont {Dantsuka}, \citenamefont {Yuyama},\
  and\ \citenamefont {Kamioka}}]{sato2006}%
  \BibitemOpen
  \bibfield  {author} {\bibinfo {author} {\bibfnamefont {A.}~\bibnamefont
  {Sato}}, \bibinfo {author} {\bibfnamefont {M.}~\bibnamefont {Maeda}},
  \bibinfo {author} {\bibfnamefont {T.}~\bibnamefont {Dantsuka}}, \bibinfo
  {author} {\bibfnamefont {M.}~\bibnamefont {Yuyama}},\ and\ \bibinfo {author}
  {\bibfnamefont {Y.}~\bibnamefont {Kamioka}},\ }\bibfield  {title} {\bibinfo
  {title} {{Temperature Dependence of the Gorter-Mellink Exponent m Measured in
  a Channel Containing He II}},\ }\href {https://doi.org/10.1063/1.2202439 }
  {\bibfield  {journal} {\bibinfo  {journal} {AIP Conference Proceedings}\
  }\textbf {\bibinfo {volume} {823}},\ \bibinfo {pages} {387} (\bibinfo {year}
  {2006})}\BibitemShut {NoStop}%
\bibitem [{\citenamefont {{CryoData Inc.}}(1998)}]{HEPAK}%
  \BibitemOpen
  \bibfield  {author} {\bibinfo {author} {\bibnamefont {{CryoData Inc.}}},\
  }\href {https://htess.com/hepak/} {\bibinfo {title} {{HEPAK, Data from: V. D.
  Arp, R. D. Mccarty and D. G. Friend - Thermophysical properties of Helium-4
  from 0.8 to 1500 K with pressures to 2000 MPa.}}} (\bibinfo {year}
  {1998})\BibitemShut {NoStop}%
\bibitem [{\citenamefont {Touloukian}\ \emph {et~al.}(1971)\citenamefont
  {Touloukian}, \citenamefont {Powell}, \citenamefont {Ho},\ and\ \citenamefont
  {Klemens}}]{TPRC2}%
  \BibitemOpen
  \bibfield  {author} {\bibinfo {author} {\bibfnamefont {Y.~S.}\ \bibnamefont
  {Touloukian}}, \bibinfo {author} {\bibfnamefont {R.~W.}\ \bibnamefont
  {Powell}}, \bibinfo {author} {\bibfnamefont {C.~Y.}\ \bibnamefont {Ho}},\
  and\ \bibinfo {author} {\bibfnamefont {P.~G.}\ \bibnamefont {Klemens}},\
  }\href@noop {} {\bibinfo {title} {{Thermophysical properties of matter - the
  TPRC data series. Volume 2. Thermal conductivity - nonmetallic solids. Data
  book}}} (\bibinfo {year} {1971})\BibitemShut {NoStop}%
\bibitem [{\citenamefont {Astone}\ \emph {et~al.}(1992)\citenamefont {Astone}
  \emph {et~al.}}]{Explorer}%
  \BibitemOpen
  \bibfield  {author} {\bibinfo {author} {\bibfnamefont {P.}~\bibnamefont
  {Astone}} \emph {et~al.},\ }\bibfield  {title} {\bibinfo {title} {Noise
  behaviour of the explorer gravitational wave antenna during $\lambda$
  transition to the superfluid phase},\ }\href
  {https://doi.org/10.1016/0011-2275(92)90300-Y } {\bibfield  {journal}
  {\bibinfo  {journal} {Cryogenics}\ }\textbf {\bibinfo {volume} {32}},\
  \bibinfo {pages} {668} (\bibinfo {year} {1992})}\BibitemShut {NoStop}%
\bibitem [{\citenamefont {Puppo}\ and\ \citenamefont {Ricci}(2011)}]{GR-Cryo}%
  \BibitemOpen
  \bibfield  {author} {\bibinfo {author} {\bibfnamefont {P.}~\bibnamefont
  {Puppo}}\ and\ \bibinfo {author} {\bibfnamefont {F.}~\bibnamefont {Ricci}},\
  }\bibfield  {title} {\bibinfo {title} {{Cryogenics and Einstein Telescope}},\
  }\href {https://doi.org/10.1007/s10714-010-1037-x } {\bibfield  {journal}
  {\bibinfo  {journal} {General Relativity and Gravitation volume}\ }\textbf
  {\bibinfo {volume} {43}},\ \bibinfo {pages} {657} (\bibinfo {year}
  {2011})}\BibitemShut {NoStop}%
\bibitem [{\citenamefont {Vinen}(2004)}]{Vinen2004}%
  \BibitemOpen
  \bibfield  {author} {\bibinfo {author} {\bibfnamefont {W.~F.}\ \bibnamefont
  {Vinen}},\ }\href
  {https://www.semanticscholar.org/paper/The-physics-of-superfluid-helium-Vinen/bb23c04171209d6ebace1ad8ed2f3718620747c5}
  {\bibinfo {title} {The physics of superfluid helium}} (\bibinfo {year}
  {2004})\BibitemShut {NoStop}%
\bibitem [{\citenamefont {Van~Sciver}(2012)}]{Hebook}%
  \BibitemOpen
  \bibfield  {author} {\bibinfo {author} {\bibfnamefont {S.~W.}\ \bibnamefont
  {Van~Sciver}},\ }\href {https://doi.org/10.1007/978-1-4419-9979-5  } {\emph
  {\bibinfo {title} {Helium Cryogenics}}},\ \bibinfo {edition} {2nd}\ ed.,\
  International Cryogenics Monograph Series\ (\bibinfo  {publisher} {Springer
  New York, NY},\ \bibinfo {year} {2012})\BibitemShut {NoStop}%
\bibitem [{\citenamefont {Landau}(1941)}]{landau}%
  \BibitemOpen
  \bibfield  {author} {\bibinfo {author} {\bibfnamefont {L.}~\bibnamefont
  {Landau}},\ }\bibfield  {title} {\bibinfo {title} {Theory of the
  superfluidity of {Helium II}},\ }\href
  {https://doi.org/10.1103/PhysRev.60.356 } {\bibfield  {journal} {\bibinfo
  {journal} {Phys. Rev.}\ }\textbf {\bibinfo {volume} {60}},\ \bibinfo {pages}
  {356} (\bibinfo {year} {1941})}\BibitemShut {NoStop}%
\bibitem [{\citenamefont {Tisza}(1938)}]{tisza}%
  \BibitemOpen
  \bibfield  {author} {\bibinfo {author} {\bibfnamefont {L.}~\bibnamefont
  {Tisza}},\ }\bibfield  {title} {\bibinfo {title} {Transport phenomena in
  {Helium II}},\ }\href {https://doi.org/https://doi.org/10.1038/141913a0    }
  {\bibfield  {journal} {\bibinfo  {journal} {Nature}\ }\textbf {\bibinfo
  {volume} {141}},\ \bibinfo {pages} {913} (\bibinfo {year}
  {1938})}\BibitemShut {NoStop}%
\bibitem [{\citenamefont {Yamada}\ and\ \citenamefont {on~behalf of~the
  KAGRA~Collaboration}(2020)}]{Yamada_2020}%
  \BibitemOpen
  \bibfield  {author} {\bibinfo {author} {\bibfnamefont {T.}~\bibnamefont
  {Yamada}}\ and\ \bibinfo {author} {\bibnamefont {on~behalf of~the
  KAGRA~Collaboration}},\ }\bibfield  {title} {\bibinfo {title} {{KAGRA}
  cryogenic suspension control toward the observation run 3},\ }\href
  {https://doi.org/10.1088/1742-6596/1468/1/012217   } {\bibfield  {journal}
  {\bibinfo  {journal} {Journal of Physics: Conference Series}\ }\textbf
  {\bibinfo {volume} {1468}},\ \bibinfo {pages} {012217} (\bibinfo {year}
  {2020})}\BibitemShut {NoStop}%
\bibitem [{\citenamefont {Busch}\ and\ \citenamefont
  {Grohmann}(2022{\natexlab{a}})}]{Busch_2022}%
  \BibitemOpen
  \bibfield  {author} {\bibinfo {author} {\bibfnamefont {L.}~\bibnamefont
  {Busch}}\ and\ \bibinfo {author} {\bibfnamefont {S.}~\bibnamefont
  {Grohmann}},\ }\bibfield  {title} {\bibinfo {title} {Conceptual layout of a
  helium cooling system for the {Einstein Telescope}},\ }\href
  {https://doi.org/10.1088/1757-899x/1240/1/012095   } {\bibfield  {journal}
  {\bibinfo  {journal} {{IOP} Conference Series: Materials Science and
  Engineering}\ }\textbf {\bibinfo {volume} {1240}},\ \bibinfo {pages} {012095}
  (\bibinfo {year} {2022}{\natexlab{a}})}\BibitemShut {NoStop}%
\bibitem [{\citenamefont {Busch}\ and\ \citenamefont
  {Grohmann}(2022{\natexlab{b}})}]{Busch2022_Elba}%
  \BibitemOpen
  \bibfield  {author} {\bibinfo {author} {\bibfnamefont {L.}~\bibnamefont
  {Busch}}\ and\ \bibinfo {author} {\bibfnamefont {S.}~\bibnamefont
  {Grohmann}},\ }\href {https://doi.org/10.5445/IR/1000153741   } {\bibinfo
  {title} {{Thermal design of the He-II suspension tube for ET-LF: Status and
  outlook}}},\ \bibinfo {howpublished} {Talk held at ECLOUD and GWDVac
  Workshops, Portoferraio, Italy} (\bibinfo {year}
  {2022}{\natexlab{b}})\BibitemShut {NoStop}%
\bibitem [{\citenamefont {Busch}\ \emph {et~al.}(2021)\citenamefont {Busch},
  \citenamefont {Koroveshi},\ and\ \citenamefont
  {Grohmann}}]{LennardGWADW2021}%
  \BibitemOpen
  \bibfield  {author} {\bibinfo {author} {\bibfnamefont {L.}~\bibnamefont
  {Busch}}, \bibinfo {author} {\bibfnamefont {X.}~\bibnamefont {Koroveshi}},\
  and\ \bibinfo {author} {\bibfnamefont {S.}~\bibnamefont {Grohmann}},\ }\href
  {https://doi.org/https://doi.org/10.5445/IR/1000132799   } {\bibinfo {title}
  {Helium-based cooling concept of the ET-LF interferometer}},\ \bibinfo
  {howpublished} {Talk held at the Gravitational Wave Advanced Detector
  Workshop (GWADW), Online} (\bibinfo {year} {2021})\BibitemShut {NoStop}%
\bibitem [{\citenamefont {Baudouy}(2011)}]{Baudouy2011}%
  \BibitemOpen
  \bibfield  {author} {\bibinfo {author} {\bibfnamefont {B.}~\bibnamefont
  {Baudouy}},\ }\bibfield  {title} {\bibinfo {title} {Low temperature thermal
  conductivity of aluminum alloy 1200},\ }\href
  {https://doi.org/10.1016/j.cryogenics.2011.09.002   } {\bibfield  {journal}
  {\bibinfo  {journal} {Cryogenics}\ }\textbf {\bibinfo {volume} {51}},\
  \bibinfo {pages} {617–620} (\bibinfo {year} {2011})}\BibitemShut {NoStop}%
\bibitem [{\citenamefont {Lemmon}\ \emph {et~al.}(2013)\citenamefont {Lemmon},
  \citenamefont {Huber},\ and\ \citenamefont {McLinden}}]{Lemmon-RP9.1}%
  \BibitemOpen
  \bibfield  {author} {\bibinfo {author} {\bibfnamefont {E.}~\bibnamefont
  {Lemmon}}, \bibinfo {author} {\bibfnamefont {M.}~\bibnamefont {Huber}},\ and\
  \bibinfo {author} {\bibfnamefont {M.}~\bibnamefont {McLinden}},\ }\href@noop
  {} {\bibinfo {title} {{NIST Standard Reference Database 23: Reference Fluid
  Thermodynamic and Transport Properties-REFPROP, Version 9.1}}} (\bibinfo
  {year} {2013})\BibitemShut {NoStop}%
\bibitem [{\citenamefont {Twu}\ \emph {et~al.}(1995)\citenamefont {Twu},
  \citenamefont {Coon},\ and\ \citenamefont {Cunningham}}]{TWU199549}%
  \BibitemOpen
  \bibfield  {author} {\bibinfo {author} {\bibfnamefont {C.~H.}\ \bibnamefont
  {Twu}}, \bibinfo {author} {\bibfnamefont {J.~E.}\ \bibnamefont {Coon}},\ and\
  \bibinfo {author} {\bibfnamefont {J.~R.}\ \bibnamefont {Cunningham}},\
  }\bibfield  {title} {\bibinfo {title} {{A new generalized alpha function for
  a cubic equation of state Part 1. Peng-Robinson equation}},\ }\href
  {https://doi.org/https://doi.org/10.1016/0378-3812(94)02601-V } {\bibfield
  {journal} {\bibinfo  {journal} {Fluid Phase Equilibria}\ }\textbf {\bibinfo
  {volume} {105}},\ \bibinfo {pages} {49} (\bibinfo {year} {1995})}\BibitemShut
  {NoStop}%
\bibitem [{\citenamefont {Yakhot}\ and\ \citenamefont
  {Orszag}(1986)}]{yakhot1986}%
  \BibitemOpen
  \bibfield  {author} {\bibinfo {author} {\bibfnamefont {V.}~\bibnamefont
  {Yakhot}}\ and\ \bibinfo {author} {\bibfnamefont {S.~A.}\ \bibnamefont
  {Orszag}},\ }\bibfield  {title} {\bibinfo {title} {{Renormalization group
  analysis of turbulence. I. Basic theory}},\ }\href
  {https://doi.org/10.1007/BF01061452 } {\bibfield  {journal} {\bibinfo
  {journal} {Journal of scientific computing}\ }\textbf {\bibinfo {volume}
  {1}},\ \bibinfo {pages} {3} (\bibinfo {year} {1986})}\BibitemShut {NoStop}%
\bibitem [{\citenamefont {Abrahamson}(2022)}]{Abrahamson2022}%
  \BibitemOpen
  \bibfield  {author} {\bibinfo {author} {\bibfnamefont {J.}~\bibnamefont
  {Abrahamson}},\ }\href@noop {} {\bibinfo {title} {{Ansys Fluent Theory
  Guide}}} (\bibinfo {year} {2022}),\ \bibinfo {note} {release 2022
  R1}\BibitemShut {NoStop}%
\bibitem [{\citenamefont {{Constancio Jr}}\ \emph {et~al.}(2020)\citenamefont
  {{Constancio Jr}}, \citenamefont {Adhikari}, \citenamefont {Aguiar},
  \citenamefont {Arai}, \citenamefont {Markowitz}, \citenamefont {Okada},\ and\
  \citenamefont {Wipf}}]{CONSTANCIOJR2020}%
  \BibitemOpen
  \bibfield  {author} {\bibinfo {author} {\bibfnamefont {M.}~\bibnamefont
  {{Constancio Jr}}}, \bibinfo {author} {\bibfnamefont {R.~X.}\ \bibnamefont
  {Adhikari}}, \bibinfo {author} {\bibfnamefont {O.~D.}\ \bibnamefont
  {Aguiar}}, \bibinfo {author} {\bibfnamefont {K.}~\bibnamefont {Arai}},
  \bibinfo {author} {\bibfnamefont {A.}~\bibnamefont {Markowitz}}, \bibinfo
  {author} {\bibfnamefont {M.~A.}\ \bibnamefont {Okada}},\ and\ \bibinfo
  {author} {\bibfnamefont {C.~C.}\ \bibnamefont {Wipf}},\ }\bibfield  {title}
  {\bibinfo {title} {Silicon emissivity as a function of temperature},\ }\href
  {https://doi.org/https://doi.org/10.1016/j.ijheatmasstransfer.2020.119863 }
  {\bibfield  {journal} {\bibinfo  {journal} {International Journal of Heat and
  Mass Transfer}\ }\textbf {\bibinfo {volume} {157}},\ \bibinfo {pages}
  {119863} (\bibinfo {year} {2020})}\BibitemShut {NoStop}%
\bibitem [{\citenamefont {Desai}(1986)}]{Desai1986}%
  \BibitemOpen
  \bibfield  {author} {\bibinfo {author} {\bibfnamefont {P.~D.}\ \bibnamefont
  {Desai}},\ }\bibfield  {title} {\bibinfo {title} {Thermodynamic properties of
  iron and silicon},\ }\href {https://doi.org/10.1063/1.555761       } {\bibfield
  {journal} {\bibinfo  {journal} {Journal of Physical and Chemical Reference
  Data}\ }\textbf {\bibinfo {volume} {15}},\ \bibinfo {pages} {967} (\bibinfo
  {year} {1986})}\BibitemShut {NoStop}%
\bibitem [{\citenamefont {Callen}\ and\ \citenamefont {Welton}(1951)}]{callen}%
  \BibitemOpen
  \bibfield  {author} {\bibinfo {author} {\bibfnamefont {H.~B.}\ \bibnamefont
  {Callen}}\ and\ \bibinfo {author} {\bibfnamefont {T.~A.}\ \bibnamefont
  {Welton}},\ }\bibfield  {title} {\bibinfo {title} {Irreversibility and
  generalized noise},\ }\href {https://doi.org/10.1103/PhysRev.83.34         }
  {\bibfield  {journal} {\bibinfo  {journal} {Phys. Rev.}\ }\textbf {\bibinfo
  {volume} {83}},\ \bibinfo {pages} {34} (\bibinfo {year} {1951})}\BibitemShut
  {NoStop}%
\bibitem [{\citenamefont {Saulson}(1994)}]{Saulsonbook}%
  \BibitemOpen
  \bibfield  {author} {\bibinfo {author} {\bibfnamefont {P.~R.}\ \bibnamefont
  {Saulson}},\ }\href@noop {} {\emph {\bibinfo {title} {Fundamentals of
  Interferometric Gravitational Wave Detectors}}}\ (\bibinfo  {publisher}
  {World Scientific},\ \bibinfo {year} {1994})\BibitemShut {NoStop}%
\bibitem [{\citenamefont {Levin}(1998)}]{Levin}%
  \BibitemOpen
  \bibfield  {author} {\bibinfo {author} {\bibfnamefont {Y.}~\bibnamefont
  {Levin}},\ }\bibfield  {title} {\bibinfo {title} {{Internal thermal noise in
  the LIGO test masses: A direct approach}},\ }\href
  {https://doi.org/10.1103/PhysRevD.57.659      } {\bibfield  {journal} {\bibinfo
  {journal} {Phys. Rev. D}\ }\textbf {\bibinfo {volume} {57}},\ \bibinfo
  {pages} {659} (\bibinfo {year} {1998})}\BibitemShut {NoStop}%
\bibitem [{\citenamefont {Gonz{\'{a}}lez}\ and\ \citenamefont
  {Saulson}(1994)}]{Gonzalez1994}%
  \BibitemOpen
  \bibfield  {author} {\bibinfo {author} {\bibfnamefont {G.~I.}\ \bibnamefont
  {Gonz{\'{a}}lez}}\ and\ \bibinfo {author} {\bibfnamefont {P.~R.}\ \bibnamefont
  {Saulson}},\ }\bibfield  {title} {\bibinfo {title} {Brownian motion of a mass
  suspended by an anelastic wire},\ }\href {https://doi.org/10.1121/1.410467        }
  {\bibfield  {journal} {\bibinfo  {journal} {The Journal of the Acoustical
  Society of America}\ }\textbf {\bibinfo {volume} {96}},\ \bibinfo {pages}
  {207} (\bibinfo {year} {1994})}\BibitemShut {NoStop}%
\bibitem [{\citenamefont {Gonz{\'{a}}lez}(2000)}]{Gonz_lez_2000}%
  \BibitemOpen
  \bibfield  {author} {\bibinfo {author} {\bibfnamefont {G.}~\bibnamefont
  {Gonz{\'{a}}lez}},\ }\bibfield  {title} {\bibinfo {title} {Suspensions
  thermal noise in the {LIGO} gravitational wave detector},\ }\href
  {https://doi.org/10.1088/0264-9381/17/21/305      } {\bibfield  {journal} {\bibinfo
   {journal} {Classical and Quantum Gravity}\ }\textbf {\bibinfo {volume}
  {17}},\ \bibinfo {pages} {4409} (\bibinfo {year} {2000})}\BibitemShut
  {NoStop}%
\bibitem [{\citenamefont {Majorana}\ and\ \citenamefont
  {Ogawa}(1997)}]{MAJORANA1997162}%
  \BibitemOpen
  \bibfield  {author} {\bibinfo {author} {\bibfnamefont {E.}~\bibnamefont
  {Majorana}}\ and\ \bibinfo {author} {\bibfnamefont {Y.}~\bibnamefont
  {Ogawa}},\ }\bibfield  {title} {\bibinfo {title} {Mechanical thermal noise in
  coupled oscillators},\ }\href
  {https://doi.org/https://doi.org/10.1016/S0375-9601(97)00458-1   } {\bibfield
  {journal} {\bibinfo  {journal} {Physics Letters A}\ }\textbf {\bibinfo
  {volume} {233}},\ \bibinfo {pages} {162} (\bibinfo {year}
  {1997})}\BibitemShut {NoStop}%
\bibitem [{\citenamefont {Cagnoli}\ \emph {et~al.}(2000)\citenamefont
  {Cagnoli}, \citenamefont {Hough}, \citenamefont {DeBra}, \citenamefont
  {Fejer}, \citenamefont {Gustafson}, \citenamefont {Rowan},\ and\
  \citenamefont {Mitrofanov}}]{CAGNOLI200039}%
  \BibitemOpen
  \bibfield  {author} {\bibinfo {author} {\bibfnamefont {G.}~\bibnamefont
  {Cagnoli}}, \bibinfo {author} {\bibfnamefont {J.}~\bibnamefont {Hough}},
  \bibinfo {author} {\bibfnamefont {D.}~\bibnamefont {DeBra}}, \bibinfo
  {author} {\bibfnamefont {M.}~\bibnamefont {Fejer}}, \bibinfo {author}
  {\bibfnamefont {E.}~\bibnamefont {Gustafson}}, \bibinfo {author}
  {\bibfnamefont {S.}~\bibnamefont {Rowan}},\ and\ \bibinfo {author}
  {\bibfnamefont {V.}~\bibnamefont {Mitrofanov}},\ }\bibfield  {title}
  {\bibinfo {title} {Damping dilution factor for a pendulum in an
  interferometric gravitational waves detector},\ }\href
  {https://doi.org/https://doi.org/10.1016/S0375-9601(00)00411-4  } {\bibfield
  {journal} {\bibinfo  {journal} {Physics Letters A}\ }\textbf {\bibinfo
  {volume} {272}},\ \bibinfo {pages} {39} (\bibinfo {year} {2000})}\BibitemShut
  {NoStop}%
\bibitem [{\citenamefont {Rowan}\ \emph {et~al.}(2000)\citenamefont {Rowan},
  \citenamefont {Cagnoli}, \citenamefont {Sneddon}, \citenamefont {Hough},
  \citenamefont {Route}, \citenamefont {Gustafson}, \citenamefont {Fejer},\
  and\ \citenamefont {Mitrofanov}}]{Rowan2000}%
  \BibitemOpen
  \bibfield  {author} {\bibinfo {author} {\bibfnamefont {S.}~\bibnamefont
  {Rowan}}, \bibinfo {author} {\bibfnamefont {G.}~\bibnamefont {Cagnoli}},
  \bibinfo {author} {\bibfnamefont {P.}~\bibnamefont {Sneddon}}, \bibinfo
  {author} {\bibfnamefont {J.}~\bibnamefont {Hough}}, \bibinfo {author}
  {\bibfnamefont {R.}~\bibnamefont {Route}}, \bibinfo {author} {\bibfnamefont
  {E.}~\bibnamefont {Gustafson}}, \bibinfo {author} {\bibfnamefont
  {M.}~\bibnamefont {Fejer}},\ and\ \bibinfo {author} {\bibfnamefont
  {V.}~\bibnamefont {Mitrofanov}},\ }\bibfield  {title} {\bibinfo {title}
  {Investigation of mechanical loss factors of some candidate materials for the
  test masses of gravitational wave detectors},\ }\href
  {https://doi.org/https://doi.org/10.1016/S0375-9601(99)00874-9 } {\bibfield
  {journal} {\bibinfo  {journal} {Physics Letters A}\ }\textbf {\bibinfo
  {volume} {265}},\ \bibinfo {pages} {5} (\bibinfo {year} {2000})}\BibitemShut
  {NoStop}%
\bibitem [{\citenamefont {Schr\"oter}(2008)}]{AnjaDiss}%
  \BibitemOpen
  \bibfield  {author} {\bibinfo {author} {\bibfnamefont {A.}~\bibnamefont
  {Schr\"oter}},\ }\href
  {https://www.db-thueringen.de/servlets/MCRFileNodeServlet/dbt_derivate_00013798/Dissertation.pdf}
  {\bibinfo {title} {{ Dissertation: Mechanical losses in materials for future
  cryogenic gravitational wave detectors }}} (\bibinfo {year}
  {2008})\BibitemShut {NoStop}%
\bibitem [{\citenamefont {Braginsky}\ \emph {et~al.}(1985)\citenamefont
  {Braginsky}, \citenamefont {Mitrofanov},\ and\ \citenamefont
  {Panov}}]{Braginsky}%
  \BibitemOpen
  \bibfield  {author} {\bibinfo {author} {\bibfnamefont {V.~B.}\ \bibnamefont
  {Braginsky}}, \bibinfo {author} {\bibfnamefont {V.}~\bibnamefont
  {Mitrofanov}},\ and\ \bibinfo {author} {\bibfnamefont {V.~I.}\ \bibnamefont
  {Panov}},\ }\href@noop {} {\emph {\bibinfo {title} {Systems with Small
  Dissipation}}}\ (\bibinfo  {publisher} {University of Chicago Press},\
  \bibinfo {year} {1985})\BibitemShut {NoStop}%
\bibitem [{\citenamefont {Uchiyama}\ \emph {et~al.}(1999)\citenamefont
  {Uchiyama} \emph {et~al.}}]{UCHIYAMA19995}%
  \BibitemOpen
  \bibfield  {author} {\bibinfo {author} {\bibfnamefont {T.}~\bibnamefont
  {Uchiyama}} \emph {et~al.},\ }\bibfield  {title} {\bibinfo {title}
  {Mechanical quality factor of a cryogenic sapphire test mass for
  gravitational wave detectors},\ }\href
  {https://doi.org/https://doi.org/10.1016/S0375-9601(99)00563-0 } {\bibfield
  {journal} {\bibinfo  {journal} {Physics Letters A}\ }\textbf {\bibinfo
  {volume} {261}},\ \bibinfo {pages} {5} (\bibinfo {year} {1999})}\BibitemShut
  {NoStop}%
\bibitem [{\citenamefont {Majorana}\ \emph {et~al.}(1992)\citenamefont
  {Majorana}, \citenamefont {Rapagnani},\ and\ \citenamefont
  {Ricci}}]{Majorana_1992}%
  \BibitemOpen
  \bibfield  {author} {\bibinfo {author} {\bibfnamefont {E.}~\bibnamefont
  {Majorana}}, \bibinfo {author} {\bibfnamefont {P.}~\bibnamefont
  {Rapagnani}},\ and\ \bibinfo {author} {\bibfnamefont {F.}~\bibnamefont
  {Ricci}},\ }\bibfield  {title} {\bibinfo {title} {Test facility for resonance
  transducers of cryogenic gravitational wave antennas},\ }\href
  {https://doi.org/10.1088/0957-0233/3/5/010} {\bibfield  {journal} {\bibinfo
  {journal} {Measurement Science and Technology}\ }\textbf {\bibinfo {volume}
  {3}},\ \bibinfo {pages} {501} (\bibinfo {year} {1992})}\BibitemShut {NoStop}%
\bibitem [{\citenamefont {Duffy}(1990)}]{DuffyAluminium90}%
  \BibitemOpen
  \bibfield  {author} {\bibinfo {author} {\bibfnamefont {W.}~\bibnamefont
  {Duffy}},\ }\bibfield  {title} {\bibinfo {title} {Acoustic quality factor of
  aluminum alloys from 50 {mK to 300 K}},\ }\href
  {https://doi.org/10.1063/1.346971} {\bibfield  {journal} {\bibinfo  {journal}
  {Journal of Applied Physics}\ }\textbf {\bibinfo {volume} {68}},\ \bibinfo
  {pages} {5601} (\bibinfo {year} {1990})}\BibitemShut {NoStop}%
\bibitem [{\citenamefont {Coccia}\ and\ \citenamefont
  {Niinikoski}(1984)}]{coccia}%
  \BibitemOpen
  \bibfield  {author} {\bibinfo {author} {\bibfnamefont {E.}~\bibnamefont
  {Coccia}}\ and\ \bibinfo {author} {\bibfnamefont {T.}~\bibnamefont
  {Niinikoski}},\ }\bibfield  {title} {\bibinfo {title} {Acoustic quality
  factor of an aluminium alloy for gravitational wave antennae below{ 1 K}},\
  }\href {https://doi.org/https://doi.org/10.1007/BF02747011} {\bibfield
  {journal} {\bibinfo  {journal} {Lett. Nuovo Cimento}\ }\textbf {\bibinfo
  {volume} {41}},\ \bibinfo {pages} {242–246} (\bibinfo {year}
  {1984})}\BibitemShut {NoStop}%
\bibitem [{\citenamefont {Astone}\ \emph {et~al.}(1993)\citenamefont {Astone}
  \emph {et~al.}}]{Explorerpiero}%
  \BibitemOpen
  \bibfield  {author} {\bibinfo {author} {\bibfnamefont {P.}~\bibnamefont
  {Astone}} \emph {et~al.},\ }\bibfield  {title} {\bibinfo {title} {Long-term
  operation of the {Rome "Explorer"} cryogenic gravitational wave detector},\
  }\href {https://doi.org/10.1103/PhysRevD.47.362} {\bibfield  {journal}
  {\bibinfo  {journal} {Phys. Rev. D}\ }\textbf {\bibinfo {volume} {47}},\
  \bibinfo {pages} {362} (\bibinfo {year} {1993})}\BibitemShut {NoStop}%
\bibitem [{\citenamefont {Zener}(1938)}]{zener}%
  \BibitemOpen
  \bibfield  {author} {\bibinfo {author} {\bibfnamefont {C.}~\bibnamefont
  {Zener}},\ }\bibfield  {title} {\bibinfo {title} {{Internal Friction in
  Solids II. General Theory of Thermoelastic Internal Friction}},\ }\href
  {https://doi.org/10.1103/PhysRev.53.90} {\bibfield  {journal} {\bibinfo
  {journal} {Phys. Rev.}\ }\textbf {\bibinfo {volume} {53}},\ \bibinfo {pages}
  {90} (\bibinfo {year} {1938})}\BibitemShut {NoStop}%
\bibitem [{\citenamefont {Cagnoli}\ and\ \citenamefont
  {Willems}(2002)}]{Cagnoli2002}%
  \BibitemOpen
  \bibfield  {author} {\bibinfo {author} {\bibfnamefont {G.}~\bibnamefont
  {Cagnoli}}\ and\ \bibinfo {author} {\bibfnamefont {P.~A.}\ \bibnamefont
  {Willems}},\ }\bibfield  {title} {\bibinfo {title} {Effects of nonlinear
  thermoelastic damping in highly stressed fibers},\ }\href
  {https://doi.org/10.1103/PhysRevB.65.174111 } {\bibfield  {journal} {\bibinfo
  {journal} {Phys. Rev. B}\ }\textbf {\bibinfo {volume} {65}},\ \bibinfo
  {pages} {174111} (\bibinfo {year} {2002})}\BibitemShut {NoStop}%
\bibitem [{\citenamefont {Nawrodt}\ \emph {et~al.}(2011)\citenamefont
  {Nawrodt}, \citenamefont {Rowan}, \citenamefont {Hough}, \citenamefont
  {Punturo}, \citenamefont {Ricci},\ and\ \citenamefont
  {Vinet}}]{Nawrodt.2011}%
  \BibitemOpen
  \bibfield  {author} {\bibinfo {author} {\bibfnamefont {R.}~\bibnamefont
  {Nawrodt}}, \bibinfo {author} {\bibfnamefont {S.}~\bibnamefont {Rowan}},
  \bibinfo {author} {\bibfnamefont {J.}~\bibnamefont {Hough}}, \bibinfo
  {author} {\bibfnamefont {M.}~\bibnamefont {Punturo}}, \bibinfo {author}
  {\bibfnamefont {F.}~\bibnamefont {Ricci}},\ and\ \bibinfo {author}
  {\bibfnamefont {J.}~\bibnamefont {Vinet}},\ }\bibfield  {title} {\bibinfo
  {title} {Challenges in thermal noise for 3rd generation of gravitational wave
  detectors},\ }\href {https://doi.org/10.1007/s10714-010-1066-5  } {\bibfield
  {journal} {\bibinfo  {journal} {General Relativity and Gravitation}\ }\textbf
  {\bibinfo {volume} {43}},\ \bibinfo {pages} {593} (\bibinfo {year}
  {2011})}\BibitemShut {NoStop}%
\bibitem [{\citenamefont {Gretarsson}\ \emph {et~al.}(2000)\citenamefont
  {Gretarsson}, \citenamefont {Harry}, \citenamefont {Penn}, \citenamefont
  {Saulson}, \citenamefont {Startin}, \citenamefont {Rowan}, \citenamefont
  {Cagnoli},\ and\ \citenamefont {Hough}}]{GRETARSSON2000108}%
  \BibitemOpen
  \bibfield  {author} {\bibinfo {author} {\bibfnamefont {A.~M.}\ \bibnamefont
  {Gretarsson}}, \bibinfo {author} {\bibfnamefont {G.~M.}\ \bibnamefont
  {Harry}}, \bibinfo {author} {\bibfnamefont {S.~D.}\ \bibnamefont {Penn}},
  \bibinfo {author} {\bibfnamefont {P.~R.}\ \bibnamefont {Saulson}}, \bibinfo
  {author} {\bibfnamefont {W.~J.}\ \bibnamefont {Startin}}, \bibinfo {author}
  {\bibfnamefont {S.}~\bibnamefont {Rowan}}, \bibinfo {author} {\bibfnamefont
  {G.}~\bibnamefont {Cagnoli}},\ and\ \bibinfo {author} {\bibfnamefont
  {J.}~\bibnamefont {Hough}},\ }\bibfield  {title} {\bibinfo {title} {Pendulum
  mode thermal noise in advanced interferometers: a comparison of fused silica
  fibers and ribbons in the presence of surface loss},\ }\href
  {https://doi.org/https://doi.org/10.1016/S0375-9601(00)00295-4  } {\bibfield
  {journal} {\bibinfo  {journal} {Physics Letters A}\ }\textbf {\bibinfo
  {volume} {270}},\ \bibinfo {pages} {108} (\bibinfo {year}
  {2000})}\BibitemShut {NoStop}%
\bibitem [{\citenamefont {Cumming}\ \emph {et~al.}(2020)\citenamefont
  {Cumming}, \citenamefont {Sorazu}, \citenamefont {Daw}, \citenamefont
  {Hammond}, \citenamefont {Hough}, \citenamefont {Jones}, \citenamefont
  {Martin}, \citenamefont {Rowan}, \citenamefont {Strain},\ and\ \citenamefont
  {Williams}}]{Cumming_2020}%
  \BibitemOpen
  \bibfield  {author} {\bibinfo {author} {\bibfnamefont {A.~V.}\ \bibnamefont
  {Cumming}}, \bibinfo {author} {\bibfnamefont {B.}~\bibnamefont {Sorazu}},
  \bibinfo {author} {\bibfnamefont {E.}~\bibnamefont {Daw}}, \bibinfo {author}
  {\bibfnamefont {G.~D.}\ \bibnamefont {Hammond}}, \bibinfo {author}
  {\bibfnamefont {J.}~\bibnamefont {Hough}}, \bibinfo {author} {\bibfnamefont
  {R.}~\bibnamefont {Jones}}, \bibinfo {author} {\bibfnamefont {I.~W.}\
  \bibnamefont {Martin}}, \bibinfo {author} {\bibfnamefont {S.}~\bibnamefont
  {Rowan}}, \bibinfo {author} {\bibfnamefont {K.~A.}\ \bibnamefont {Strain}},\
  and\ \bibinfo {author} {\bibfnamefont {D.}~\bibnamefont {Williams}},\
  }\bibfield  {title} {\bibinfo {title} {Lowest observed surface and weld
  losses in fused silica fibres for gravitational wave detectors},\ }\href
  {https://doi.org/10.1088/1361-6382/abac42  } {\bibfield  {journal} {\bibinfo
  {journal} {Classical and Quantum Gravity}\ }\textbf {\bibinfo {volume}
  {37}},\ \bibinfo {pages} {195019} (\bibinfo {year} {2020})}\BibitemShut
  {NoStop}%
\bibitem [{\citenamefont {Piergiovanni}\ \emph {et~al.}(2009)\citenamefont
  {Piergiovanni}, \citenamefont {Punturo},\ and\ \citenamefont
  {Puppo}}]{pppeffect}%
  \BibitemOpen
  \bibfield  {author} {\bibinfo {author} {\bibfnamefont {F.}~\bibnamefont
  {Piergiovanni}}, \bibinfo {author} {\bibfnamefont {M.}~\bibnamefont
  {Punturo}},\ and\ \bibinfo {author} {\bibfnamefont {P.}~\bibnamefont
  {Puppo}},\ }\href {https://tds.virgo-gw.eu/ql/?c=2199} {\bibinfo {title}
  {{The thermal noise of the Virgo+ and Virgo Advanced Last Stage Suspension
  (The PPP Effect)}}} (\bibinfo {year} {2009})\BibitemShut {NoStop}%
\bibitem [{\citenamefont {Somiya}(2011)}]{somiya}%
  \BibitemOpen
  \bibfield  {author} {\bibinfo {author} {\bibfnamefont {K.}~\bibnamefont
  {Somiya}},\ }\href
  {https://www.academia.edu/44046909/Suspension_thermal_noise_reduction_in_a_cryogenic_interferometer}
  {\bibinfo {title} {Suspension thermal noise reduction in a cryogenic
  interferometer}} (\bibinfo {year} {2011})\BibitemShut {NoStop}%
\bibitem [{\citenamefont {Bondu}\ \emph {et~al.}(1998)\citenamefont {Bondu},
  \citenamefont {Hello},\ and\ \citenamefont {Vinet}}]{BONDU1998227}%
  \BibitemOpen
  \bibfield  {author} {\bibinfo {author} {\bibfnamefont {F.}~\bibnamefont
  {Bondu}}, \bibinfo {author} {\bibfnamefont {P.}~\bibnamefont {Hello}},\ and\
  \bibinfo {author} {\bibfnamefont {J.}~\bibnamefont {Vinet}},\ }\bibfield
  {title} {\bibinfo {title} {Thermal noise in mirrors of interferometric
  gravitational wave antennas},\ }\href
  {https://doi.org/https://doi.org/10.1016/S0375-9601(98)00450-2     } {\bibfield
  {journal} {\bibinfo  {journal} {Physics Letters A}\ }\textbf {\bibinfo
  {volume} {246}},\ \bibinfo {pages} {227} (\bibinfo {year}
  {1998})}\BibitemShut {NoStop}%
\bibitem [{\citenamefont {Komori}\ \emph {et~al.}(2018)\citenamefont {Komori},
  \citenamefont {Enomoto}, \citenamefont {Takeda}, \citenamefont {Michimura},
  \citenamefont {Somiya}, \citenamefont {Ando},\ and\ \citenamefont
  {Ballmer}}]{PhysRevD.97.102001}%
  \BibitemOpen
  \bibfield  {author} {\bibinfo {author} {\bibfnamefont {K.}~\bibnamefont
  {Komori}}, \bibinfo {author} {\bibfnamefont {Y.}~\bibnamefont {Enomoto}},
  \bibinfo {author} {\bibfnamefont {H.}~\bibnamefont {Takeda}}, \bibinfo
  {author} {\bibfnamefont {Y.}~\bibnamefont {Michimura}}, \bibinfo {author}
  {\bibfnamefont {K.}~\bibnamefont {Somiya}}, \bibinfo {author} {\bibfnamefont
  {M.}~\bibnamefont {Ando}},\ and\ \bibinfo {author} {\bibfnamefont {S.~W.}\
  \bibnamefont {Ballmer}},\ }\bibfield  {title} {\bibinfo {title} {Direct
  approach for the fluctuation-dissipation theorem under nonequilibrium
  steady-state conditions},\ }\href
  {https://doi.org/10.1103/PhysRevD.97.102001} {\bibfield  {journal} {\bibinfo
  {journal} {Phys. Rev. D}\ }\textbf {\bibinfo {volume} {97}},\ \bibinfo
  {pages} {102001} (\bibinfo {year} {2018})}\BibitemShut {NoStop}%
\bibitem [{\citenamefont {Gorter}\ and\ \citenamefont
  {Mellik}(1949)}]{Gorter-Mellink-1949}%
  \BibitemOpen
  \bibfield  {author} {\bibinfo {author} {\bibfnamefont {C.~J.}\ \bibnamefont
  {Gorter}}\ and\ \bibinfo {author} {\bibfnamefont {J.}~\bibnamefont
  {Mellik}},\ }\bibfield  {title} {\bibinfo {title} {On the irreversible
  processes in liquid helium ii},\ }\href
  {https://doi.org/10.1016/0031-8914(49)90105-6       } {\bibfield  {journal}
  {\bibinfo  {journal} {Physica}\ }\textbf {\bibinfo {volume} {15}},\ \bibinfo
  {pages} {285} (\bibinfo {year} {1949})}\BibitemShut {NoStop}%
\bibitem [{\citenamefont {Feynman}(1955)}]{Feynman-1955}%
  \BibitemOpen
  \bibfield  {author} {\bibinfo {author} {\bibfnamefont {R.~P.}\ \bibnamefont
  {Feynman}},\ }\bibinfo {title} {Application of quantum mechanics to liquid
  helium},\ in\ \href@noop {} {\emph {\bibinfo {booktitle} {Progress in Low
  Temperature Physics, edited by Gorter C. J.}}},\ Vol.~\bibinfo {volume} {1}\
  (\bibinfo  {publisher} {North-Holland Publications},\ \bibinfo {address}
  {Amsterdam},\ \bibinfo {year} {1955})\ pp.\ \bibinfo {pages}
  {17--53}\BibitemShut {NoStop}%
\bibitem [{\citenamefont {Koroveshi}\ \emph {et~al.}(2022)\citenamefont
  {Koroveshi}, \citenamefont {Grohmann}, \citenamefont {Rapagnani},\ and\
  \citenamefont {Mangano}}]{Koroveshi_Elba2}%
  \BibitemOpen
  \bibfield  {author} {\bibinfo {author} {\bibfnamefont {X.}~\bibnamefont
  {Koroveshi}}, \bibinfo {author} {\bibfnamefont {S.}~\bibnamefont {Grohmann}},
  \bibinfo {author} {\bibfnamefont {P.}~\bibnamefont {Rapagnani}},\ and\
  \bibinfo {author} {\bibfnamefont {V.}~\bibnamefont {Mangano}},\ }\href
  {https://doi.org/10.5445/IR/1000153742       } {\bibinfo {title} {{Experimental
  plans to validate the He-II based payload cooling concept}}},\ \bibinfo
  {howpublished} {Talk held at ECLOUD and GWDVac'22 Workshops, Portoferraio,
  Italy} (\bibinfo {year} {2022})\BibitemShut {NoStop}%
\bibitem [{\citenamefont {Fisher}\ and\ \citenamefont
  {Renken}(1964)}]{betaTi_Fisher}%
  \BibitemOpen
  \bibfield  {author} {\bibinfo {author} {\bibfnamefont {E.~S.}\ \bibnamefont
  {Fisher}}\ and\ \bibinfo {author} {\bibfnamefont {C.~J.}\ \bibnamefont
  {Renken}},\ }\bibfield  {title} {\bibinfo {title} {{Single-Crystal Elastic
  Moduli and the hcp \ensuremath{\rightarrow} bcc Transformation in Ti, Zr, and
  Hf}},\ }\href {https://doi.org/10.1103/PhysRev.135.A482            } {\bibfield
  {journal} {\bibinfo  {journal} {Phys. Rev.}\ }\textbf {\bibinfo {volume}
  {135}},\ \bibinfo {pages} {A482} (\bibinfo {year} {1964})}\BibitemShut
  {NoStop}%
\bibitem [{\citenamefont {{Di Pace}}\ \emph {et~al.}(2022)\citenamefont {{Di
  Pace}} \emph {et~al.}}]{ETfacilities-2022}%
  \BibitemOpen
  \bibfield  {author} {\bibinfo {author} {\bibfnamefont {S.}~\bibnamefont {{Di
  Pace}}} \emph {et~al.},\ }\bibfield  {title} {\bibinfo {title} {{Research
  Facilities for Europe’s Next Generation Gravitational-Wave Detector
  Einstein Telescope}},\ }\href
  {https://doi.org/https://doi.org/10.3390/galaxies10030065         } {\bibfield
  {journal} {\bibinfo  {journal} {Galaxies}\ }\textbf {\bibinfo {volume}
  {10}},\ \bibinfo {pages} {1} (\bibinfo {year} {2022})}\BibitemShut {NoStop}%
\end{thebibliography}
\providecommand{\noopsort}[1]{}\providecommand{\singleletter}[1]{#1}%

\end{document}